\documentclass[12pt]{article}
\usepackage{a4}
\usepackage{graphicx}
\usepackage{epsf}
\usepackage{amssymb}
\usepackage{cite}
\newcommand{\be}{\begin{equation}}
\newcommand{\ee}{\end{equation}}
\newcommand{\bea}{\begin{eqnarray}}
\newcommand{\eea}{\end{eqnarray}}
\newcommand{\pt}{\partial}

\begin{document}
\def\C{{\mathbb{C}}}
\def\R{{\mathbb{R}}}
\def\s{{\mathbb{S}}}
\def\T{{\mathbb{T}}}
\def\Z{{\mathbb{Z}}}
\def\W{{\mathbb{W}}}
\def\Bbb{\mathbb}
\def\BZ{\Bbb Z} \def\BR{\Bbb R}
\def\BW{\Bbb W}
\def\BM{\Bbb M}
\def\BC{\Bbb C} \def\BP{\Bbb P}
\def\CP{\BC\BP}
\begin{titlepage}
\title{Thermodynamic Geometry and Phase Transitions in Kerr-Newman-AdS Black Holes}
\author{}
\date{
Anurag Sahay, Tapobrata Sarkar, Gautam Sengupta
\thanks{\noindent E--mail:~ ashaya, tapo, sengupta @iitk.ac.in}
\vskip0.4cm
{\sl Department of Physics, \\
Indian Institute of Technology,\\
Kanpur 208016, \\
India}}
\maketitle
\abstract{
\noindent
We investigate phase transitions and critical phenomena  in Kerr-Newman-Anti de Sitter black holes in the framework of the geometry of their
equilibrium thermodynamic state space.  The scalar curvature of these state space Riemannian geometries is computed in various ensembles.
The scalar curvature diverges at the critical point of second order phase transitions for these systems. Remarkably, however, we show that the state space
scalar curvature also carries information about the liquid-gas like first order phase transitions and the consequent instabilities and phase
coexistence for these black holes. This is encoded in the turning point behavior and the multi-valued branched structure
of the scalar curvature in the neighborhood of these first order phase transitions. We re-examine this first for the conventional
Van der Waals system, as a preliminary exercise. Subsequently, we study the Kerr-Newman-AdS black holes for a grand canonical and two
``mixed'' ensembles and establish novel phase structures. The state space scalar curvature bears out our assertion
for the first order phase transitions for both the known and the new phase structures, and closely resembles the Van der Waals system.}
\end{titlepage}

\section{Introduction}

Black Hole thermodynamics has emerged as an arena for testing candidate theories of quantum gravity, and
has thus witnessed intensely focussed interest in recent times (see e.g. \cite{td1},\cite{td2},\cite{td3} and references therein). 
It has been recognized from a semi classical analysis that they are thermodynamic objects with a
Bekenstein-Hawking entropy and a Hawking temperature \cite{td4},\cite{td5}. 
Although an exact microscopic statistical basis underlying this thermodynamic description requires a fully consistent
quantum theory of gravity, yet this has yielded a rich and varied structure of phase transitions and critical phenomena 
(for a small sampling of this extensive field of research, see, e.g. \cite{td6},\cite{td7},\cite{td8},\cite{td9}). 

It is however
well known that black hole thermodynamics is considerably different from usual thermodynamic systems. The macroscopic entropy is non extensive and for
conventional asymptotically flat black holes, negativity
 of specific heats in certain regions of the parameter space renders the canonical ensemble unstable. Furthermore the third law is invalid as zero temperature extremal
 black holes exhibit macroscopic degeneracy and consequently a nonzero entropy. However in spite of these departures, many of which have been addressed
 extensively in the literature, the
 study of black hole thermodynamics has led to important insights. For diverse black holes, specific heats seem to change signs from one region to another and
  exhibits divergences similar to the phenomena of phase transitions in conventional thermodynamic systems. Although the
 issue of phase transitions in black holes is still far from being completely settled, conventional thermodynamic analysis has elucidated a rich phase structure and 
 critical phenomena
 very similar to that of usual thermodynamic systems. Some of the critical indices for certain black hole and black ring solutions have also been determined from 
 standard scaling arguments \cite{td10},\cite{td11},\cite{td12}, \cite{td12a}. It must however be
 pointed out that a lack of exact knowledge about the microscopic statistical framework underlying black hole thermodynamics
 renders the issue of the choice of ensembles to be an extremely
subtle issue. In fact it has been established that the phase structure of black holes depend crucially on the choice of the ensemble in contrast to 
conventional thermodynamic systems \cite{td13},\cite{td14}.

The focus on this field has become more intense over the last decade in the light of developments in string theory and a better
grasp of the underlying microscopic statistical basis. The AdS-CFT correspondence
between a bulk gravity theory in AdS spaces and a gauge theory on the boundary of the AdS space has been one of the main theoretical advances in
string theory. In particular this has led to a correspondence between
Hawking-Page phase transitions in asymptotically AdS black holes and confinement-deconfinement phase transitions \cite{malda} 
in the boundary field theory which powered an intense focus on the thermodynamics of AdS black holes, their phase structures and critical phenomena.
The fact that AdS spaces act like a box with a specific length scale and reflecting walls renders the thermodynamics to be stable and a stronger basis is 
provided by the correspondence with the boundary field theory. A comprehensive analysis
has resulted in a rich phase structure for asymptotically AdS black holes which is very similar to conventional thermodynamic
systems. For example, it has been observed that charged Reissner-Nordstrom-AdS (RN-AdS) black holes show a first order liquid-gas like 
phase transition in the fixed charge canonical
ensemble analogous to the Van der Waals-Maxwell gas whereas a Hawking-Page phase transition is observed in the grand canonical ensemble \cite{td13}.

In a different context, a geometrical perspective of thermodynamics has been an area of interest for several decades. Following an early treatment due to Tisza 
\cite{tisza} and further extended by Callen \cite{callen}, the extrinsic geometric description of thermodynamics has been elucidated. However an 
intrinsic geometrical realization of
thermodynamic behaviour was rather elusive. Weinhold \cite{weinhold}, in an early investigation, attributed a Riemannian metric to the equilibrium state space of a
thermodynamic system in terms of the Hessian matrix
 of the derivatives of the internal energy as a function of the entropy and other extensive variables. However this did not have a clear notion of a distance
 in the state space from a physical perspective. Ruppeiner \cite{rupp} later introduced a similar metric in the equilibrium thermodynamic state space 
 based on the Hessian matrix of the entropy density as a function of the
 internal energy and other extensive variables. A physical interpretation of this geometric structure involved the consideration of classical fluctuation theory
 together with the laws of thermodynamics. The probability distribution of thermodynamic fluctuations between two equilibrium states in the Gaussian
 approximation could then be connected to the notion of the invariant distance between them in the 
state space. Remarkably enough, the curvature of the state space could be related to interactions in the underlying statistical system
and scaled as the correlation volume, diverging at critical points of second order phase transitions. This relation with classical fluctuation theory and the 
divergence of the state space scalar curvature at critical points has made
thermodynamic geometry an interesting approach to study thermodynamic systems in diverse areas. 
Since its inception this geometrical framework has been applied extensively to study systems
such as Van der Waals gas, Fermi gas, Ising Model, paramagnets etc. and has provided
a consistent and direct method to study second order  phase transitions and critical phenomena in these systems. However this geometric approach seemed
insensitive to the interesting phenomena of first order phase transitions involving a finite correlation length and a finite discontinuity in the specific heat and entropy.
One of our aims in this paper would be to address this crucial issue in the formalism of thermodynamic geometries in the context of KN-AdS black holes.

The first allusion to black hole thermodynamics in the framework of their state space geometry was made by Ferrara et al \cite{ferrara} in their seminal work on black 
hole attractors in the context of extremal black holes in string theory and its relation to the underlying moduli space. Following this,
there has been considerable amount of interest in this field starting with the work of \cite{aman},\cite{caibtz} in studying the thermodynamic geometries of 
diverse non extremal and extremal black holes. Several classes of black holes were shown to posess
curved state space geometries, indicating
an interacting microscopic statistical basis. More interestingly the divergences of the curvature scalar encoded the critical points of second order phase 
transitions in these systems \cite{cai1}.

In this connection, we had earlier studied a family of non extremal BTZ and BTZ-CS black holes \cite{ts1}, and shown that these state space geometries were flat. 
Thermal corrections to the canonical entropy induced negligibly small curvature to the state space. 
Subsequently, we studied several examples of charged extremal black holes in string theory, and 
showed that their state space geometries were flat. They were also shown to be insensitive also to higher derivative corrections in the large charge limit. 
This exhibited the stability of such extremal BPS configurations and the regularity of the curvature scalar in the large charge limit ruled out 
any classical phase transition. More significantly we established that the signature of the state space geometries were crucially dependent on the
signature of the moduli space metric. \cite{ts2}.

As we have mentioned, the thermodynamics of asymptotically AdS black holes show a rich structure of phase transitions and critical phenomena including a
liquid-gas like first order phase transition analogous to the Van der Waals gas and a critical point of second order phase transition. It was a natural
question then to investigate if this rich phase structure was encoded in the scalar curvature of the equilibrium state space thermodynamic
geometry of these black holes. Several attempts to investigate these issues had been made earlier. Although the critical points of second order
phase transitions were encoded in the divergence of the state space scalar curvature, it was seemingly insensitive to any signature of first order phase
transitions. So it is a natural issue to investigate if the scalar curvature also encodes first order phase transitions for general thermodynamic systems.  
As applied to conventional thermodynamic systems, the answer to this question seems to be in the negative,
as apparently the state space scalar curvature does not exhibit any special behavior near such a transition. An early attempt in this context was
made by \cite{diosi} to analyse the state space geometry of the Van der Waals gas and classify the phases
in terms of the geodesic behavior. A later analysis \cite{ruppvw} of the same system yielded again the critical point of second order phase transition at the end of the
``vapour pressure curve'' without any significant special behavior of the scalar curvature near the first order phase transition curve.

In this article, we take a first step towards addressing this important issue in the context of the phase transitions and critical behavior 
of Kerr-Newman AdS (KN-AdS) black holes in the framework of the thermodynamic geometry of their state space. We reproduce the critical points of second order phase transitions for these black holes, from the divergences of state space scalar curvature.  Remarkably, we are also able to establish that the state space
scalar curvature encodes the discontinuities of the first order phase transitions in terms of its turning point behavior and multi-valued branched structure as a function of the thermodynamic variables for the system. The first clue to this behavior arises from an analysis of the conventional first order liquid-gas phase transition in the
Van der Waals gas. Studying the state space scalar curvature of this system as a function of the thermodynamic variables,
we observe that the curvature diverges near the first order point but shows a turning point behavior at infinity and gives rise to
a multi-valued branched structure similar but not identical to the behavior of the corresponding Gibbs free energy in the grand canonical ensemble. The turning point and the multi-valued behavior of the scalar curvature disappears at the end of the ``vapor pressure curve'' where it is divergent but single valued. From the usual wisdom of thermodynamic geometries this happens because the scalar curvature goes like the inverse of the singular part of the free energy associated with the long range correlations
but is insensitive to the short range correlations. Near the critical point of the second order phase transition the singular part of the free energy decreases and is exactly zero at the critical point.
Hence the divergence of the scalar curvature occurs exactly at the point where the free energy shows a kink. This seemingly points to a qualification of the original proposal by Ruppeiner
about the behavior of the state space scalar curvature near phase transitions. That the curvature scalar diverges at the critical point of second order phase
transition but is single valued is well known. We show here that the curvature scalar does have a very interesting behaviour near first order phase transition
points as well. In the latter case, in regions of phase coexistence, the curvature shows multi-valued branch structures, with turning points at infinity, and physicality implies that the
curvature changes branch close to the first order phase transition point. We will qualify these statements in the remainder of the paper.

Having clarified our assertion, we then proceed to study the phase structure of the various KN-AdS black hole solutions for diverse ensembles in the framework of their state space thermodynamic geometries, which has been previously studied in \cite{calda}. 
In this case, the conventional phase structure envisions a grand canonical ensemble which allows both the charge and the angular momentum to fluctuate, or
a canonical ensemble where both the extensive parameters ( charge and angular momentum) are held fixed. However we further study the case of mixed 
 ensembles in which only
one of the two extensive thermodynamic parameters (charge or angular momentum) is allowed to fluctuate and the other is held fixed (such ensembles were considered, in the case of asymptotically flat black holes in \cite{td7}, and also in \cite{rupp2}). Studying the behavior of the corresponding
modified Gibbs free energy we unravel a rich novel phase structure
in both these cases involving first order phase transitions identical to the liquid-gas transition in the Van der Waals-Maxwell gas ending in a critical point of second order phase transition.
Study of the scalar curvature of the state space of the system seems to
bear out our qualification of the original assertion of Ruppeiner, namely that the scalar curvature encodes information about the first order phase transition in addition to its divergence at the critical point of second order phase transition at the end of the vapor pressure curve.

The article is organized as follows, in section 2, we present a brief  review of thermodynamic geometry and its application to study phase transitions
and critical phenomena in conventional thermodynamic systems and black holes. As a preliminary exercise, we revisit the Van der Waals gas
and re-examine its phase behavior in the framework of thermodynamic geometry
and obtain the first clue of the state space scalar curvature encoding information about the first order liquid-gas phase transition.
Section 3 deals with the thermodynamics, phase structure and critical phenomena in the most general rotating charged KN-AdS black holes for
diverse ensembles including mixed ensembles and we establish new phase behavior for these black holes.  In the last section 4, we present a summary of
our results and discuss possible future directions of research.

\section{Thermodynamic Geometry and Phase Transitions}

In this section we briefly review the essential features of thermodynamic geometries especially their application to study phase transitions and critical phenomena in thermodynamic systems particularly  black holes. The issue of an intrinsic geometrical structure underlying the thermodynamic behavior of systems is a long standing one.
However although an extrinsic geometric description was evident the issue of an intrinsic geometric structure remained elusive.
The first such geometric structure in equilibrium thermodynamics
was introduced by Weinhold \cite{weinhold} through an inner product in the space of
equilibrium thermodynamic macrostates defined by the minima of the internal
energy function $U= U(S/T, V/T, \mu_i/T)$ as the Hessian
\bea
h_{ij}=\pt_i\pt_j U
\eea
As mentioned in the introduction,
the quantities $\mu_i, T, V, S$ are the chemical potentials,
temperature, volume and entropy respectively and the volume or any other
parameter is held fixed to provide a physical scale and to restrict the occurrence of
negative eigenvectors of the metric. Although such a Riemannian geometric
structure was interesting, no physical significance could be ascribed to
it. The inner product on the state space was later reformulated by
Ruppeiner \cite{rupp} in the entropy representation as the negative of the
Hessian matrix of the entropy density $s=s(u,n)$ with respect to the densities of the other extensive variables. The
thermodynamic macrostates underlying the equilibrium state space being now
described by the maxima of the entropy function $S =S(U, V, N)$. Explicitly
the Ruppeiner metric in the state space was given as,
\bea
g_{ij}=-\pt_i\pt_j s( u,n)
\label{metric}
\eea
and was conformal to the Weinhold metric with the inverse temperature as
the conformal factor.  The negative sign was necessary to ensure positive
definiteness of the metric, as the entropy is a maximum in the equilibrium
state.  It could be shown that the Riemannian structure defined by the
Ruppeiner metric was closely related to classical thermodynamic fluctuation
theory \cite{rupp2} and critical phenomena.  The probability
distribution of thermodynamic fluctuations in the equilibrium state space
was characterised by the invariant interval of the corresponding
thermodynamic geometry in the Gaussian approximation as
\begin{equation}
W(x) = A {\rm exp}\left[ -\frac {1}{2} g_{ij}(x) dx^i dx^j\right]
\end{equation}
where $A$ is a constant and $x^i$ are the extensive variables. The inverse metric may be shown to be the second
moment of fluctuations or the pair correlation functions and given as
$g^{ij}= < X^i X^j>$ where $X^i$ are the intensive thermodynamic variables
conjugate to $x^i$.  The Riemannian structure may be expressed in terms of the densities of
any suitable thermodynamic potential arrived at by Legendre transforms
which corresponds to general coordinate transformations of the equilibrium
state space metric. One should emphasize here that the invariant interval in the state space
has the dimensions of volume for conventional thermodynamic systems unlike the
usual dimension of length in Riemannian geometries. This is a consequence of using
densities of thermodynamic poetntials in the expression for the state space metric.

For a standard two dimensional thermodynamic state space defined by the
extensive variables $ (x^1, x^2)$, application of these geometric
notions to conventional thermodynamic systems suggest that a non zero scalar
curvature indicates an underlying interacting statistical system. It could  be
shown that the scalar curvature $R\sim\kappa_2\xi^d$ where $\xi$ is the
correlation length, $d$ is the physical dimensionality of the system and
$\kappa_2$ is a dimensionless constant of order one.  This naturally suggests that
the scalar curvature is a measure of the effective interactions in the underlying statistical system.
Near a critical point of second order phase transition the correlation length $\xi$ diverges
and hence the scalar curvature $R$ also diverges at the critical point.  A detailed analysis
shows that the scalar curvature is actually proportional to the inverse of the singular
part of the free energy arising from ``long range correlations.'' This goes to zero at the
critical point and hence $R$ diverges.
The Ruppeiner formalism of  thermodynamic geometry has been applied to diverse condensed
matter systems with two dimensional state spaces and is completely consistent with the scaling and hyperscaling relations involving critical phenomena and have reproduced the corresponding critical indices.

Note that for black hole systems, the line element in the equilibrium thermodynamic state space is dimensionless in contrast to 
usual extensive thermodynamic systems. As a consequence, the state space scalar curvature $R$ is dimensionless here. 
Hence, the usual interpretation of $R$ as the correlation volume is not applicable. However, we will proceed with the
understanding that $R$ is a measure of interactions at the horizon and will continue to refer to $R$ as the ``correlation volume.'' 

Although the divergences of the state space scalar curvature encodes the critical points of second order phase transitions it
seems insensitive to the interesting thermodynamic phenomena of first
order phase transitions. Instances of such first order phase transitions involving coexistence of two or more phases
and multiple length scales, occur quite frequently for conventional thermodynamic systems
the most common being a liquid-gas phase transition. The simplest example of  such a first order phase transition is the
Van der Waals-Maxwell gas where a ``vapor pressure curve'' along which both the liquid and gas phase exists ends at
critical point of second order phase transition beyond which no phase distinctions are possible. An early analysis by
Ruppeiner of this system showed no special behavior of the state space scalar curvature near the first order phase transition point.
Although the curvature showed a clear divergence at the second order critical point and was consistent with the usual scaling relations and the known critical indices. Several other authors later analysed the state space for the Van der Waals gas using smoothly connected geodesics as an indication of a coherent phase and also the degeneracy and change in the sign of the thermodynamic metric
across the spinodial vapor pressure curve, but were however open to several criticisms.
Quite obviously it is far from satisfactory that  an important aspect of thermodynamic behavior namely the occurrence of first order phase transitions had no signature in the intrinsic geometry of the equilibrium state space. Or the state space
scalar curvature was seemingly insensitive to first order phase transitions. So certainly this aspect needed further investigation.

We have reexamined the issue of first order phase transitions in the framework of thermodynamic geometries in the context of our study of the phase structure of charged rotating AdS black holes. One of the main general results of our studies is that the state space scalar curvature actually encodes information about the first order phase transitions through the multivalued branched structure and a typical turning point behavior at infinity with respect to other thermodynamic variables at a first order phase transition. Our result seems to be sufficiently general and universal in the examples that we have studied. The first clue to this arises from revisiting the Van der Waals gas state space and studying the behavior of the scalar curvature with respect to other thermodynamic variables in the grand canonical ensemble. Our analysis shows a very typical turning point behavior at infinity and a multivalued branch structure similar to the behavior of the appropriate free energy and that of the corresponding specific heat or compressibilities/capacitances. This behavior extends to the phase structure of the KN-AdS black hole
where we show first order phase transitions in mixed ensembles of fixed angular momentum or fixed charge. The full grand canonical treatment earlier by Caldarelli et al \cite{calda} had led to a Hawking Page phase transition where we find
a single valued divergence of the state space scalar curvature.

It should be mentioned here that as alluded in the introduction black hole thermodynamics and phase transitions are somewhat
different from usual thermodynamic systems. It follows naturally from this that the thermodynamic geometry as a framework for
studying phase transitions in black holes involves several subtle issues. Although the general overall thermodynamic behavior
is strongly similar to the usual notions there are indeed several points of departure. The most important amongst these is the fact that
we have little or no idea still of the underlying microstates or the statistical system for black holes. So the issue of a suitable ensemble
underlying the analysis of the phase structure is an extremely critical. In fact it has been adequately established now from many different analysis that the nature of the phase structure and critical phenomena for black holes is strongly dependent on the choice of a suitable ensemble. So it is always prudent to analyse and contrast 
the phase behavior in various different ensembles.  Furthermore there is no concept of a volume for black holes as thermodynamic systems so unlike conventional 
systems the thermodynamic metric is directly given by the Hessian matrix of the appropriate thermodynamic potential and not its density as for usual systems. In 
addition black holes by themselves are not extensive systems so all analysis of black thermodynamics tacitly assume that the black hole is a subsystem of  a larger 
infinite system to ensure thermodynamic extensivity. With these comments, let us proceed to analyse the Van der Waals model. 
Its also important
to emphasize at this juncture that unlike usual thermodynamic systems, black holes may be described by a exact fundamental relation
between its thermodynamic variables through the Smarr formula.

\subsection{The Van der Waals Model}

We begin with a review of the main features of the Van der Waals model. The thermodynamic geometry of this system has
been well studied in the literature (see, e.g \cite{ruppvw},\cite{santoro},\cite{janyszek1}). However,
as pointed out in the introduction, we will be able to comment on
certain features of this system which we believe are novel results that have not been reported in the past. We begin with the
expression for the Helmholtz free energy of the system \cite{landau}
\begin{equation}
F = N\left(-c_vT{\rm ln}T - \zeta T + \epsilon\right) - NT{\rm ln}\left(\frac{e\left(V - Nb\right)}{N}\right) -
\frac{N^2a}{V}
\end{equation}
where, as in standard terminology, $a$ and $b$ are the Van der Waals constants, $c_v$ refers to the molar specific heat
at constant volume, and $\zeta$ and $\epsilon$ are constatns. $N$ refers to the number of particles with volume
$V$ so that the number density is $\rho = \frac{N}{V}$ at a temperature $T$. With the substitution
$c_v = \frac{3}{2}$ as is appropriate for the Van der Waals model, the molar Helmholtz potential is
\begin{equation}
f = -\frac{3}{2}T{\rm ln}T - \zeta T + \epsilon - T\left(1 - {\rm ln}\rho + {\rm ln}(1 - b\rho)\right) - a\rho
\end{equation}
The pressure is given by
\begin{equation}
P = \rho^2\left(\frac{\partial f}{\partial \rho}\right)_T = \frac{\rho\left(a\rho - a\rho^2 b - T\right)}{b\rho - 1}
\label{pressure}
\end{equation}
The molar Gibbs potential (i.e the chemical potential) is given by
\begin{equation}
g = f + \frac{P}{\rho}
\end{equation}
The molar entropy is given by
\begin{equation}
s = -\left(\frac{\partial f}{\partial T}\right)_{\rho} =  \frac{3}{2}{\rm ln}T + \frac{5}{2} + \zeta - {\rm ln}\rho + {\rm ln}\left(1 - b\rho\right)
\end{equation}
Now, without loss of generality, we set the constants $\zeta = \epsilon = 1$, and with this substitution, the expression for the Ruppeiner curvature for the
system is given by
\begin{equation}
R = \frac{\left[4a^2\rho^2b\left(3 + \rho^2b^3 - 3\rho b^2\right) +T\left(a +5a\rho b + 3a\rho^3b^3 +3Tb
- 9Ta\rho^2b^2\right) - 4a^2\rho\right]}{3\left(2a\rho + 2a\rho^3b^2 - T - 4a\rho^2b\right)^2\left(b\rho - 1\right)^{-1}}
\label{ruppvw}
\end{equation}
Note that in calculating $R$, the components of the thermodynamic metric involves the free energy density, i.e free energy per unit volume.
From the expression for the molar specific heat at constant pressure,
\begin{equation}
c_p = \frac{1}{2}\frac{6a\rho^3b^2 + 6a\rho - 5T -12a\rho^2b}{2a\rho + 2a\rho^3b^2 - T - 4a\rho^2b}
\end{equation}
we see that the Ruppeiner curvature diverges in the same way as $c_p$. This is expected, because we are allowing the density to fluctuate envisioning a description in the grand canonical ensemble.
Hence, the appropriate potential in this case is the molar Gibbs potential which remains constant across a first
order phase transition.

Our main interest would be to look at the behaviour of the Ruppeiner curvature at phase transition points of the system.
In what follows, we set the Van der Waals
constants as $a=8, b=2$. These are just convenient values. Any other value will not change our results in any qualitative way.
The critical temperature and pressure for this choice of parameters can be obtained from the $P-V$ isotherms \cite{landau}.
These values are found to be
\begin{equation}
T_c = 1.1851,~~~P_c = 0.0741
\label{critvw}
\end{equation}
For temperatures below the critical value, the Van der Waals system undergoes a first order phase transition. It is instructive
to plot the Ruppeiner curvature along an isotherm, as a function of the system pressure, in order to compare with the variation of the molar entropy
with pressure. This can be done by using the expression
for the pressure in eq. (\ref{pressure}), and the number density $\rho$ is treated as a parameter. We will find it convenient to discuss the behaviour
of the Ruppeiner curvature along various isotherms,
vis a vis that of the molar entropy along the same, $\rho$ again being the parameter. \footnote{There is no qualitative difference in our description
involving isotherms, and one involving curves of constant pressure (isopotential curves). For black holes, we will find it more convenient to
investigate the Ruppeiner  curvature along isopotential curves as a function of temperature.}

\begin{figure}[!ht]
\begin{minipage}[b]{0.5\linewidth}
\centering
\includegraphics[height=3in,width=3in]{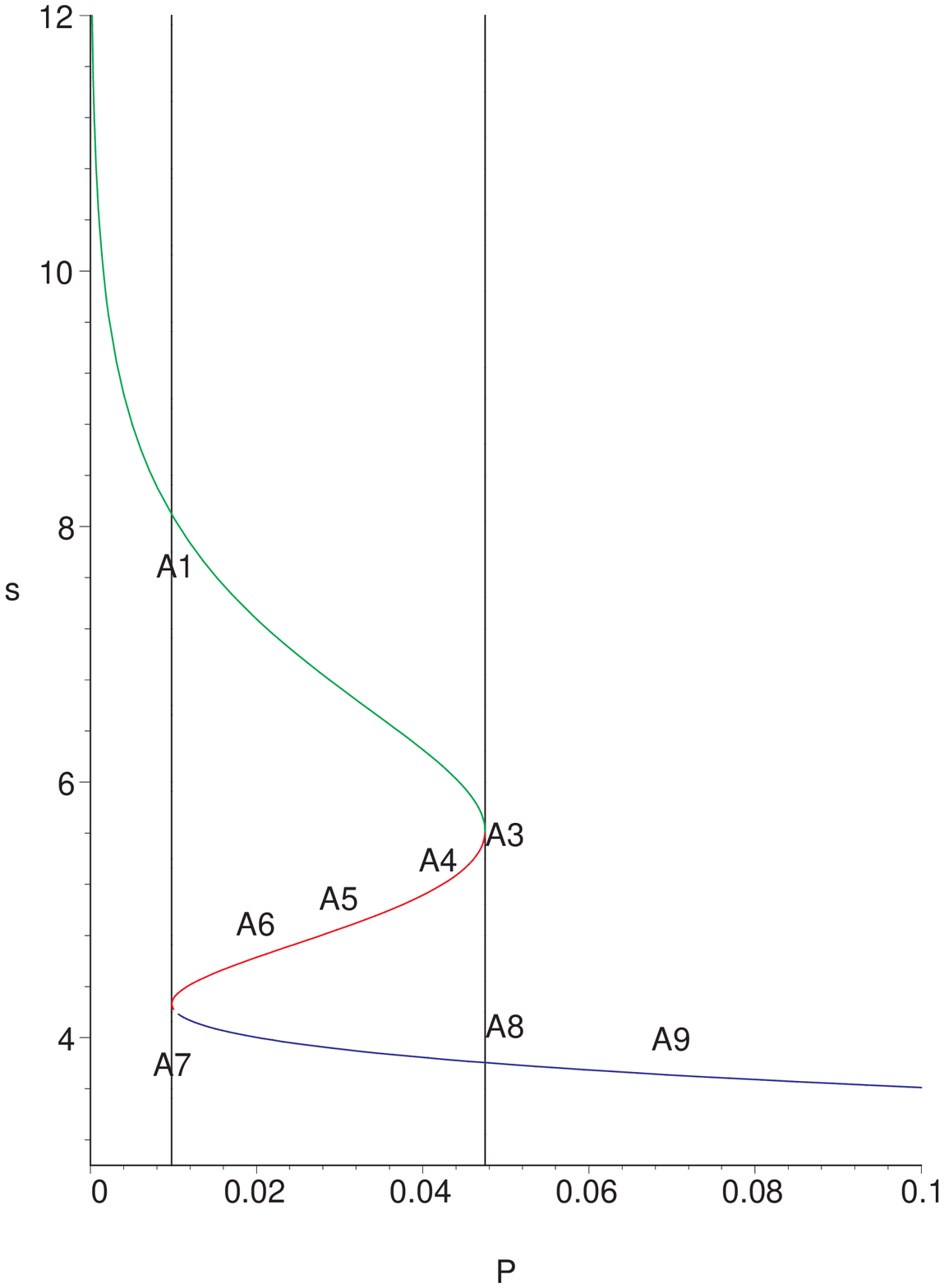}
\caption{Molar entropy vs pressure for $T = 1.02$}
\label{fig0a}
\end{minipage}
\hspace{0.6cm}
\begin{minipage}[b]{0.5\linewidth}
\centering
\includegraphics[height=3in,width=3in]{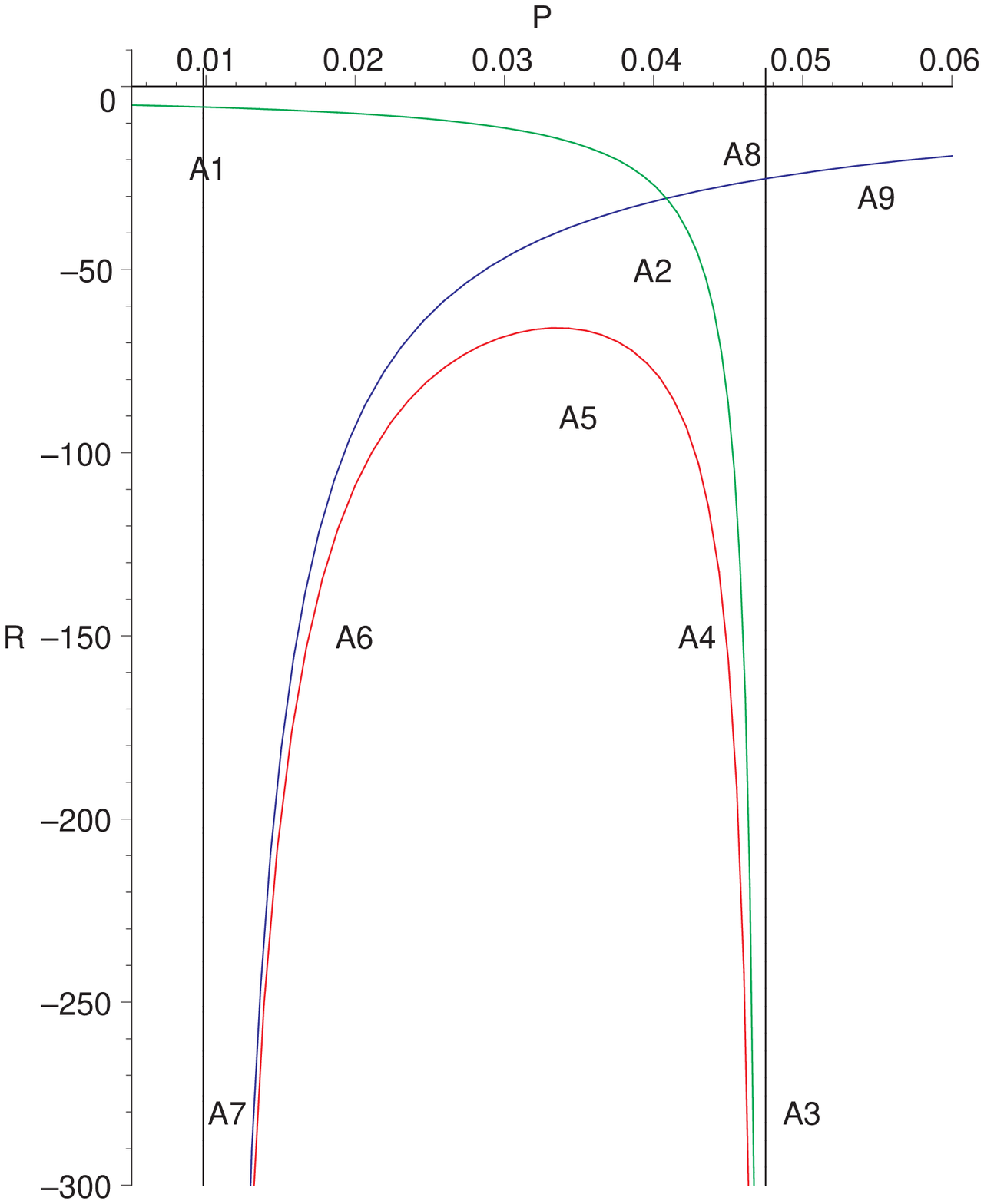}
\caption{Ruppeiner curvature vs pressure for $T = 1.02$}
\label{fig0b}
\end{minipage}
\end{figure}

In fig. (\ref{fig0a}) and fig. (\ref{fig0b}), we illustrate the parametric plot of the molar entropy vs pressure and the Ruppeiner curvature vs pressure,
both along an isotherm corresponding to $T = 1.02$.

The Ruppeiner curvature in this case, captures many features similar to those seen in black hole systems.
For comparison purposes, we first review the well known behaviour of the molar entropy as a function of pressure. 
The Maxwell relation
\begin{equation}
\left(\frac{\partial s}{\partial P}\right)_T = - \left(\frac{\partial v}{\partial T}\right)_P
\end{equation}
where $s$ is the molar entropy and $v$ is the volume per mole,
implies that the partial derivative of the molar entropy with pressure along an isotherm has to be negative in order to maintain physicality.
From fig. (\ref{fig0a}), there are thus two locally stable branches of the system in the $s - P$ plane, one along the points $A1~A3$,
and the other along the points $A7~A8~A9$.\footnote{Note also that along the unstable branch, the line element of the equilibrium state space geometry
is negative, and hence is not relevant to thermodynamic geometry.} Let us denote these as branch 1 and branch 3. These have been marked in green and
blue respectively. Inbetween these, marked in red, is an unstable branch 2 of the system,
along the points $A4~A5~A6$. We have drawn the asymptotes at the turning points of the
entropy, which occur at $P(A1)~{\rm or}~P(A7) = 0.0475$ and $P(A3)~{\rm or}~P(A8)  = 0.0098$. These are marked in black in the figure.

As the system pressure is increased from zero, the system ``travels'' along branch 1, until it reaches the point $A1$, where the entropy becomes
multivalued. The system continues along
branch 1, till the point where its free energy exceeds that of branch 3. At this point, branch 1 becomes metastable, and branch 2 becomes globally
stable, and the system may undergo a first order phase transition from branch 1
to branch 3, driven by the fluctuations.
Thereafter, upon further increase of pressure, the system follows branch 3 of the entropy. Let us now compare this with the behaviour
of the Ruppeiner curvature $R$ along the same isotherm, which is shown in fig. (\ref{fig0b}). In this figure, we have labelled the corresponding points
in fig. (\ref{fig0a}). The branches of $R$ are coded with the same colors as those of the molar entropy (remember that we have used the same
parameter $\rho$ for both the plots). \footnote{The exact values of the molar entropy or Ruppeiner curvature at these points will not be important for us,
but these can be easily read off.}

Note that, as can be seen from fig. (\ref{fig0b}), the Ruppeiner curvature has three branches as expected from the plot of the molar entropy vs pressure. 
Indeed, the Ruppeiner curvature by itself cannot
predict the stability or instability of a given branch, but comparison with the known behaviour of the system entropy or Gibbs free energy as a function
of the system pressure, this classification can indeed be made. Hence, in fig. (\ref{fig0b}), the green line that passes through the points
$A1,~A2,~A3$ is the first stable branch (branch 1) as we increase the parameter $\rho$ from zero. The unstable branch, marked in red (branch 2),
passing through $A4,~A5,~A6$ begins thereafter, and diverges at $A7$. The next stable branch (branch 3) begins from negative infinity (at $A7$) and
asymptotes to zero along $A7,~A8,~A9$. The values of pressure corresponding to the divergences in the Ruppeiner curvature corresponds to those at
the turning points of the molar entropy in fig. (\ref{fig0a}).
Branch 1 diverges at $P = 0.0475$.
(Compare the similar points in fig. (\ref{fig0a}). The unstable branch 2 culminates in the asymptotic point $A7$, that has
the value of pressure $P = 0.0098$. \footnote{The fact that the curvature
blows up at values of pressure corresponding to the asymptotes shown in black can be seen from eq. (\ref{ruppvw}).}

Let us qualitatively discuss the features of the Ruppeiner curvature.
If we increase the system pressure from zero, the curvature becomes multivalued starting from the point A1, and remains so in the phase coexistence region
between the two asymptotes in fig. (\ref{fig0b}).
At the pressure corresponding to a first order phase transition, the curvature will ``jump'' from one stable branch to the other and then continue along the latter
as the system pressure is further increased.
We emphasize that although there are divergences in the expression for the Ruppeiner curvature at the turning points of the molar entropy,
the system moving along the isotherm will not access these equilibrium states and consequently the corresponding divergences of the
state space scalar curvature are rendered unphysical.

\begin{figure}[!ht]
\centering
\includegraphics[width=3in,height=2.5in]{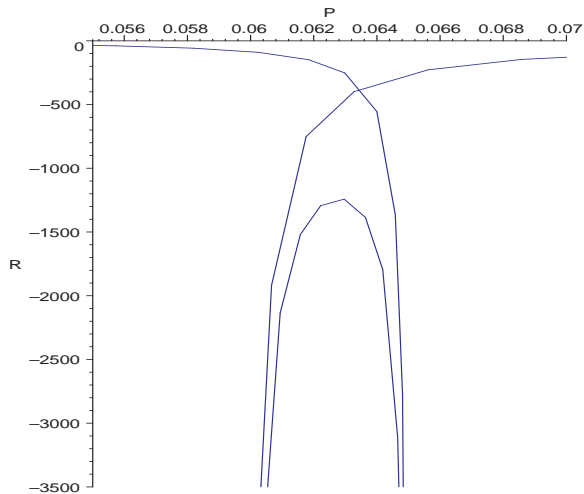}
\caption{Molar entropy vs pressure for $T = 1.14$}
\label{fig0c}
\end{figure}

For the moment, we note that for isotherms with higher temperatures (but below $T_c$),  the qualitative behaviour of the Ruppeiner curvature remains the same.
In fig. (\ref{fig0c}), we show a parametric plot of the Ruppeiner curvature as a function of the system pressure along the isotherm $T = 1.14$.
As can be seen, this indeed retains the features alluded to in the last paragraph.

\begin{figure}[!ht]
\begin{minipage}[b]{0.5\linewidth}
\centering
\includegraphics[width=2.5in,height=3in]{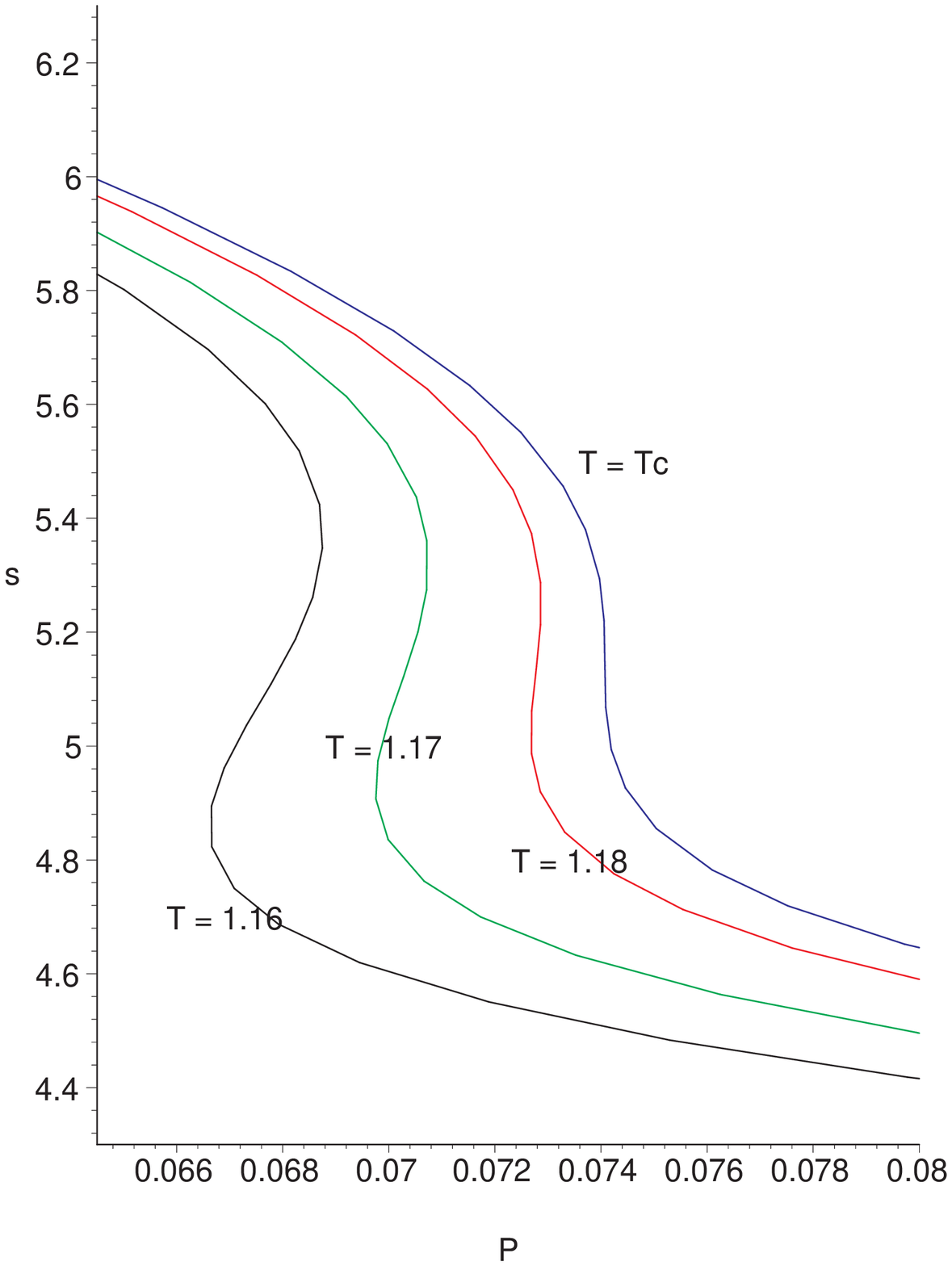}
\caption{Molar entropy vs pressure for temperatures approaching $T_c$.}
\label{fig0d}
\end{minipage}
\hspace{0.6cm}
\begin{minipage}[b]{0.5\linewidth}
\centering
\includegraphics[width=3.0in,height=3in]{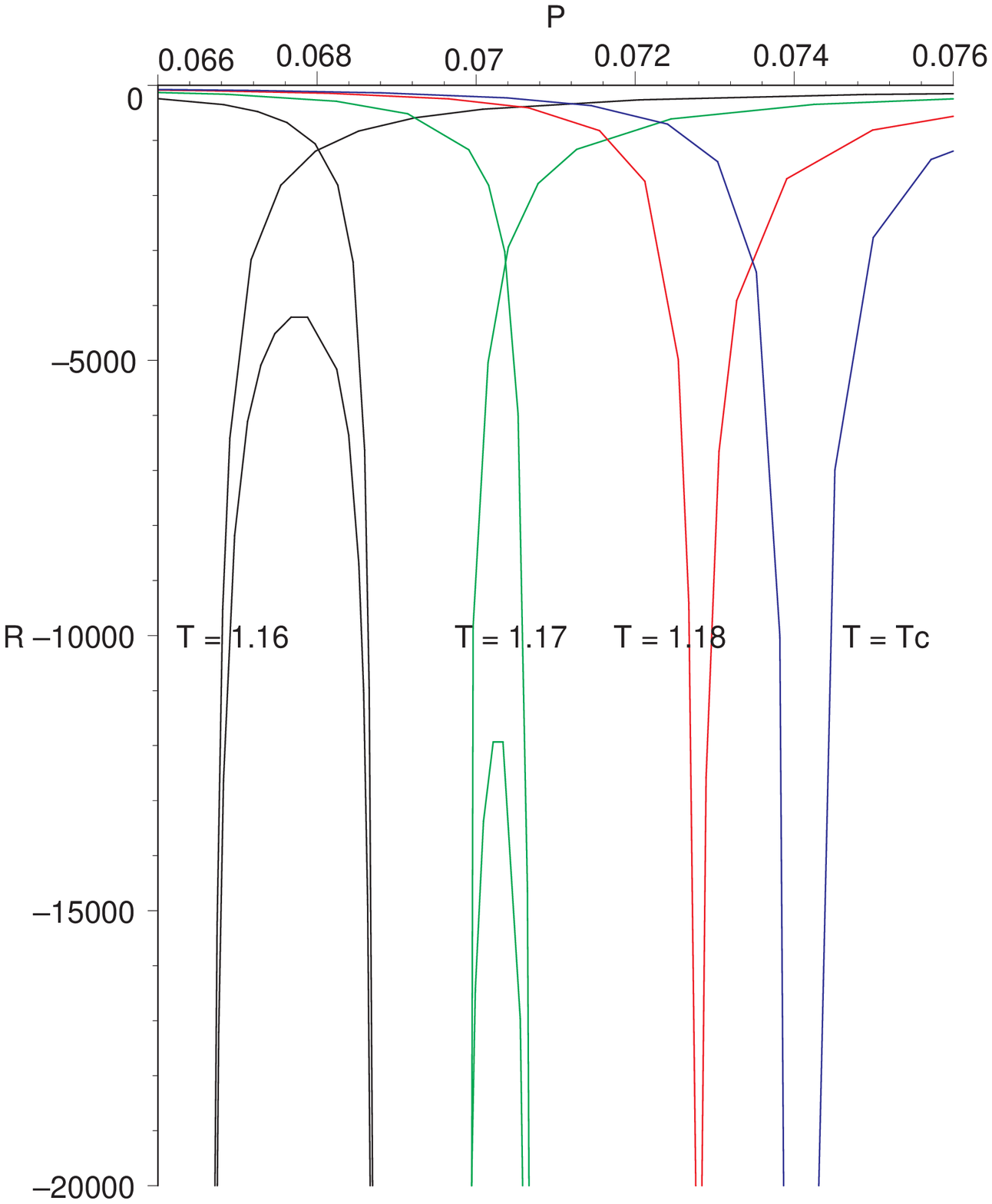}
\caption{Ruppeiner curvature vs pressure for temperatures approaching $T = T_c$}
\label{fig0e}
\end{minipage}
\end{figure}

We now discuss the isotherms near $T = T_c$ (eq. (\ref{critvw}).
We find that the behaviour of the Ruppeiner curvature closely follows that of the known behaviour
of the molar entropy. Namely, as we approach $T = T_c$, the unstable branch of the curvature gets pushed to infinity, and the two stable branches join
at infinity,  i.e the curvature goes to infinity. This happens precisely at the critical value of the pressure, $P = P_c$ in eq. (\ref{critvw}).
In figs. (\ref{fig0d}) and (\ref{fig0e}), we have shown the snapshots of the molar entropy vs pressure and Ruppeiner curvature vs. pressure as
one approaches the critical temperature.

From fig. (\ref{fig0d}), it is seen that the unstable branch in the molar entropy disappears at $T=T_c$, and the entropy becomes single valued
thereafter. From fig. (\ref{fig0e}), we see that the Ruppeiner curvature follows the same behaviour, although in a different manner. As we approach
$T_c$,  the two asymptotes of fig. (\ref{fig0b}) begin to come closer.
\footnote{It can be checked that this is true for the isotherm corresponding to $T = 1.18$ in fig. (\ref{fig0e}), marked in red, as well. The unstable branch for this
isotherm will be seen for very large negative values of the curvature.} Finally, at $T = T_c$, the unstable branch disappears and the two stable
branches ``separate,'' (i.e do not have any overlap) signaling a  second order phase transition. The Ruppeiner curvature diverges at this point.

\begin{figure}[!ht]
\centering
\includegraphics[width=5in,height=3.5in]{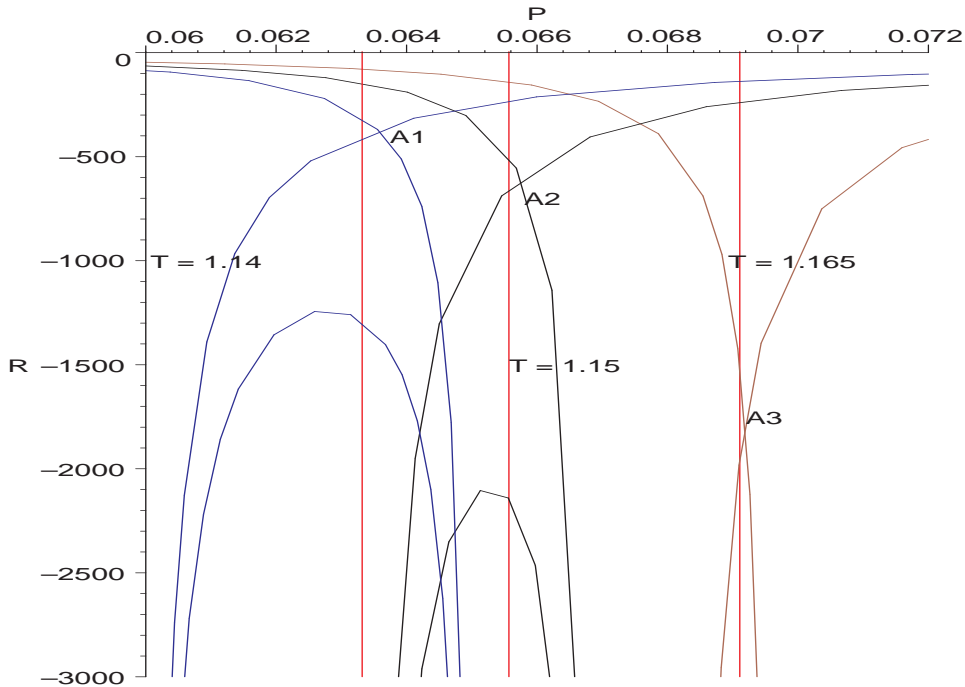}
\caption{A plot of Ruppeiner curvatures vs pressure as we approach the critical isotherm.}
\label{fig0f}
\end{figure}

Let us now come back to the question of where exactly the system changes branch in the Ruppeiner curvature. This issue can be addressed by
examining the behaviour of the Gibbs free energy as a function of the system pressure.
In fig. (\ref{fig0f}), we have marked asymptotes corresponding to the values of the pressure at which the Gibbs free energy corresponding to the
given isotherms changes branch (see, eg. \cite{callen}). It is to be noted that the point of intersection of the stable branches of the Ruppeiner curvature
occurs at slightly different values of pressure compared to the value of the pressure at which the Gibbs free energy transits from one stable branch
to the other. As can be seen from the figure, this difference decreases as the critical isotherm is reached. At the critical temperature $T_c$, there
is no intersection of branches of the Ruppeiner curvature, both of which go to infinity at $P_c$. At that temperature, the Gibbs free energy of the
two stable branches become identical at $P_c$, as is well known. Since the (absolute value of the) Ruppeiner curvature is proportional to the
correlation volume, our results would therefore suggest that there is jump in correlation volume at the first order phase transition. This is indeed
a novel feature, although we do not have full analytical control of this phenomenon at the moment. However, it seems that the curvature
jumps from one stable branch to the other at the value dictated by the Gibbs free energy, whereafter it continues its journey along the other
stable branch. We believe that this has not been reported earlier in the literature, and constitutes one of the main results of this subsection. We will
see that essentially the same features appear in black hole thermodynamics as well, to which we will presently turn to. As a final comment, note
that the Van der Waals system is only an approximate model. The Ruppeiner curvature should have a more definite meaning in the context
of black holes, as the fundamental relation is exactly known, and hence a more complete thermodynamic description is available.

\section{KN-AdS Black Holes}

In this section, we will elucidate the behaviour of thermodynamic geometry for Kerr-Newman-AdS black holes. These are charged, rotating black
holes which are asymptotically AdS. The thermodynamics and phase structure of these black hole configurations has been extensively studied in the literature 
(see, e.g. \cite{calda}), and we will only mention the details relevant for our analysis here.  In contrast to the asymptotically flat case, the angular velocity 
of rotating black holes in AdS space does not 
vanish at infinity. The angular velocity that enters the thermodynamic description of the black hole is then the difference between the angular velocity at the
horizon and that at infinity, which turns out to be identical to the angular velocity of the rotating Einstein universe at the AdS boundary. This is
also expected from the AdS-CFT correspondence. In the same way, the electric potential that enters the analysis is also measured with respect
to potential at the boundary of the AdS space, which is non zero in this case. In our analysis, we will use the thermodynamic quantities, keeping
in mind that this subtraction has already been made. \footnote{In our analysis these quantities will be calculated after suitable
rescaling of the black hole parameters by factors of the AdS radius, and will be denoted by lower case letters. For example, the angular velocity and
electric potential are denoted by $\omega$ and $\phi$ in our analysis. In the notation of \cite{calda}, these quantities are denoted by
$\Omega$ and $\Phi$ respectively.}

In the absence of a complete knowledge of the microstates of the system, one naturally studies the KN-AdS black hole in various ensembles,
defined by fixing the charges and the corresponding potentials.
In its conventional form, Ruppeiner geometry naturally alludes to open systems which envisions
a grand canonical ensemble, where we allow
fluctuations in all the charges that appear in the description of the thermodynamic state space.
Fixing a particular charge implies that the curvature is insensitive to fluctuations in the same. In the KN-AdS black hole, we have two thermodynamic
charges, the electric charge and the angular momentum (the magnetic charge is set to zero).
If we fix both, then in the Ruppeiner formalism, the thermodynamic state space is
naturally flat, i.e the curvature is trivially zero. This is of course different from fixing the charges in the expression for the curvature. The latter
would imply that we still allow for fluctuations in the charges. Hence, a fully canonical ensemble is uninteresting as far as the Ruppeiner
formalism is concerned.

However, for KN-AdS black holes, it is possible to define a {\it mixed} ensemble, where one of the thermodynamic charges, say $j$ is
held fixed, while the black hole is allowed to exchange the other charge ($q$) with its environment which is held at a fixed potential.
This implies that for the electric charge , we are in a grand canonical ensemble, in
that the system can exchange electric charge with a surrounding bath, but the system does not exchange angular momentum
which is held at a fixed value, and hence is in a canonical ensemble with respect to this variable. We will refer to this as the ``fixed angular momentum
ensemble'' or the ``fixed $j$ ensemble'' in what follows. Analogously, we can talk about an ensemble where we fix the angular velocity and the electric charge.  
This will be referred to as the ``fixed charge ensemble'' or the ``fixed $q$ ensemble.'' 
The astute reader might question the stability of the system in these ensembles. 
Before we proceed, let us make some brief comments on the stability issues in the mixed ensembles. 

The themodynamic potentials appropriate to the fixed $q$ and fixed $j$ ensembles can be obtained in a straightforward manner by partial Legendre transforms 
of the scaled Gibbs potential $g$ (the scaled variables are appropriately defined in the next subsection). For the fixed $j$ ensemble, we first construct the 
potential 
\begin{equation}
A=g+{\omega}j
\end{equation}
the differential of which is $dA=-sdt-qd{\phi}+{\omega}dj$. On restricting to the 
constant $j$ sections of the parameter space we get, at constant $j$,
\begin{equation}
\left(dA\right)_j=-sdt-qd{\phi}
\end{equation}
which then is the appropriate potential for black holes with fixed angular momentum 
and fluctuating charge. One can proceed in an exactly analogous manner by defining the potential 
\begin{equation}
B=g+{\phi}q
\end{equation}
for the fixed $q$ ensemble. 

Following the discussion in \cite{calda}, we can briefly comment on the procedure to obtain the potentials $A$ and $B$ from the Euclidean action. 
The authors in \cite{calda} use the counterterm method of \cite{bala} to obtain the action $I(t,\omega,\phi)$, corresponding to a fixed potential 
boundary condition in the grand canonical ensemble and the action, $\tilde{I}(t,\omega,q)$, corresponding to a fixed charge boundary condition 
in the canonical ensemble. Subsequently they obtain the Gibbs potential,  $g$, and the Helmohltz potential, $f$, from these actions as
\begin{equation}
g(t,\omega,\phi)=\frac{I}{\beta}~;~~~ f(t,j,q)=\frac{\tilde{I}}{\beta}+{\omega}j,
\end{equation}
where $\beta$ is the inverse of temperature. Note that a Legendre transform is required on the action $\tilde{I}$ in order that angular momentum may also be held 
fixed in the canonical ensemble. One can check that if the action $I$, rather than $\tilde{I}$, is Legendre transformed, we obtain the potentials $A$ and $B$ 
correspondig to 
the ``mixed'' ensembles. Namely,
\begin{equation}
A(t,j,\phi)=\frac{I}{\beta}+{\omega}j;~~~B(t,\omega,q)=\frac{\tilde{I}}{\beta},
\end{equation}

There remains now to discuss the issue of global stability in various ensembles. The global stability of the black hole solutions is clearly understood only in the grand 
canonical ensemble where the action, $I$, can be equivalently obtained using the background subtraction method of Gibbons and Hawking \cite{gibbons}. 
Thus, in this case, the 
zero of the Gibbs potential corresponds to thermal AdS with a constant pure gauge potential, thereby implying that the black hole is globally stable against AdS 
only when its Gibbs potential is negative. However, when either of the charges $j$ or $q$ are held fixed, the AdS space cannot serve as a reference 
background as it is no more a 
 solution of Einstein's equation with the given fixed charge boundary conditions. In such cases we should compare the free energy of the black hole with that of a hot gas 
 of particles with a fixed $q$ or $j$. For example, in the fixed $j$ ensemble, we could possibly consider as a reference background a thermal AdS with a constant potential 
 containing a gas of uncharged particles carrying a fixed total angular momentum. However, we shall not go into detaisl on these issues. We shall just mention that, 
 as discussed in \cite{carlip} (and the references therein), even if these black holes are in a metastable ``supercooled'' state, where they show interesting
 phase behaviour, there is a possibility that these metastable states are long lived.

To begin with, we will study the Ruppeiner geometry of KN-AdS black holes in the grand canonical ensemble, where both the electric potential
and the angular velocity are held fixed. As in the Van der Waals model, we will be interested mainly in exploring the behaviour of the Ruppeiner
curvature at phase transitions.

\subsection{KN-AdS black hole in the Grand Canonical ensemble}

The thermodynamics of
these black holes have been studied extensively in \cite{calda}. We begin with the generalised Smarr formula for the KN-AdS black holes :
\begin{equation}
M = \left[\frac{A}{16\pi} + \frac{\pi}{A}\left(4J^2 + Q^4\right) + \frac{Q^2}{2} + \frac{J^2}{l^2}
+ \frac{A}{8\pi l^2}\left(Q^2 + \frac{A}{4\pi} + \frac{A^2}{32\pi^2l^2}\right)\right]^{1 \over 2}
\label{smarrkn}
\end{equation}
where $A = 4S$, $S$ being the entropy. $J$  and $Q$ are the angular momentum and the electric charge respectively.
It is convenient to absorb the AdS radius in our formulae and redefine
\begin{equation}
m = \frac{M}{l},~~~s = \frac{S}{l^2},~~~q = \frac{Q}{l},~~~j = \frac{J}{l^2}
\end{equation}
One can calculate different thermodynamic quantities like the temperature, potentials etc. via the first law of thermodynamics,
\begin{equation}
dm = tds + \omega dj + \phi dq
\end{equation}
where the scaled intensive quantities are
\begin{equation}
t = lT,~~~\omega = l\Omega,~~~\phi=\Phi
\end{equation}
Although the expressions are standard,
we reproduce some of them for convenience. The conjugate quantities, i.e the electric potential and the angular velocity, as a function of the
charges (i.e $s$, $q$, $j$) are
\begin{eqnarray}
\phi = \frac{\pi^{\frac{1}{2}} q \left(s^2 + s\pi + q^2\pi^2\right)} {s^{1 \over 2}\left[s^4 + 2s^3\pi + s^2\pi^2\left(1 + 2q^2\right)  + 2q^2\pi^3s +
4\pi^3j^2\left(\pi + s\right)\right]^{1 \over 2}}\nonumber\\
\omega = \frac{\pi^{\frac{3}{2}}j\left(\pi + s\right)} {s^{1 \over 2}\left[s^4 + 2s^3\pi + s^2\pi^2\left(1 + 2q^2\right)  + 2q^2\pi^3s +
4\pi^3j^2\left(\pi + s\right)\right]^{1 \over 2}}
\label{potkn}
\end{eqnarray}
The temperature of the black hole is given by
\begin{equation}
t = \frac{3s^4 + 4s^3\pi + s^2\pi^2\left(1 + 2q^2\right) - 4\pi^4 j^2 - \pi^4q^4}{4\pi^{3 \over 2}s^{3 \over 2}
\left[s^4 + 2s^3\pi + s^2\pi^2\left(1 + 2q^2\right)  + 2q^2\pi^3s +
4\pi^3j^2\left(\pi + s\right)\right]^{1 \over 2}}
\label{tempkn}
\end{equation}
We can invert the relations in eq. (\ref{potkn}), and express the electric charge and angular momentum in terms of the electric potential, the angular
velocity, and the entropy. This will be useful for us in studying the behaviour of the system in the grand canonical ensemble. The results are 
\begin{eqnarray}
q &=& \frac{\left[\phi\pi s\left(\pi + s - \omega^2s\right)\left(\pi + s\right)\right]^{1 \over 2}}{\omega^2s\pi - \pi s - \pi^2}\nonumber\\
j &=& \frac{\omega s^{\frac{3}{2}}\left(\pi\phi^2 + s - \omega^2s + \pi\right)\left[\left(s + \pi\right)\left(s + \pi - \omega^2s\right)\right]^{1 \over 2}}
{2\pi^{\frac{3}{2}}\left(\omega^2s - \pi - s\right)^2}
\label{chargepot}
\end{eqnarray}

\begin{figure}[!ht]
\begin{minipage}[b]{0.5\linewidth}
\centering
\includegraphics[width=2.5in,height=3in]{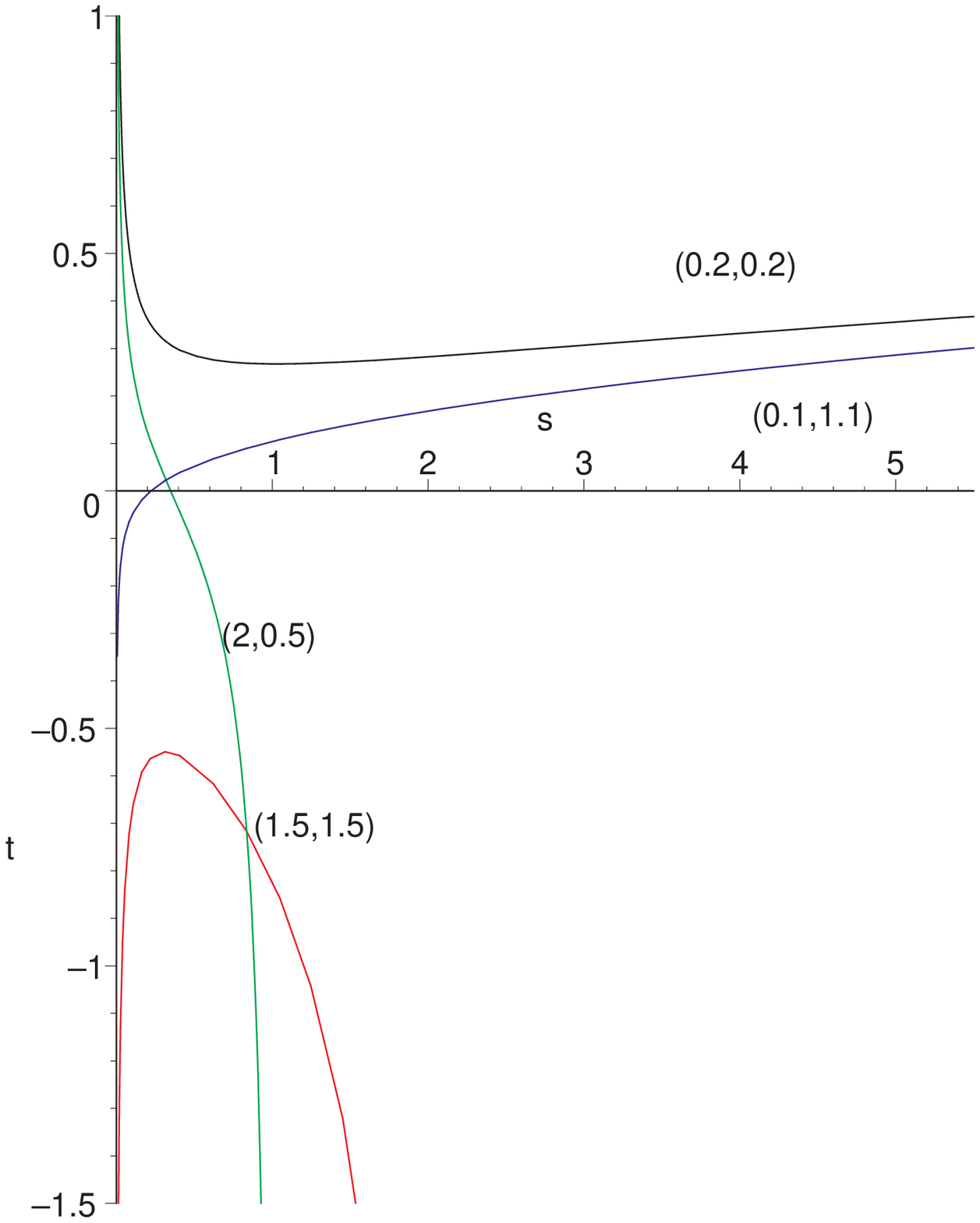}
\caption{Temperature vs entropy for various values of $\left(\omega, \phi\right)$.}
\label{kn1}
\end{minipage}
\hspace{0.6cm}
\begin{minipage}[b]{0.5\linewidth}
\centering
\includegraphics[width=2.5in,height=3in]{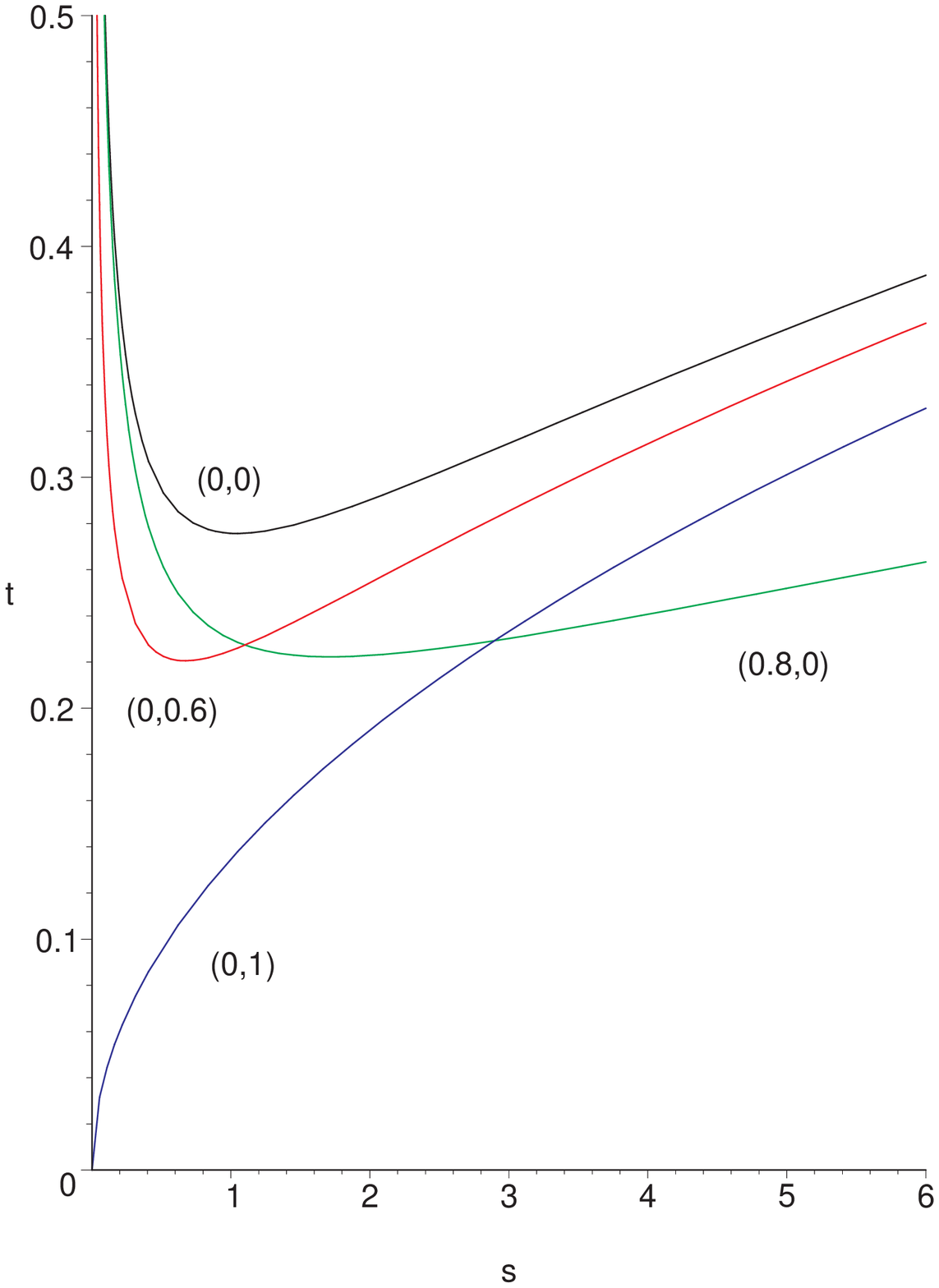}
\caption{Temperature vs entropy for various values of $\left(\omega, \phi\right)$. (contd.)}
\label{kn2}
\end{minipage}
\end{figure}

The temperature in eq. (\ref{tempkn}) can now be expressed entirely in terms of the entropy and the potentials 
$\phi$ and $\omega$, and a plot of temperature vs entropy
gives us information about the stability of the black hole in different regions of parameter space. For convenience, we have plotted this in two
separate graphs, that are illustrated In figs (\ref{kn1}) and (\ref{kn2}). These graphs are self explanatory. eg., from fig. (\ref{kn1}), we see that
for $\left(\omega,\phi\right) < \left(1,1\right)$ (the black curve),  we have stable black holes beyond a certain value of the entropy, the system undergoing a Hawking
Page phase transition in 
between. For $\left(\omega,\phi\right) > \left(1,1\right)$ (the red curve), no black holes
exist at non negative temperature. Keeping $\omega < 1$ and $\phi > 1$, we have a stable black hole branch beginning with an
extremal black hole (the blue curve), and with $\omega > 1$ and $\phi < 1$, we have an unstable black hole branch at all temperatures starting
from extremality (the
green curve). The last case, however, is problematic, as there might be superradiant modes, and a meaningful thermodynamic description of
these black holes may not be possible.
Similar analyses can be done with the graphs in fig. (\ref{kn2}), where we have restricted to $\omega < 1$. In this graph,
$\left(\omega,\phi\right) = \left(0,0\right)$ (the black curve) represents the familiar AdS-Schwarzschild
black hole solution which undergoes a Hawking Page phase transition at a non-zero value of the entropy. The same features hold when we increase
$\omega$ with $\phi$ fixed to zero, or as we increase $\phi$ with $\omega$ fixed at zero. With $\omega=0$ and $\phi = 1$ (the blue curve), we encounter
the RN-AdS black hole (with unit electric potential), which is stable at all values of the temperature.

We now embark upon an analysis of the phase structure of the KN-AdS black hole from the point of view of thermodynamic geometry. 
The line element for KN-AdS black holes is given by 
\begin{equation}
dl^2 = g_{ij}dx^idx^j
\end{equation}
where $g_{ij}$ has been defined in eq. (\ref{metric}) and the coordinates $x^k$ are the scaled variables $m,j,q$.   

Let us point out
here that in general, some of the regions in parameter space might be unphysical in the conventional sense that the fluctuations might be
unstable \cite{rupp3},\cite{rupp4}. Hence we begin with a few words on the stability of the system. 

\begin{figure}[!ht]
\centering
\includegraphics[width=3in,height=2.5in]{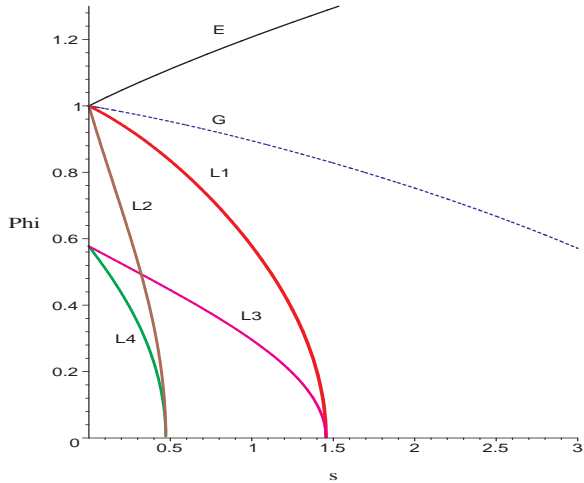}
\caption{Stability curves in the ${\phi}$-$s$ plane with ${\omega}$ fixed at $0.7$.}
\label{fig0c}
\end{figure}

As emphasized in \cite{rupp}, a necessary and sufficient condition for the line element in the thermodynamic state space to be positive-definite is that all the leading 
principal minors of the metric elements, including the determinant, be positive. This positivity constraint on the principal minors follows from requirement of 
thermodynamic stability, \emph{i.e}, the Le Chatelier's principle. In fig. 23, we depict various regions in the ${\phi}$-$s$ plane bounded by the stability curves 
$L1$, $L2$, $L3$, and $L4$. In addition, we have also depicted the extremal curve $E$ ,with physical regions lying on and below it, and the Hawking-Page curve 
$G$, which is the line of zeroes of Gibbs free energy of the black hole.  The stability curve $L4$  corresponds to the zeroes of the first minor $p_1=g_{mm}$, which is 
negative below and positive above the curve. Depending on whether the sequence of co-ordinates is ${\{m,j,q\}}$ or ${\{m,q,j\}}$ either of the curves $L2$ or $L3$ 
will correspond to the zeroes of the second leading minor, $p_2=(g_{mm}g_{jj}-g_{mj}g_{jm})$ or 
$p_2'=(g_{mm}g_{qq}-g_{mq}g_{qm})$. Again, this minor is negative below and positive above $L2$ or $L3$. 
Finally, the curve $L1$ corresponds to the zeroes of the determinant of the full metric, with the positive regions lying above the curve. It is now quite evident from the 
figure that the quadratic form will be positive definite for regions bounded by the curve $L1$ from below and the extremal curve $E$ from above. The interpretation of 
these curves in terms of thermodynamic stability is as follows. Moving across the curve $L4$ from left to right the specific heat $c_{qj}$ changes sign from negative to
positive through an infinite discontinuity. Across the curves $L2$ and $L3$, the corresponding susceptibilities 
$({\partial}j/{\partial}{\omega})|_{\phi}$ and $({\partial}q /{\partial}{\phi})|_{\omega}$ 
change sign from positive to negative by passing through zero. We may add that $L2$ and $L3$ are also the lines of divergence of 
the specific heats $c_{j\phi}$ and 
$c_{q\omega}$ respectively, below which the corresponding ``mixed'' specific heats are negative. Across the curve $L1$ the specific heat $c_{\phi\omega}$ 
changes sign from negative to positive through an infinite discontinuity. Besides, on crossing $L1$ the previously mentioned susceptibilities change sign again 
and become positive through an infinite discontinuity. Thus, we can conclude that only in the region outside the curve $L1$ is the black hole on the whole locally 
thermally stable in the grand canonical ensemble, even if there exist pockets of mechanical and electrical stability inside $L1$. Finally, we may add that the 
black hole becomes globally stable against thermal AdS on crossing to the right of the Hawking-Page curve.

As we have mentioned, in the Ruppeiner analysis, we are naturally in the grand canonical ensemble, if we keep all the potentials fixed and let the corresponding
charges fluctuate. Naturally, the variables appearing in the Ruppeiner curvature represent those quantities that are fluctuating. In the grand canonical
ensemble the Ruppeiner curvature has been calculated in the literature \cite{mirza}. We will present some of the details here, from a slightly different
perspective.

\begin{figure}[!ht]
\begin{minipage}[b]{0.5\linewidth}
\centering
\includegraphics[width=2.5in,height=2.7in]{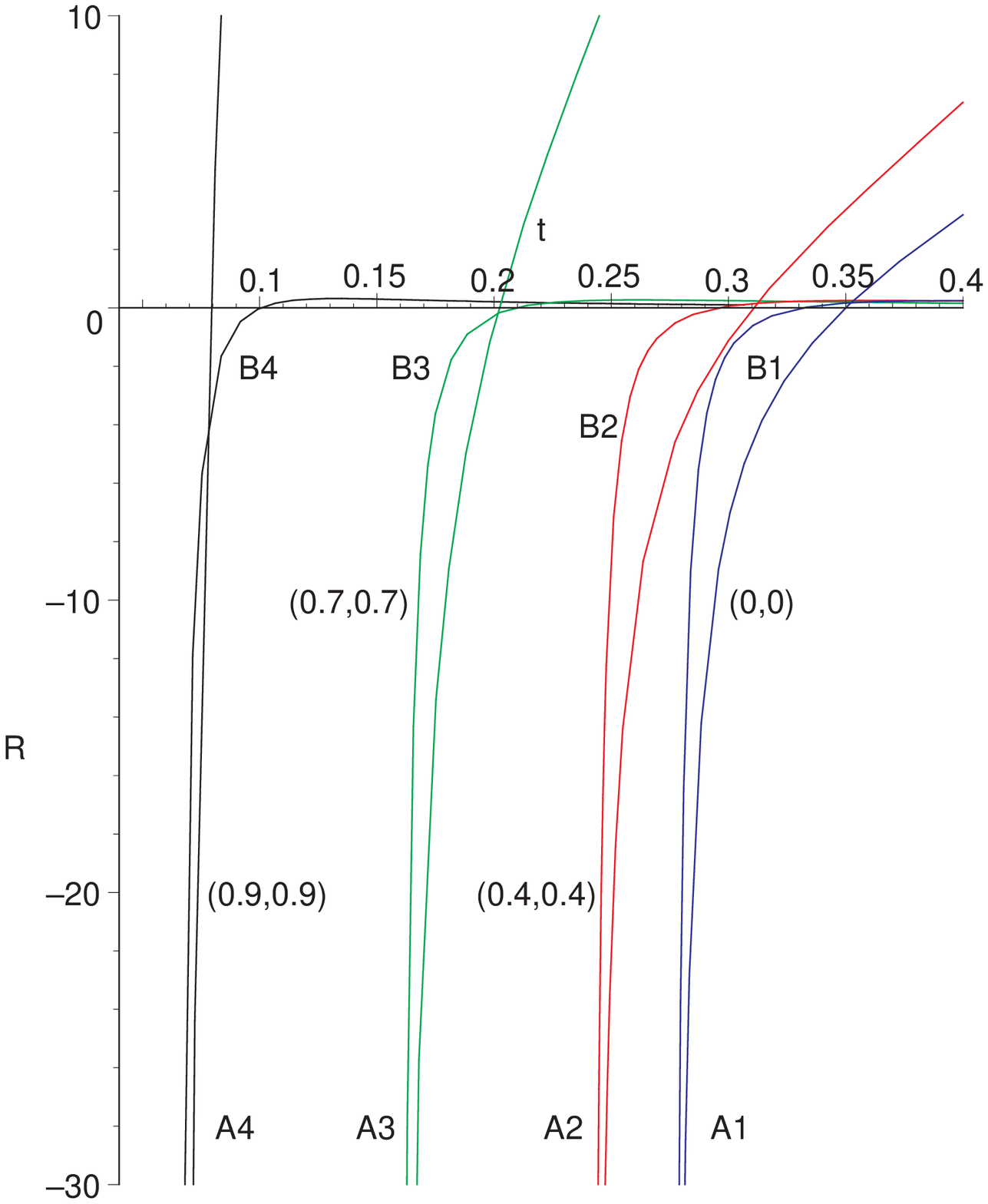}
\caption{ Ruppeiner curvature vs temperature for various values of $\left(\omega, \phi\right)$.}
\label{kn3}
\end{minipage}
\hspace{0.6cm}
\begin{minipage}[b]{0.5\linewidth}
\centering
\includegraphics[width=2.5in,height=2.7in]{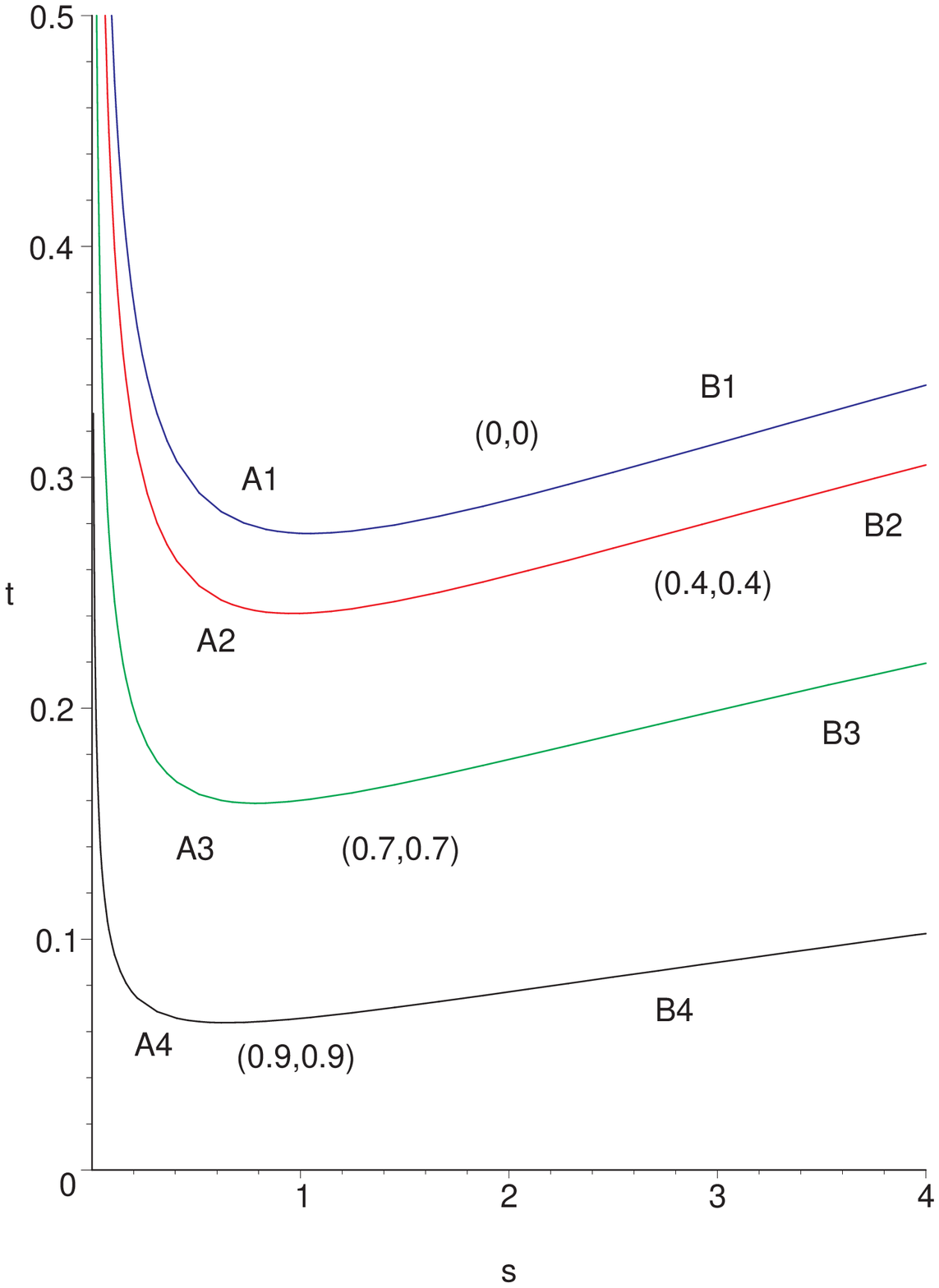}
\caption{Temperature vs entropy for same values of $\left(\omega, \phi\right)$}
\label{kn4}
\end{minipage}
\end{figure}

We find it convenient to study the Ruppeiner curvature as a function of the temperature of  the system, for given values of the (fixed) potentials,
$\omega$ and $\phi$, with entropy $s$ being treated as a parameter that we vary over suitable ranges. 
In fig. (\ref{kn3}), we have presented the plots of the Ruppeiner curvature vs temperature for various values of
$\left(\omega,\phi\right)$, with both the potentials taking values less than unity. \footnote{For other values of $\omega$ and $\phi$, the Ruppeiner
curvature shows standard behaviour, in the absence of multiple branches of the physical solution.}
For convenience, we have chosen $\omega = \phi$, although the behaviour of the curvature does not qualitatively change
for unequal choices of these parameters. For comparison purposes, we have, in fig. (\ref{kn4}), shown the temperature vs entropy plots for the
chosen values of the potentials of fig. (\ref{kn3}). Let us briefly explain the figures. First of all, note that in these cases, there are two branches of the solution, and hence
the Ruppeiner curvature is also double valued for all values of the temperature.\footnote{As before the unstable branches are irrelevant to thermodynamic
geometry.} In fig. (\ref{kn4}), the points $A1\cdots A4$ are those for which the $\left(\frac{
\partial s}{\partial t}\right)$ goes to infinity. These are analogous to the Davies points, but since the portions of the graphs to the left of these
points lie on the unstable branch of the system, no physical meaning can be attributed to them. The system progresses, beyond a certain
value of the entropy along the ``stable'' branches, and we have plotted the representative points as $B1 \cdots B4$. In fig. (\ref{kn3}),
corresponding points are shown. As the entropy is increased from zero, the Ruppeiner curvature proceeds along the unstable branch and goes to
negative infinity at the ``Davies'' points ($A1 \cdots A4$) and then rises along the stable branch, through points $B1 \cdots B4$ as the entropy is
further increased.

The behaviour of the Ruppeiner curvature, is as expected, not particularly interesting in this case, in the absence of clearly defined phase transitions.
However, it is interesting to note that this is an example where the curvature changes sign. The precise physical meaning of the
curvature going to zero is still not understood, and the issue has been debated in the literature. We are in a position to make a few preliminary comments
on this, in this example. It is believed that the Ruppeiner curvature $R$ changing sign is expected to signal some kind of
``instability'' in the system. In this case, the only transition in the system we have is the Hawking Page transition. With this in mind, we compare $R$ with the the
Gibbs free energy near the Hawking Page transition point. The Gibbs free energy is readily computed in terms of the charges to be
\begin{equation}
g = \frac{s^2\pi^2 + 4\pi^4j^2 - \pi^4q^4 - 2s^2q^2\pi^2 - s^4}{4\pi^{\frac{3}{2}}s^{\frac{1}{2}}
\left(s^4 + 2s^3\pi + s^2\pi^2\left(2q^2 + 1\right) + 2\pi^3s\left(q^2 + 2j^2\right) + 4\pi^4j^2 + \pi^4q^4\right)}
\label{gibbs1}
\end{equation}
Eq. (\ref{gibbs1}) can now be expressed in terms of the potentials using eq. (\ref{chargepot}). The expression is lengthy and we will not reproduce it here.
Rather, we will give a graphical comparison between $g$ and the $R$ near the Hawking Page transition point.

\begin{figure}[!ht]
\begin{minipage}[b]{0.5\linewidth}
\centering
\includegraphics[width=2.5in,height=2.7in]{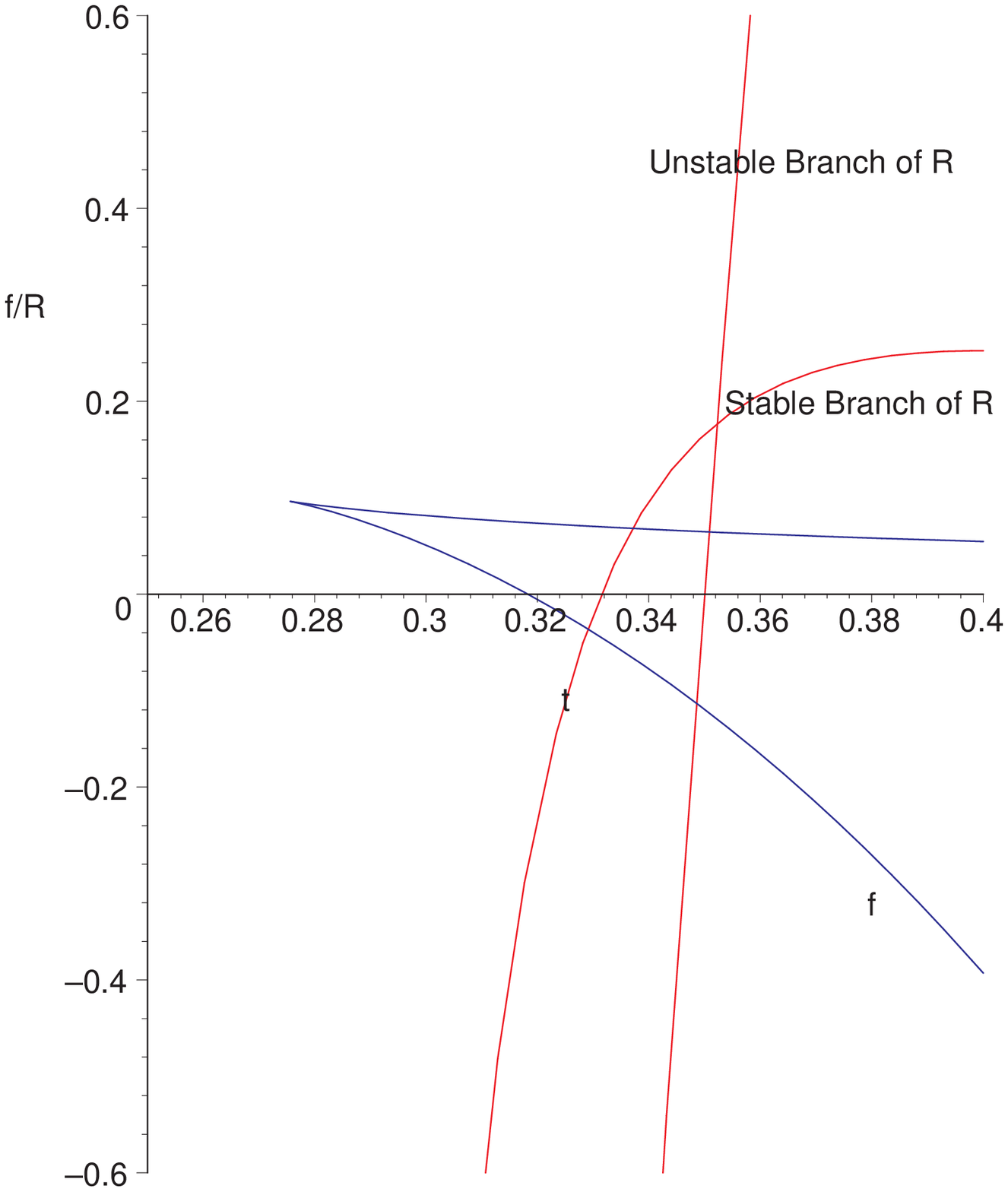}
\caption{ Ruppeiner curvature and Gibbs free energy for $\left(\omega, \phi\right) = \left(0,0\right)$ as a function of temperature.}
\label{kn5}
\end{minipage}
\hspace{0.6cm}
\begin{minipage}[b]{0.5\linewidth}
\centering
\includegraphics[width=2.5in,height=2.7in]{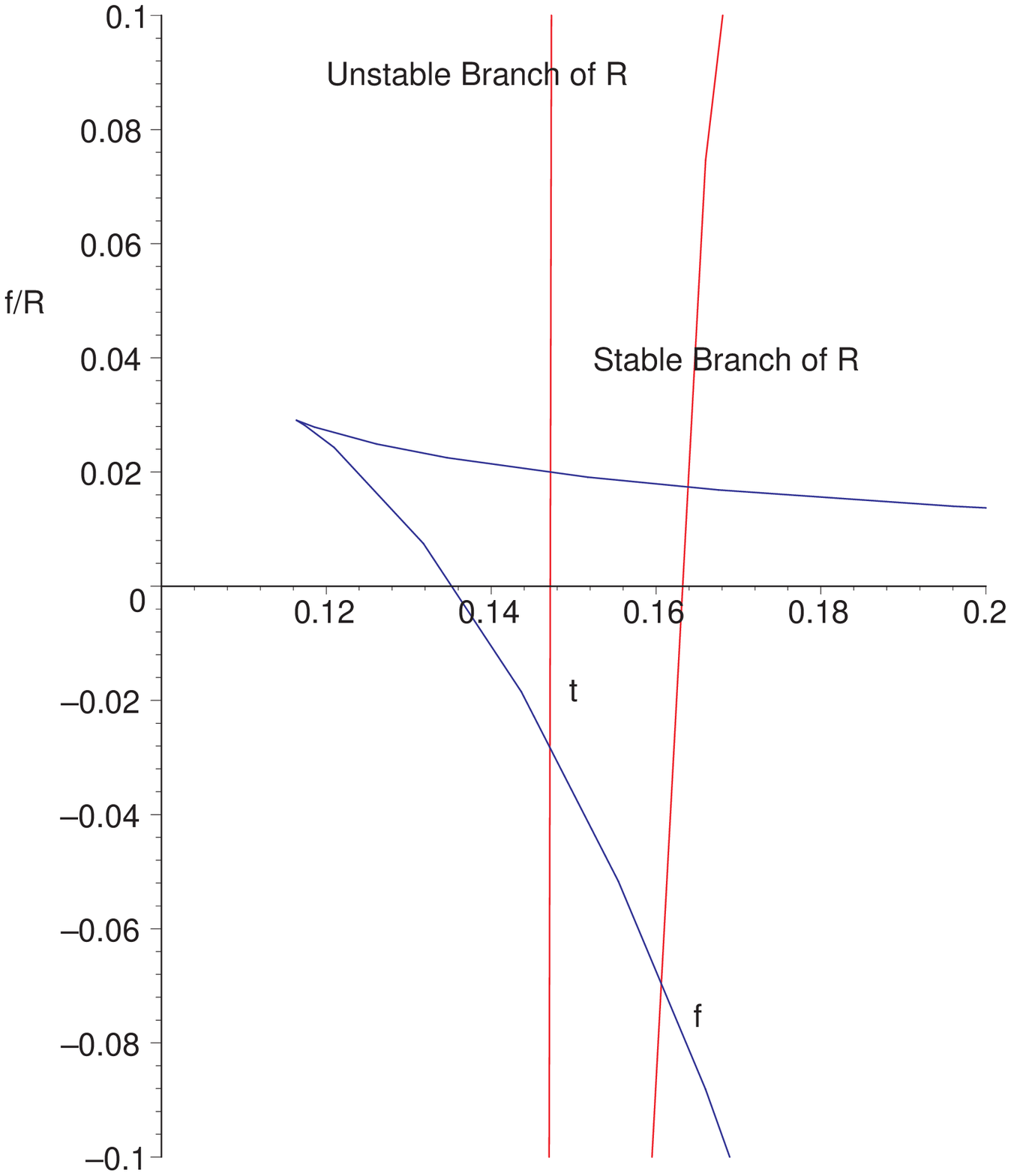}
\caption{Ruppeiner curvature and Gibbs free energy for $\left(\omega, \phi\right) = \left(0.8,0.8\right)$ as a function of temperature.}
\label{kn6}
\end{minipage}
\end{figure}

In figs. (\ref{kn5}) and (\ref{kn6}), we compare the behaviour of the Gibbs free energy with the Ruppeiner curvature near the Hawking Page
transition. We have magnified the scale in the figures, so as to provide a colser view of the difference between the behaviour of the two.
In both these figures, $R$ is plotted in red, while the free energy is plotted in blue. We see that $R$  has a slight deviation
(at the second decimal place) from the free energy at the Hawking Page transition. Indeed, it is possible to calculate this deviation exactly, from
the solutions of $R=0$ or $g=0$.
We find that for $\left(\omega, \phi\right) = \left(0,0\right)$ of fig. (\ref{kn5}), whereas the Hawking Page transition occurs at
$t = 0.3183$, the stable branch of the Ruppeiner curvature crosses zero at $t = 0.3314$.
The situation is similar for $\left(\omega, \phi\right) = \left(0.8,0.8\right)$, in fig. (\ref{kn6}). Here, the Hawking Page transition
temperature is calculated to be at  $t = 0.1356$, whereas the stable branch of $R$ crosses zero at $t = 0.1623$.  We note that
for the case $\omega=0$ but non zero $\phi$, the temperature at which the Ruppeiner curvature changes sign becomes almost indistinguishable
from that at which the Gibbs free energy changes sign (signalling the Hawking Page transition
). Let us mention here that the
zero crossing of the Ruppeiner curvature in fact exactly coincides with the Hawking Page transition for the RN-AdS black holes.

\subsection{KN-AdS black hole in the fixed $j$ ensemble}

We now turn to the description of KN-AdS black holes in the mixed ensemble described before. We will keep the angular momentum fixed (i.e in calculating
the Ruppeiner curvature, the angular momentum is not treated as a variable) and further set the electric potential to a constant value. Then, the system
can be thought as being in equilibrium with a charge reservoir (with which it exchanges charge while keeping the electric potential fixed) and hence
is in a grand canonical ensemble as far as the electric charge is concerned. However, having fixed the angular momentum, the system is in a canonical
ensemble in that it does not exchange angular momentum with its surrounding, while the angular velocity may vary. As already mentioned, we will call this the
``fixed $j$'' ensemble. As far as thermodynamics is concerned, such ensembles are perfectly admissible.

\begin{figure}[!ht]
\begin{minipage}[b]{0.5\linewidth}
\centering
\includegraphics[width=2.5in,height=2.7in]{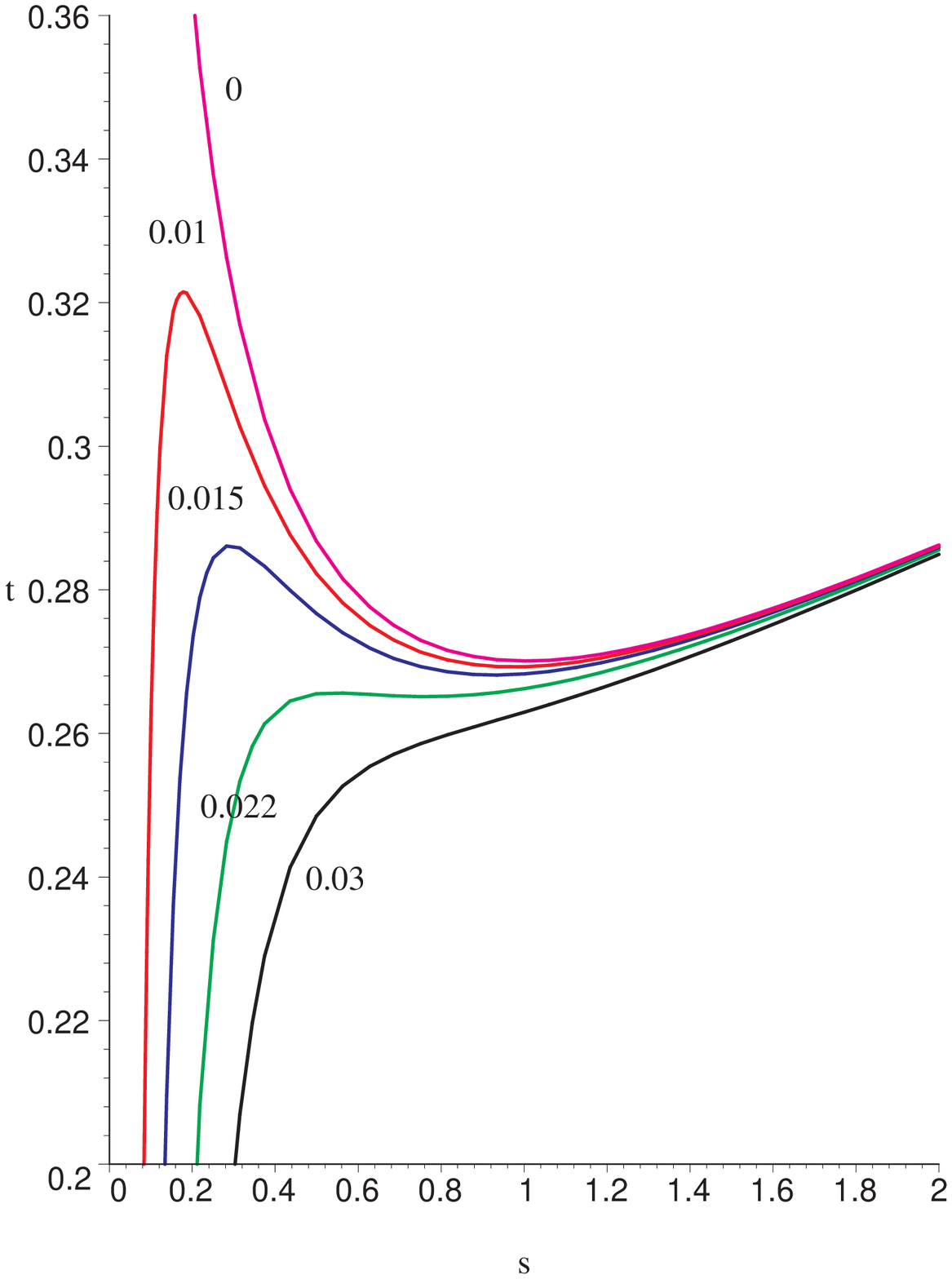}
\caption{Temperature vs entropy for $\phi=0.2$ and various fixed values of $j$}
\label{kn7}
\end{minipage}
\hspace{0.6cm}
\begin{minipage}[b]{0.5\linewidth}
\centering
\includegraphics[width=2.5in,height=2.7in]{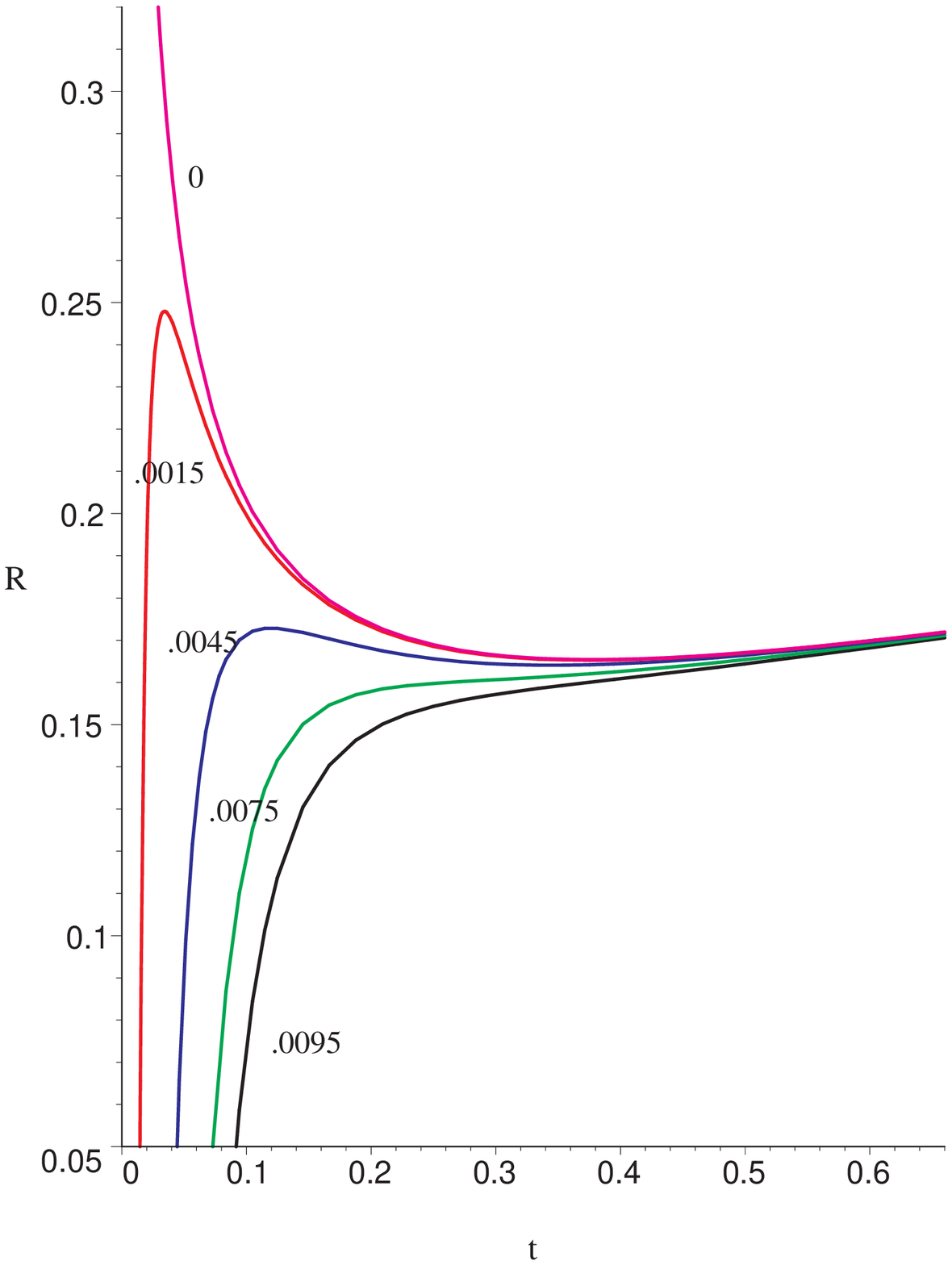}
\caption{Temperature vs entropy for $\phi=0.8$ and various fixed values of $j$}
\label{kn8}
\end{minipage}
\end{figure}

For the fixed $j$ ensemble, it is appropriate to study the thermodynamics by expressing the charge $q$ in terms of $\phi$ in
eq. (\ref{potkn}). The solution is multivalued, and the correct solution is chosen by matching with the $j=0$ limit of the RN-AdS black hole. Once this
is obtained, we can express the temperature in eq. (\ref{tempkn}) in terms of the entropy, the (constant) angular momentum and the (fixed) electric
potential. The result is lengthy, and we can only discuss our results graphically.
In figs. (\ref{kn7}) and (\ref{kn8}),
we have plotted the behaviour of temperature with entropy, for some representative values of $\phi = 0.2$ and $\phi = 0.8$ respectively, for certain
values of the angular momentum.

In fig. (\ref{kn7}), the pink, red, blue green and black lines corresponds to $j=0$, $j = 0.01$, $0.015$, $0.021$ and $0.03$ respectively, while for fig. (\ref{kn8}), the 
corresponding values are  $j=0$, $j = 0.001$, $0.002$, $0.005$, and $0.007$. For each non zero value of $j$, we see from the figures that there is a stable branch that 
starts from extremality, with a non zero entropy. As the entropy is increased, there is a turning point in the $t - s$ plane where the specific heat diverges, and an unstable
branch appears with negative specific heat. This continues into a second stable branch after a second turning point where the specific heat again diverges.
It can be seen from the figures that such behaviour is present only below a certain value of the angular momentum $j$ for a given value of the electric
potential $\phi$. Hence, we see that there is a critical value of $j$, for a given (fixed) $\phi$ below which the
system will exhibit a {\it first order} phase transition (from a small black hole phase to a large black hole phase), as the entropy is  multi valued below
this $j$. At the critical value of $j$, the system undergoes a second order phase transition, and the coexistance curve disappeaers.

\begin{figure}[!ht]
\begin{minipage}[b]{0.5\linewidth}
\centering
\includegraphics[width=2.7in,height=3.0in]{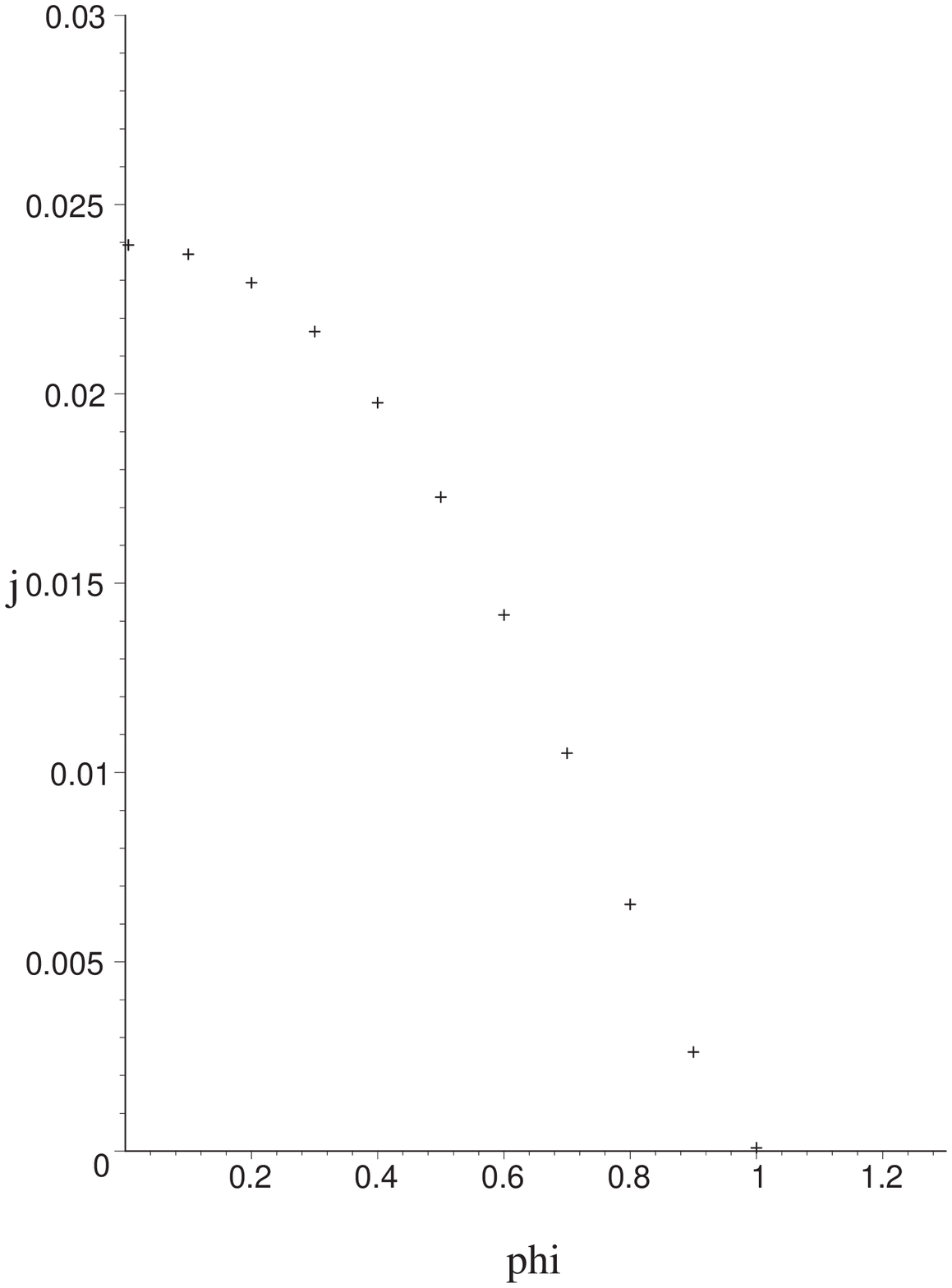}
\caption{Critical value of $j$ vs $\phi$ for the KN-AdS black hole in the fixed $j$ ensemble.}
\label{kn9}
\end{minipage}
\hspace{0.6cm}
\begin{minipage}[b]{0.5\linewidth}
\centering
\includegraphics[width=2.7in,height=3.0in]{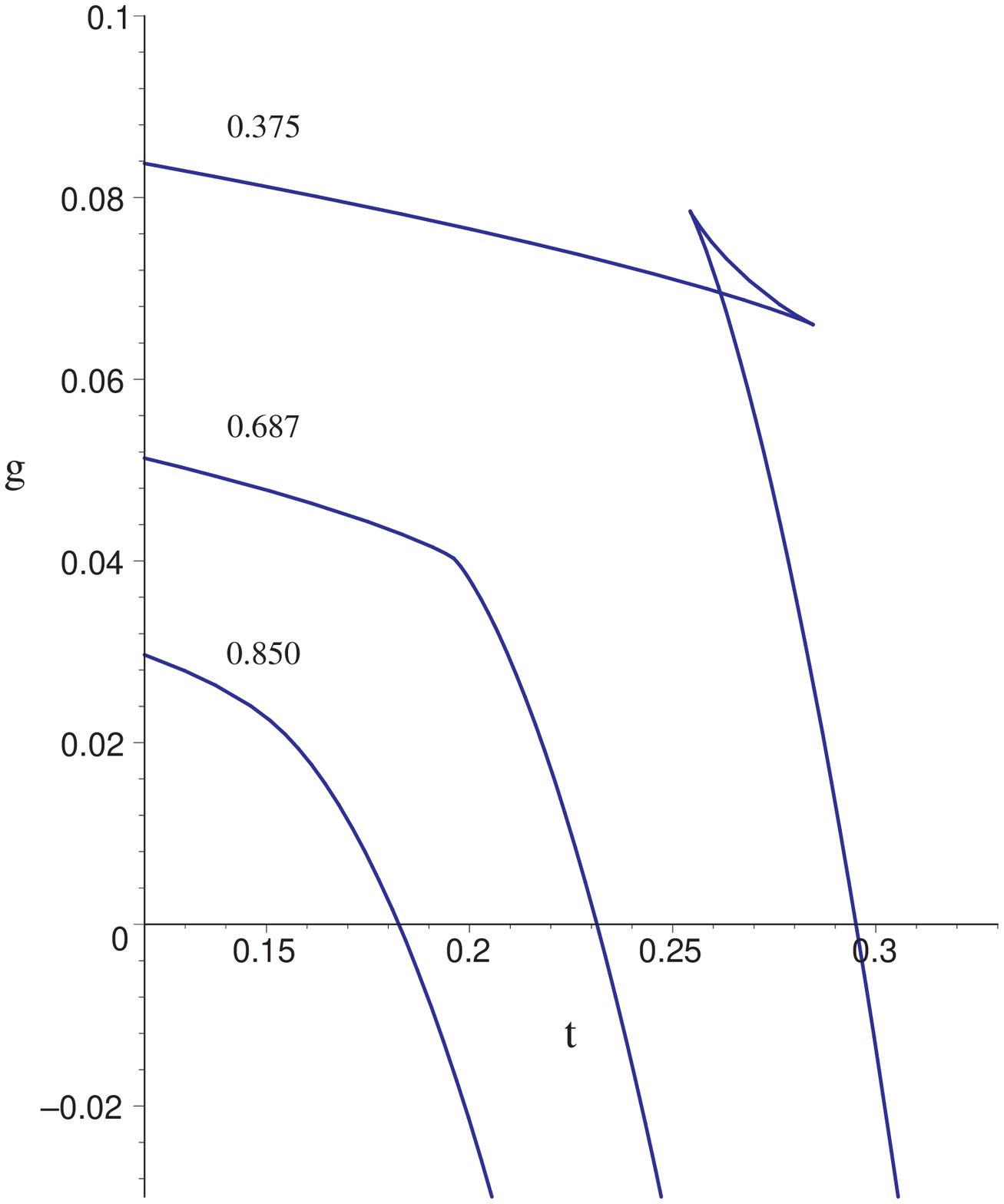}
\caption{Free energy plots for $j=0.011$ ensemble along isopotentials with $\phi$ fixed at 0.375, 0.687 (${\phi}_c$) and 0.850.}
\label{kn10}
\end{minipage}
\end{figure}

The critical value of the angular momentum at which this second order phase transition occurs can be calculated, but an exact analytic expression is
difficult to obtain. In fig. (\ref{kn9}), we present the graph of the critical value of $j$ vs the electric potential $\phi$ plotted as a set of
points calculated numerically. Fig. (\ref{kn10}) shows the behaviour of the Gibbs free energy for a subcritical, critical, and supercritcal value of $\phi$ 
for this ensemble with $j = 0.011$. With the aid of fig (\ref{kn9}), we can further improve our understanding of figs. (\ref{kn7}) and (\ref{kn8}).
In the former, we have set $\phi = 0.2$.
From fig. (\ref{kn9}), it can be checked that for this value of $\phi$, the critical value of $j$ is $0.2293$. Indeed, for $j = 0.03$ (the black curve in fig. (\ref{kn7})),
we find that the entropy is single valued. Similarly, for $\phi = 0.8$, the critical value of the angular momentum is $j = 0.0065$. At $j = 0.007$
(the black line in fig. (\ref{kn8}),  the entropy is single valued again, signalling no unstable regions. As expected, below these critical values of $j$, the system
has two stable branches connected by an unstable branch. Note that in fig. (\ref{kn10}), the Gibbs free energy of the system is positive. As mentioned
earlier, however, the issue of global stability is not yet fully settled, and we proceed with the understanding that we are possibly dealing with long
lived metastable states.

\begin{figure}[!ht]
\begin{minipage}[b]{0.5\linewidth}
\centering
\includegraphics[width=2.7in,height=3.0in]{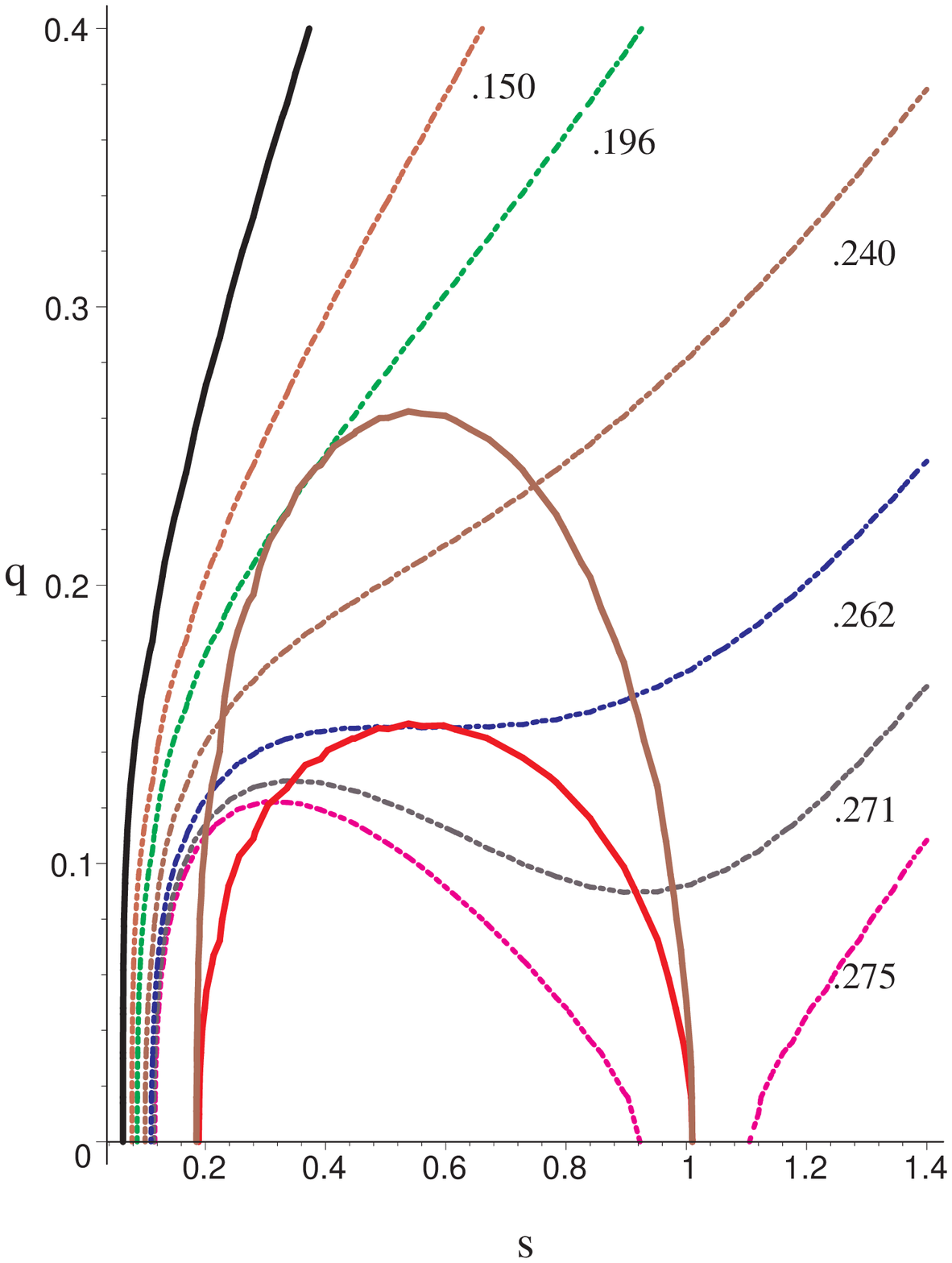}
\caption{Phase plot of isotherms in the $q-s$ plane for fixed $j$ ensemble,with $j=0.011$. The numbers denote the various values
of temperature. }
\label{kn11}
\end{minipage}
\hspace{0.6cm}
\begin{minipage}[b]{0.5\linewidth}
\centering
\includegraphics[width=2.5in,height=3.0in]{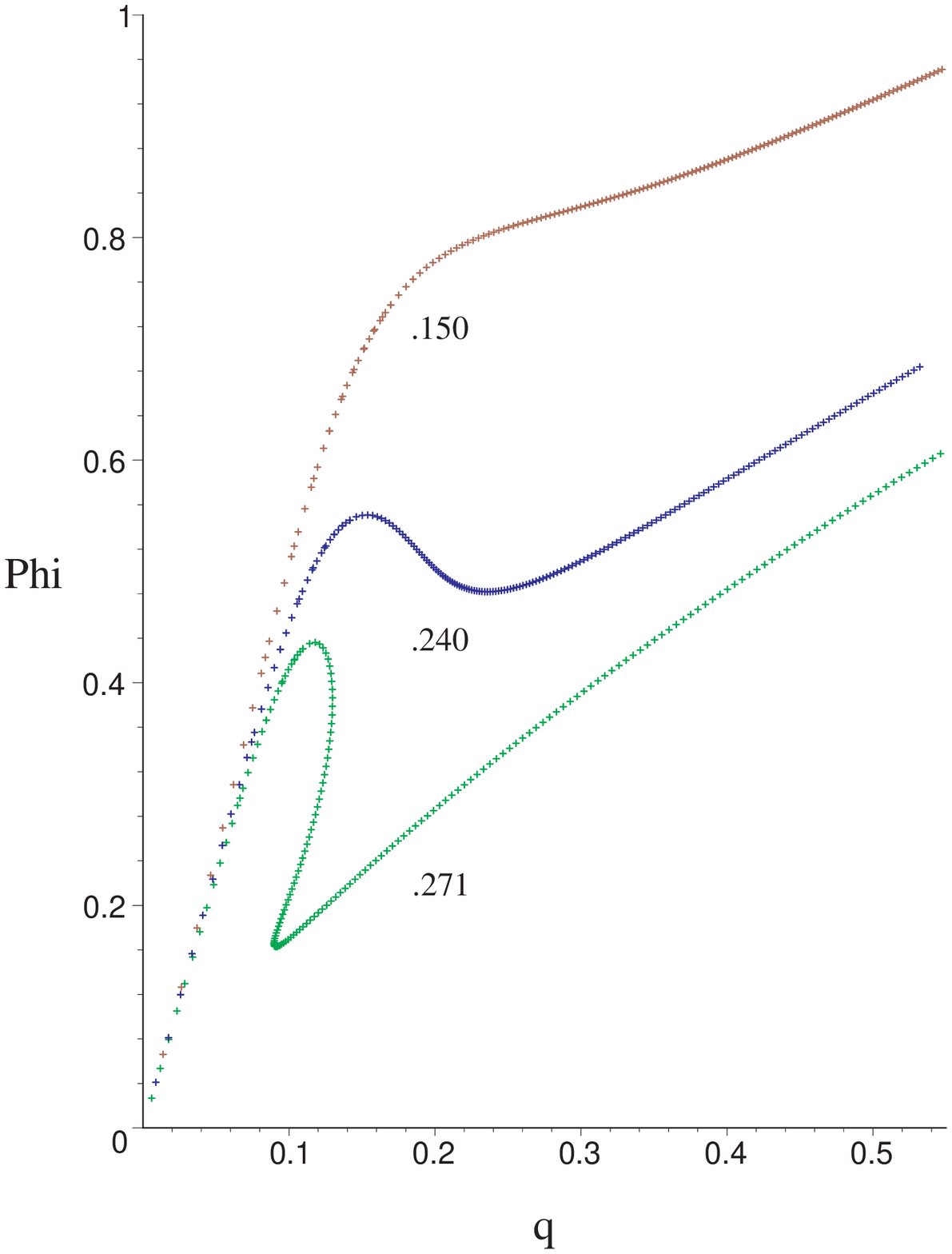}
\caption{Isotherms in the $q- \phi$ plane in the fixed $j$ ensemble, with $j=0.011$. The numbers denote some values of $t$ in fig. (\ref{kn11}).}
\label{kn12}
\end{minipage}
\end{figure}

We will now make some general comments about the phase structure in this ensemble. 
In fig. (\ref{kn11}),
we have plotted, in the $q-s$ plane, isotherms for the fixed value of $j = 0.011$. In fig. (\ref{kn12}), we have shown the isotherms in the
$\phi - q$ plane corresponding to some values of the temperature in fig. (\ref{kn11}). Let us first focus on fig. (\ref{kn11}). To understand this
figure, we use
he specific heats at constant potential and constant charge
\begin{equation}
c_{\phi} =
t\left(\frac{\partial s}{\partial t}\right)_{\phi}, ~~
c_q = t\left(\frac{\partial s}{\partial t}\right)_{q}
\label{specificj}
\end{equation}
Clearly, $c_{\phi}$ and $c_q$ are the restrictions of the specific heats $c_{j\phi}$ and $c_{jq}$ to constant $j$ sections of the parameter space. These can be 
effectively used to understand the phase behaviour. Let us see if we can substantiate this. 

In fig. (\ref{kn11}), the brown and red semi circular lines denote the curves along which the specific heats $c_{\phi}$ and $c_q$ of
eq. (\ref{specificj}) blow up. We
will refer to these loosely as the $c_{\phi}$ curve and the $c_q$ curve respectively. The black line denotes the values of $q$ and $s$ at extremality,
with $j = 0.011$.
It can be shown by direct calculation that the capacitance $c = \left(\frac{\partial q}{\partial \phi}\right)_{t}$ of the system is infinite on all
points of the boundary of the $c_{\phi}$  curve and is zero on all points of the boundary of the $c_q$ curve. The exact expressions are lengthy, and
in order to justify the above remarks, we simply state that the numerator
of the capacitance is proportional to the denominator of $c_q$ and the denominator of the capacitance is proportional to that of $c_{\phi}$. We find that
\begin{equation}
\frac{N_{\rm cap}}{D_{c_q}} = s^{\frac{1}{2}}\left( 4\pi^4j^2 +  \pi^4q^4+ 2q^2\pi^3s  + 4j^2\pi^3s + s^2\pi^2 +2s^2q^2\pi^2 + 2s^3\pi + s^4\right)^{\frac{1}{2}}
\end{equation}
where $N_{\rm cap}$ is the numerator of the capacitance and $D_{c_q}$ is the denominator of the specific heat $c_q$. A similar calculation yields
\begin{equation}
\frac{D_{\rm cap}}{D_{c_{\phi}}} = \pi^{\frac{1}{2}}
\end{equation}
where $D_{\rm cap}$ is the denominator of the capacitance and $D_{c_{\phi}}$ is the denominator of the specific heat $c_{\phi}$.

It can also be checked that the capacitance $c$ is negative at all points on the $q - s$ plane that lie outside the (brown) $c_{\phi}$ curve, and changes sign
through an infinite discontinuity at the points on that curve, where $c_{\phi}$ blows up. It further becomes zero at the points where $c_q$ diverges
and is positive at all points in the $q-s$ plane inside the $c_q$ curve,
bounded by the red line in fig. (\ref{kn11}). The isotherms of fig. (\ref{kn12}) can now be easily understood. For the $t=0.271$ isotherm, we see
from fig. (\ref{kn11}) that as we increase the entropy from its extremal value, the capacitance is initially positive, becomes infinite and then assumes
negative values in the region between the $c_{\phi}$ and $c_q$ curves, then becomes zero at the boundary of the $c_q$ curve. Thereafter, it
changes sign and becomes positive. It then repeats this behaviour in the reverse order until it exits from the boundary of the $c_{\phi}$ curve.
The isotherms corresponding to $t = 0.240$ and $t = 0.140$ should be self explanatory.

\begin{figure}[!ht]
\begin{minipage}[b]{0.5\linewidth}
\centering
\includegraphics[width=2.7in,height=3.0in]{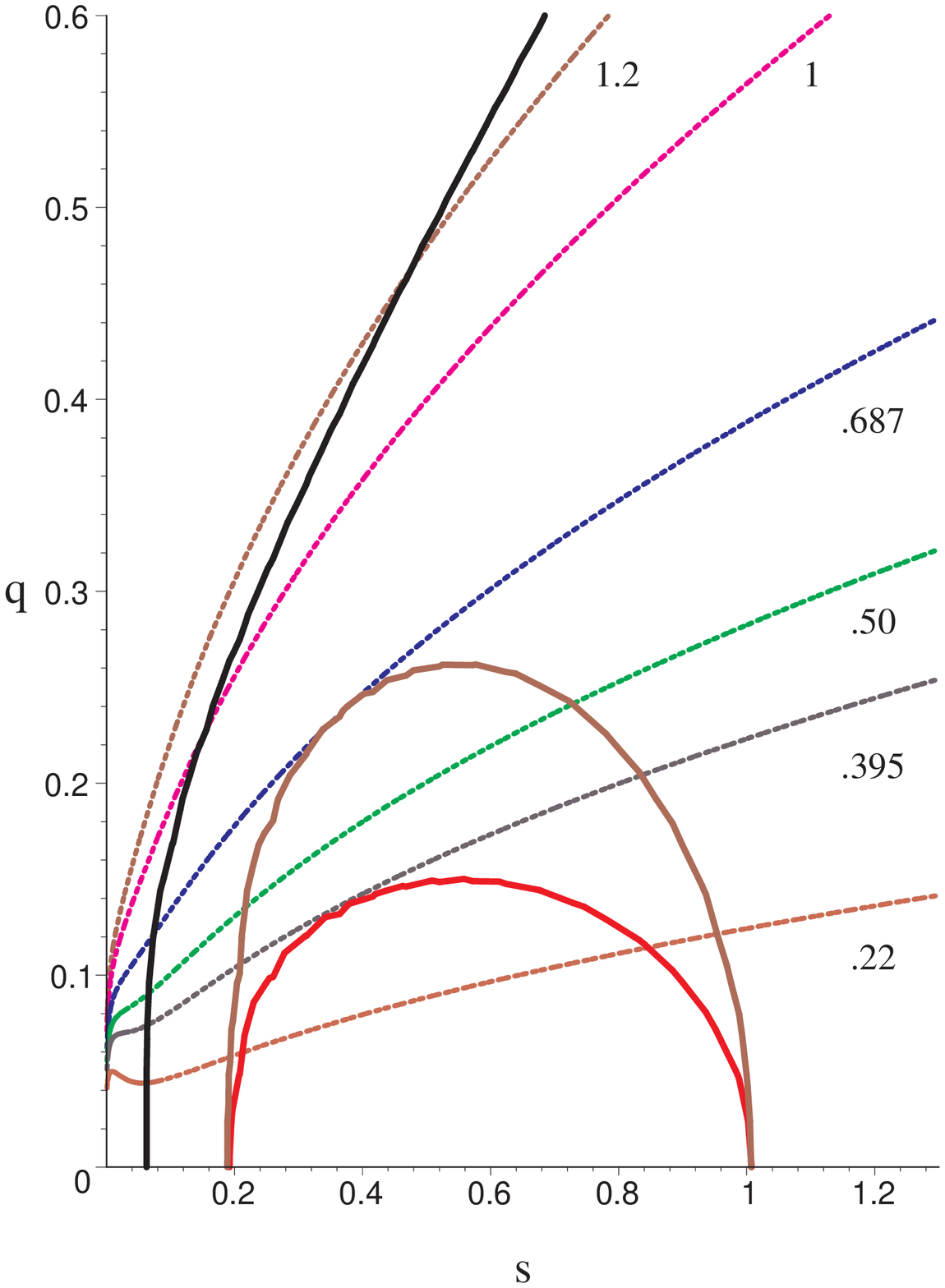}
\caption{Isopotential phase plot in the $q-s$ plane in the fixed $j$ ensemble, with $j=0.011$, for various values of $\phi$.}
\label{kn13}
\end{minipage}
\hspace{0.6cm}
\begin{minipage}[b]{0.5\linewidth}
\centering
\includegraphics[width=2.7in,height=3.0in]{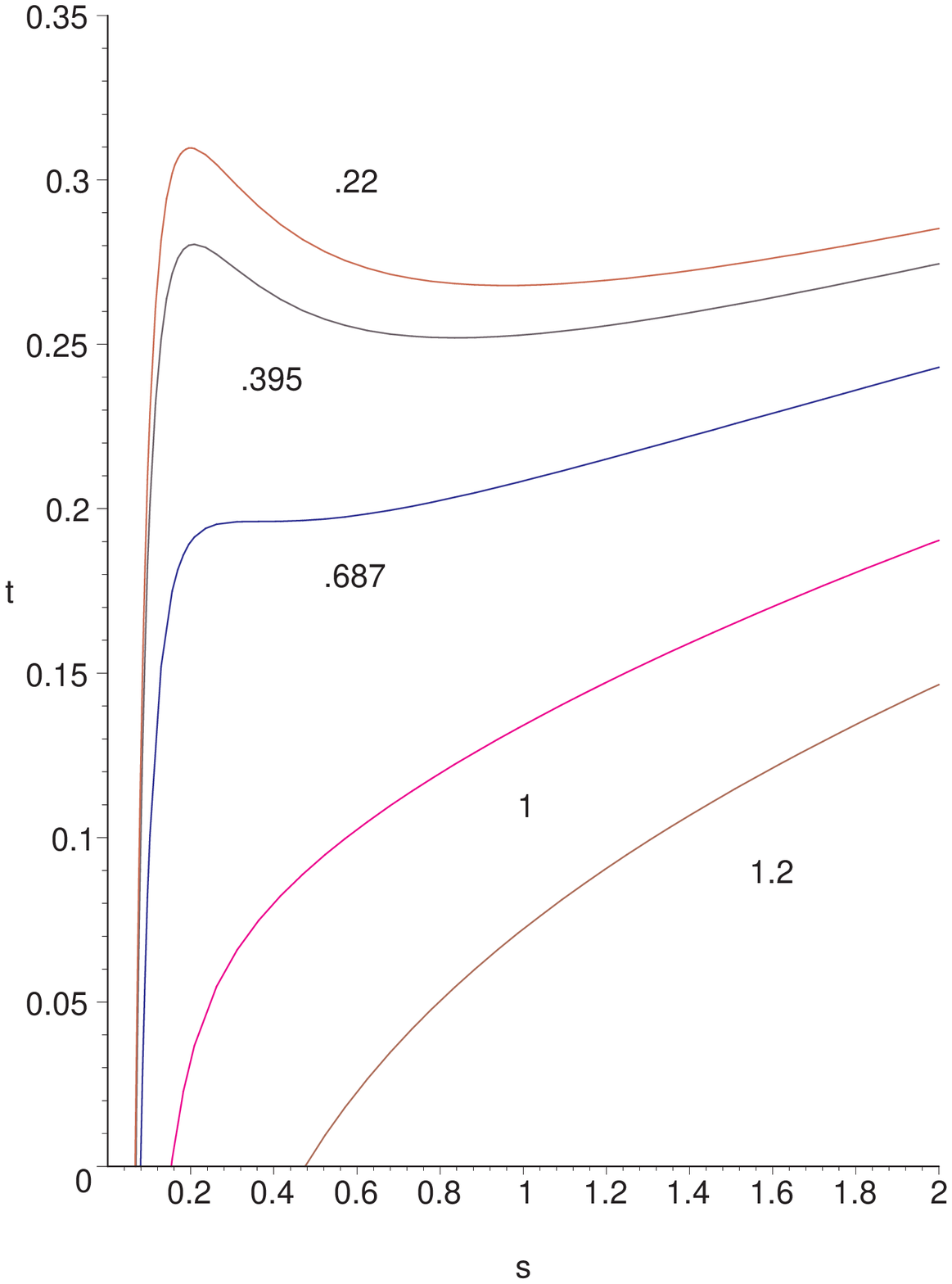}
\caption{Isopotential plots for $t$ vs $s$ for some fixed values of $\phi$ of fig. (\ref{kn15}), in the fixed $j$ ensemble with $j = 0.011$}
\label{kn14}
\end{minipage}
\end{figure}

To complete the analysis, we now turn to the isopotential graphs.
We refer to fig. (\ref{kn13}). In this figure, we have plotted the isopotential (fixed $\phi$) curves in the $q-s$ plane corresponding to a chosen value of
$j = 0.011$. The brown and red lines still denote the $c_{\phi}$ and $c_q$ curves as in fig. (\ref{kn11}), where $c_{\phi}$ and $c_q$ are the
specific heats as before (the $c_q$ curve will, however, not be important for us in this analysis). The black line denotes the exremal black hole, with
zero temperature for $j = 0.011$. As we have mentioned, in this $q-s$ parameter space, the specific heat $c_{\phi}$ is positive for all
values of $\left(q,s\right)$ that lies outside the $c_{\phi}$ curve, and is negative for the region inside
the $c_{\phi}$ curve; it changes sign via a discontinuity on the curve. The reader would note that this graph also clearly indicates the possibility
of first order phase transitions in the mixed ensemble. There is a region between the extremal line and the $c_{\phi}$ curve, where the
specific heat $c_{\phi}$ is positive. For appropriate values of $\phi$, $c_{\phi}$ becomes positive after passing through two infinites. In the
$t - s$ plane, this behaviour clearly indicates a first order phase transition between a large black hole and a small black hole. We have shown some
of these plots in fig. (\ref{kn14}). The critical value of $\phi = 0.687$ for $j = 0.011$ denotes the end of the first order curve, and a second order
phase transition occurs for these values of $j$ and $\phi$. In fig. (\ref{kn14}), this corresponds to the blue line which, as can be checked,  has a
point of inflection, and this is the second order phase transition point. Above this line, the system undergoes first order phase transitions, as seen from the
multivaluedness of the temperature with entropy. Below this, there are no phase transitions and the system resides in the unique large black hole phase.
Also note that for any given value of the angular momentum $j$, there are regions where the angular velocity $\omega > 1$. Removing these regions
in parameter space will not qualitatively affect our analysis above. 

We now elaborate upon the Ruppeiner curvature, $R$. Since in this ensemble the angular momentum $j$ is held fixed, the thermodynamic state space is two 
dimensional, with the charge $q$ and and the internal energy $m$ as the two independent fluctuating variables. The curvature  obtained from the $2-d$ Riemannian 
metric is a function of $q$ and $s$, with $j$ appearing as an ensemble parameter. \footnote{Even though $m$ and $q$ are the independent fluctuating variables 
we find it convenient to perform all our calculations with $s$ and $q$ as the independent 
\emph{paramaters}. This has to do with the difficulty involved in inverting the Smarr formula for $m$. }

It may be verified that the line element is positive definite in the regions where $c_{\phi}$ is positive definite, which is our region of interest for the Ruppeiner 
curvature $R$.
The expression for $R$ being very lengthy it is best studied graphically. Let us first make a couple of preliminary
 observations on  $R$. Firstly, on setting $j$  to zero, the expression for $R$ reduces to that of RN AdS black hole, which is as expected because the $j=0$ ensemble 
 is the RN AdS black hole in grand canonical ensemble. Secondly, we find that apart from a factor of $s$, the denominator of $R$ can be factored into two polynomials, 
 one of them being the numerator of the expression for the temperature $t$ which we call the ``extremal polynomial'' in the sequel, and the other being the 
 square of the denominator of the expression for $c_{\phi}$. It can also be checked that in the limit of $s$ going to infinity $R$ is zero.

The behaviour of the curvature regarding its sign is more complex in this ensemble. Nevertheless, the magnitude of curvature should have a definite 
meaning regardless of its sign. As mentioned earlier, we will continue to refer to $R$ as a correlation volume \cite{rupp2}. 

\begin{figure}[!ht]
\begin{minipage}[b]{0.5\linewidth}
\centering
\includegraphics[width=2.7in,height=3.0in]{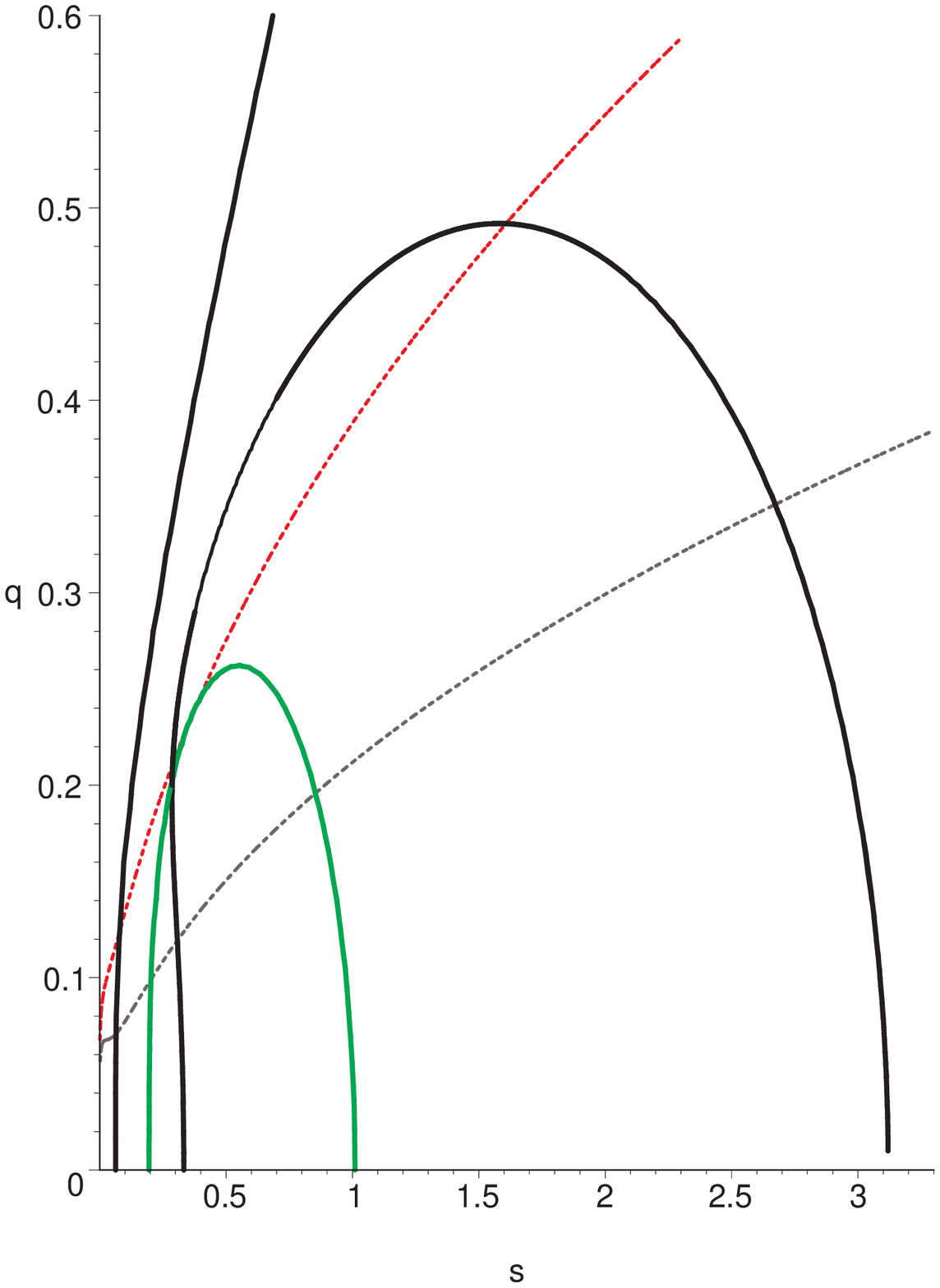}
\caption{Plot of zeroes and infinities of $R$ in the $j=0.011$ ensemble. Also shown are isopotentials with $\phi$=0.6873(${\phi}_c$) and 0.375 respectively}
\label{kn15}
\end{minipage}
\hspace{0.6cm}
\begin{minipage}[b]{0.5\linewidth}
\centering
\includegraphics[width=2.7in,height=3.0in]{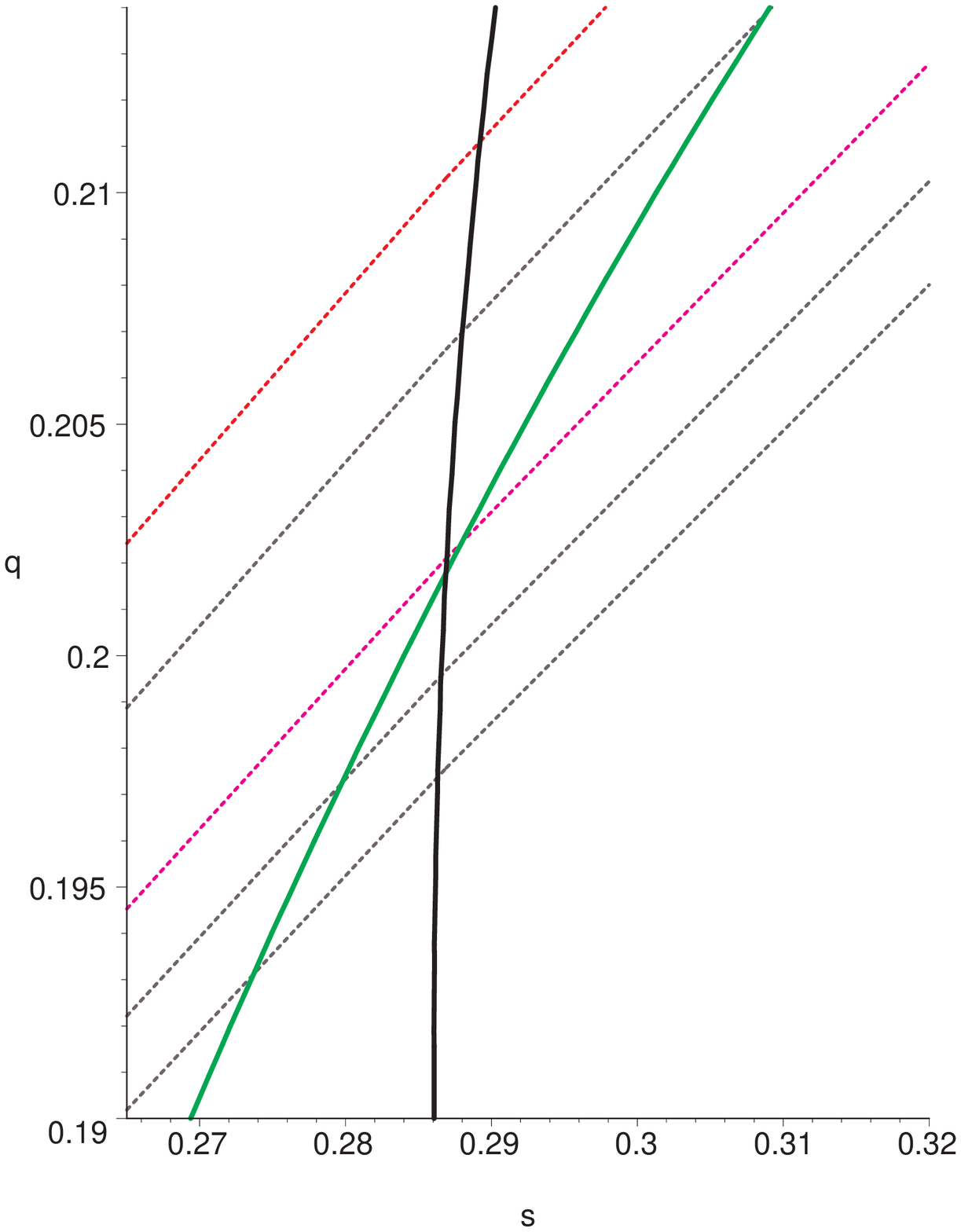}
\caption{Close up of the `negative window' of fig $12$  with the $\phi$=0.6873(${\phi}_c$), 0.675, 0.660, 0.652 and 0.645 curves drawn respectively from top to bottom}
\label{kn16}
\end{minipage}
\end{figure}

 Figure(\ref{kn15}) represents the zeroes and infinities of $R$ in the $q-s$ parameter space for the $j=0.011$ ensemble. In the figure the black semicircle represents the 
 zeroes of the numerator of $R$. We call this the N - curve for future reference. The black curve to the left is made of the zeroes of the ``extremal polynomial''  
 (that appears in the denominator of $R$), 
 and the green $c_{\phi}$ curve consists of the zeroes of the $c_{\phi}$ polynomial (which also appears in the denominator of $R$, as we have remarked before). 
 In the region of parameter space inside the $c_{\phi}$ curve, the system is thermally unstable 
 owing to negative $c_{\phi}$ . It can be verified that $R$ is negative in the naked singularity region to the left of the extremal curve. 
 
 Given this observation we can now see that the Ruppeiner curvature changes sign through an infinite discontinuity and becomes positive on crossing to 
 the right of the extremal curve, while below the N - curve,
 it becomes negative by passing through zero. Along the $c_{\phi}$ curve it diverges without changing its sign as the $c_{\phi}$ polynomial appears as a square 
 in the denominator in the expression for $R$. In the same figure two isopotential curves have been drawn, the red one being the critical curve $(\phi=0.687)$ 
 which we see is tangential to the $c_{\phi}$ 
 curve and the brown one below being a subcritical one $(\phi=0.37)$ which cuts across the $c_{\phi}$ curve. 
 
 One can follow the variation of $R$ along the sub 
 critical curve as it negotiates various regions in the $q-s$ parameter space. Typically, along such curves, $R$ will become infinite,
pass through zero, and then go to negative infinity, as the isopotential enters and exits the unstable region bounded by the $c_{\phi}$ curve. 
 In the case of the critical curve, a closer look at the region surrounding the $c_{\phi}$ curve reveals an interesting structure (fig.(\ref{kn16})). Note that 
 the N - curve cuts the $c_{\phi}$ curve \emph{below} the critical point. 
 Thus it leaves out a small range of $\phi$ values where the isopotentials would already have acquired a negative curvature value before entering the 
 unstable region. By abuse of terminology, we will call this the ``negative $\phi$ window.'' 
 The values of $\phi$ for which the isotherms intersect the N - curve below the $c_{\phi}$ curve will be called the ``positive $\phi$ window.'' In this 
 latter window, the isopotential curves are similar to the subcritical one just discussed.

 \begin{figure}[!ht]
\begin{minipage}[b]{0.5\linewidth}
\centering
\includegraphics[width=2.7in,height=3.0in]{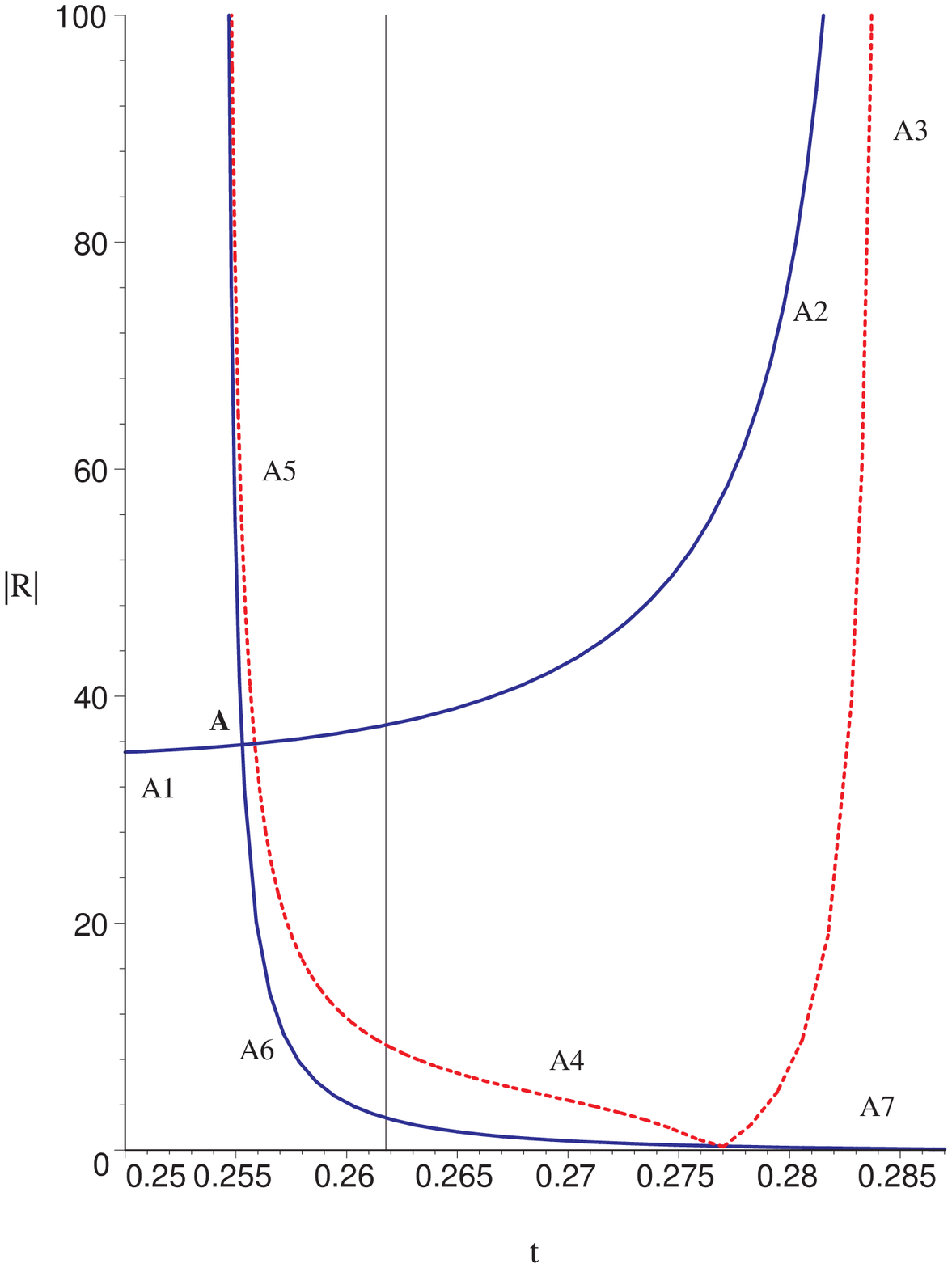}
\caption{Plot of absolute value of Ruppeiner curvature vs temperature for the $j=0.011$ ensemble, with its potential fixed at $\phi$=0.375.}
\label{kn17}
\end{minipage}
\hspace{0.6cm}
\begin{minipage}[b]{0.5\linewidth}
\centering
\includegraphics[width=2.7in,height=3.0in]{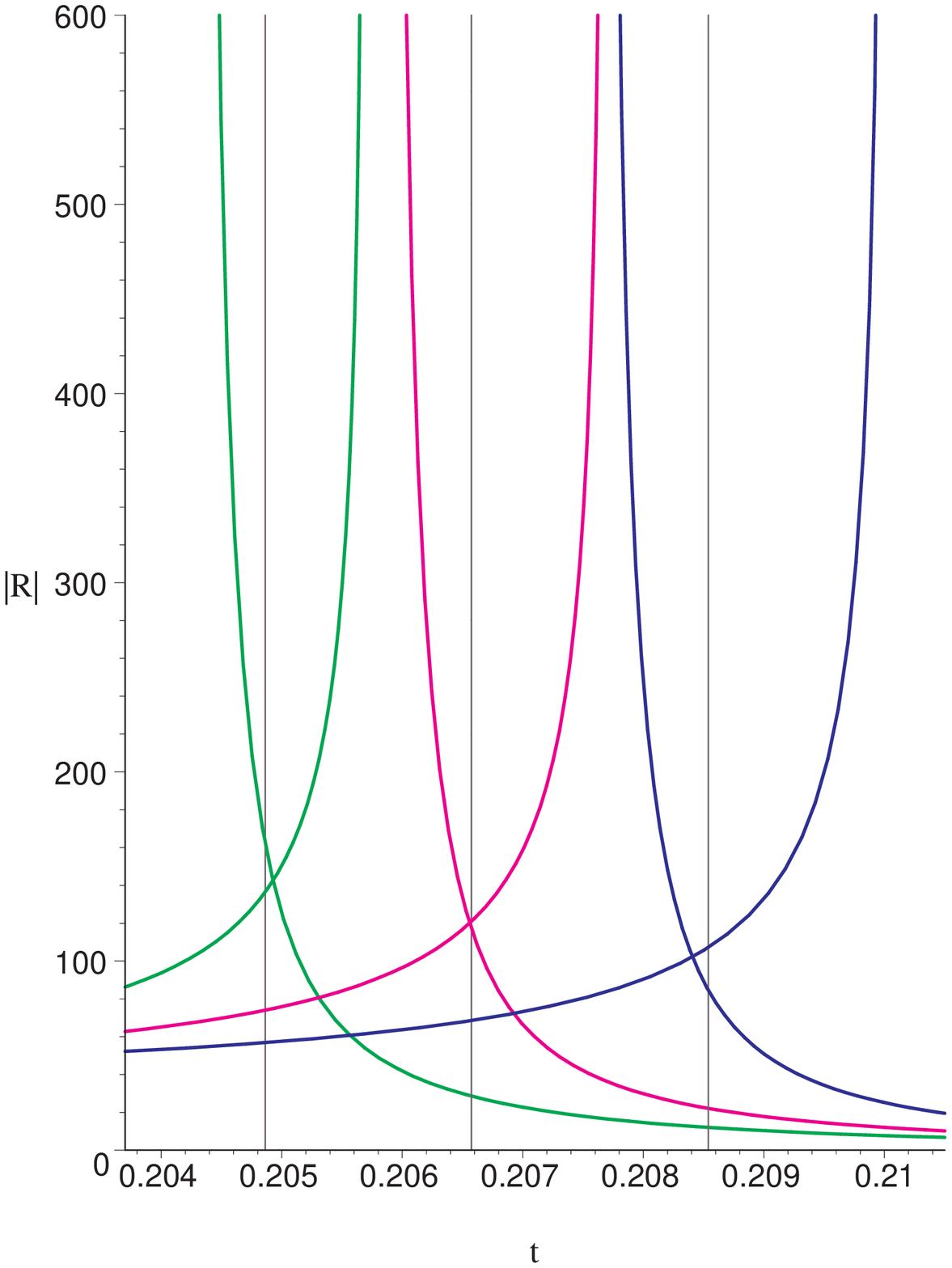}
\caption{Absolute value of Ruppeiner curvature vs temperature for $j=0.011,\phi=0.645, 0.652$, and $0.658$ from right to left.}
\label{kn18}
\end{minipage}
\end{figure}

In order to obtain the isopotential plots for the Ruppeiner curvature as a function of the temperature, we first express the charge in terms of the electric potential. 
This way, we obtain $R$ and $t$ as functions of the entropy, $s$, and the electric potential, $\phi$, with the angular momentum $j$ again appearing as the 
ensemble parameter. The plots can now be arrived at by fixing $\phi$ and varying $s$ as the common parameter along the curve, which we have varied over a 
suitable range, starting from its extremal value. We shall investigate the negative and positive windows of $\phi$ separately, with a few representative isopotential plots corresponding to each case. 

We first describe the positive $\phi$ window. It is apparent that for this range of $\phi$ values the curvature always changes sign from positive to negative as
 the black hole undergoes a first order phase transition from the small to the large black hole phase.  In order to compare the magnitudes of curvature across the phase boundary 
 we shall find it convenient to draw $|R|$ vs $t$ plots. Figure(\ref{kn17}) is one such plot for $j=0.011$ ensemble with $\phi$ fixed at $0.375$. We shall follow the curve 
 in the sense of increasing entropy. As we increase entropy beyond its extremal value ,$|R|$, which has a divergent behaviour at the extremal limit, proceeds from the left of the figure along the (blue) stable small black hole branch $A1A2$ until it diverges at $t=0.254$. Thereafter, it turns back and moves along the (red, dotted) 
 unstable branch $A3A4A5$ until it again diverges at the turning point $t=0.285$. From there on, it proceeds forward along the (blue) stable large black hole 
 branch $A6A7$, progressively diminishing in magnitude with temperature. For comparison, we refer the reader to a similar isopotential curve in the $t-s$ plot 
 of fig(\ref{kn14}). As expected, one can check that the turning point and multivalued behaviour of $|R|$ exactly corresponds with the turning point behaviour of 
 entropy and its multivaluedness with respect to temperature in the $s-t$ plane. 
 
 We now turn our attention to the vertical line at $t=0.262$ in figure (\ref{kn17}) which is the first order transition temperature. At this point,  the free energy 
 at constant $\phi$ changes branch from the small to the large black hole and vice versa (see fig(\ref{kn10})). The vertical free energy line thus naturally divides the 
 small black hole branch into a stable one on the left and a metastable one the right. Similarly, it divides the large black hole branch into a stable one on the right 
 and a metastable one on the left. Our conclusion is that the black hole, owing to the constraint of minimum free energy, jumps from one stable branch to the other 
 and never accesses the unphysical branch. Short lived metastable states, though, are not ruled out. Observe that the temperature at which the two stable branches 
 cross each other($t=0.255$, point \textbf(A)) is slightly different from the transition temperature. Thus, at the first order phase transition, for the given set of 
 parameters, the black hole ``correlation volume'' appears to decrease discontinuously to about a tenth of its value on the small black hole branch, if we 
 associate $|R|$ with the ``correlation volume''. As we move up the positive $\phi$ window, the free energy line shifts to the left and approaches 
 the point of intersection of the stable branches. It crosses it at about $\phi=0.652$ and then continues leftwards. This has been shown in figure(\ref{kn18}) for 
 three isopotentials, with their respective unstable branches omitted for clarity of presentation. This can be directly interpreted in terms of the correlation volume. 
 Thus, we conclude that the difference in correlation volume across the phase boundary progressively decreases as we approach $\phi=0.652$ from below, 
 beyond which it starts to \emph{increase} from the small black hole to the large black hole phase.

\begin{figure}[!ht]
\begin{minipage}[b]{0.5\linewidth}
\centering
\includegraphics[width=2.5in,height=2.7in]{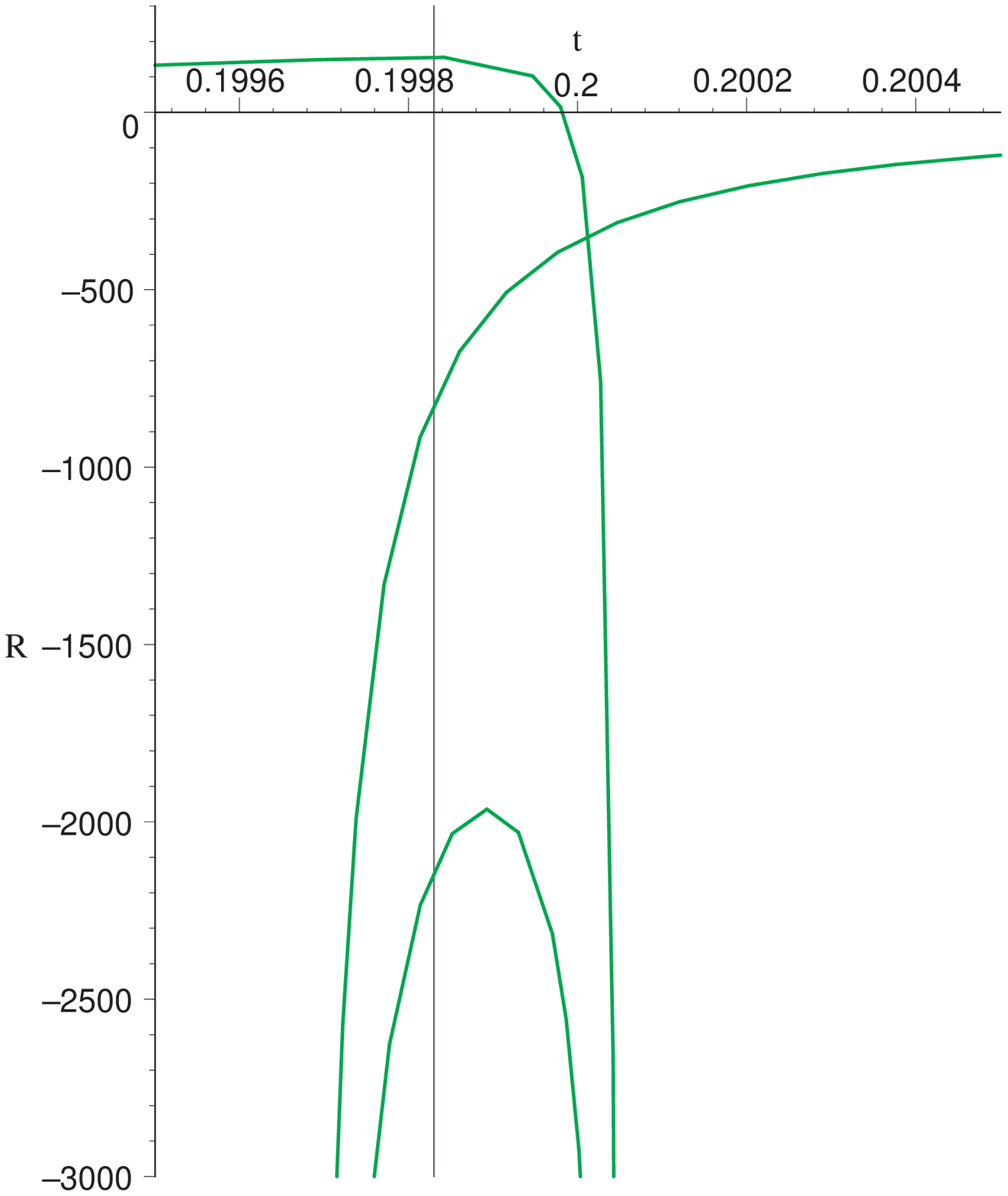}
\caption{Plot of $R$ vs $t$ in $j=0.011$ ensemble for $\phi=0.675$}
\label{kn19}
\end{minipage}
\hspace{0.6cm}
\begin{minipage}[b]{0.5\linewidth}
\centering
\includegraphics[width=2.5in,height=2.7in]{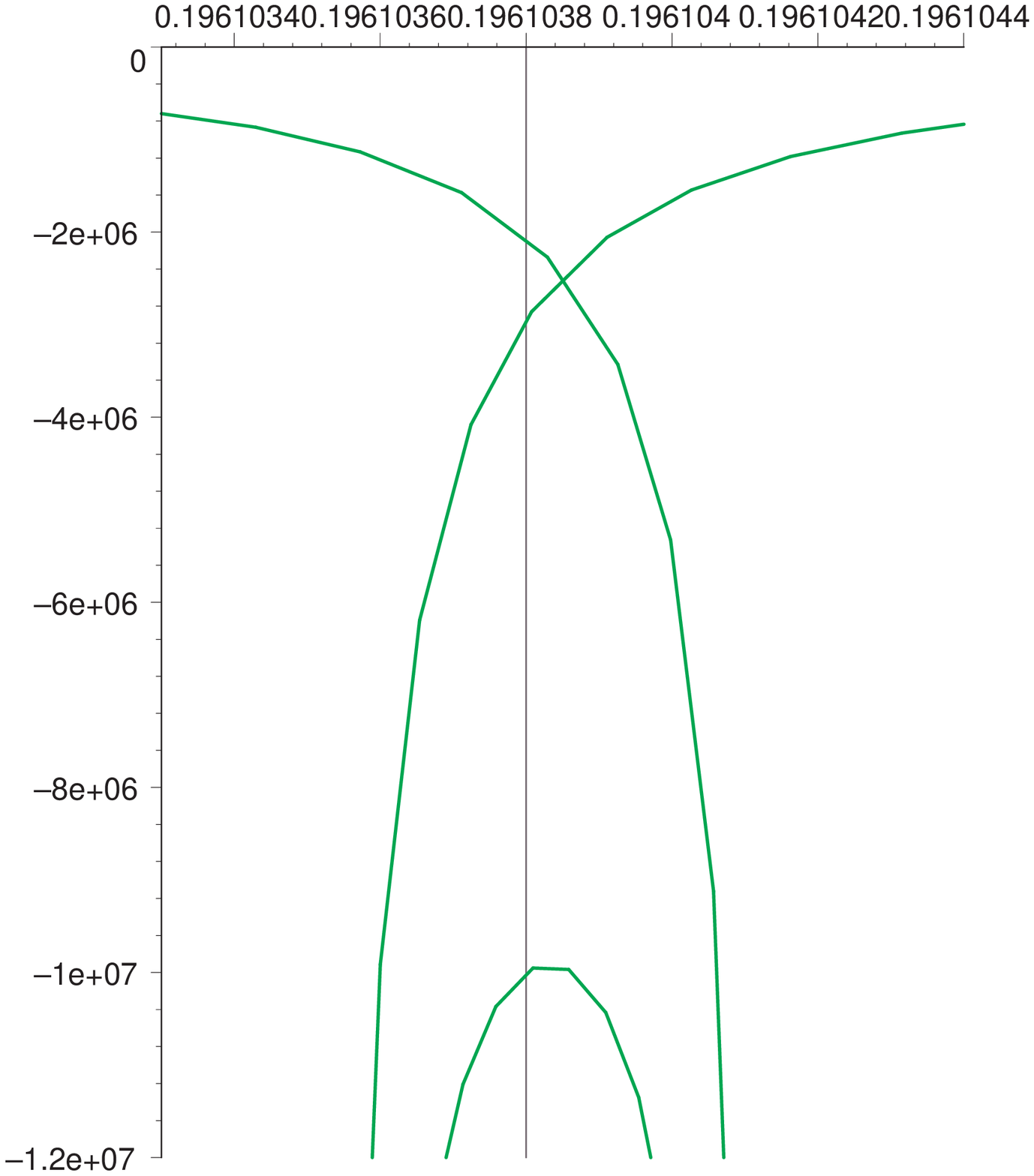}
\caption{Plot of $R$ vs $t$ in $j=0.011$ ensemble for $\phi=0.687$}
\label{kn20}
\end{minipage}
\end{figure}

\begin{figure}[!ht]
\begin{minipage}[b]{0.5\linewidth}
\centering
\includegraphics[width=2.5in,height=2.7in]{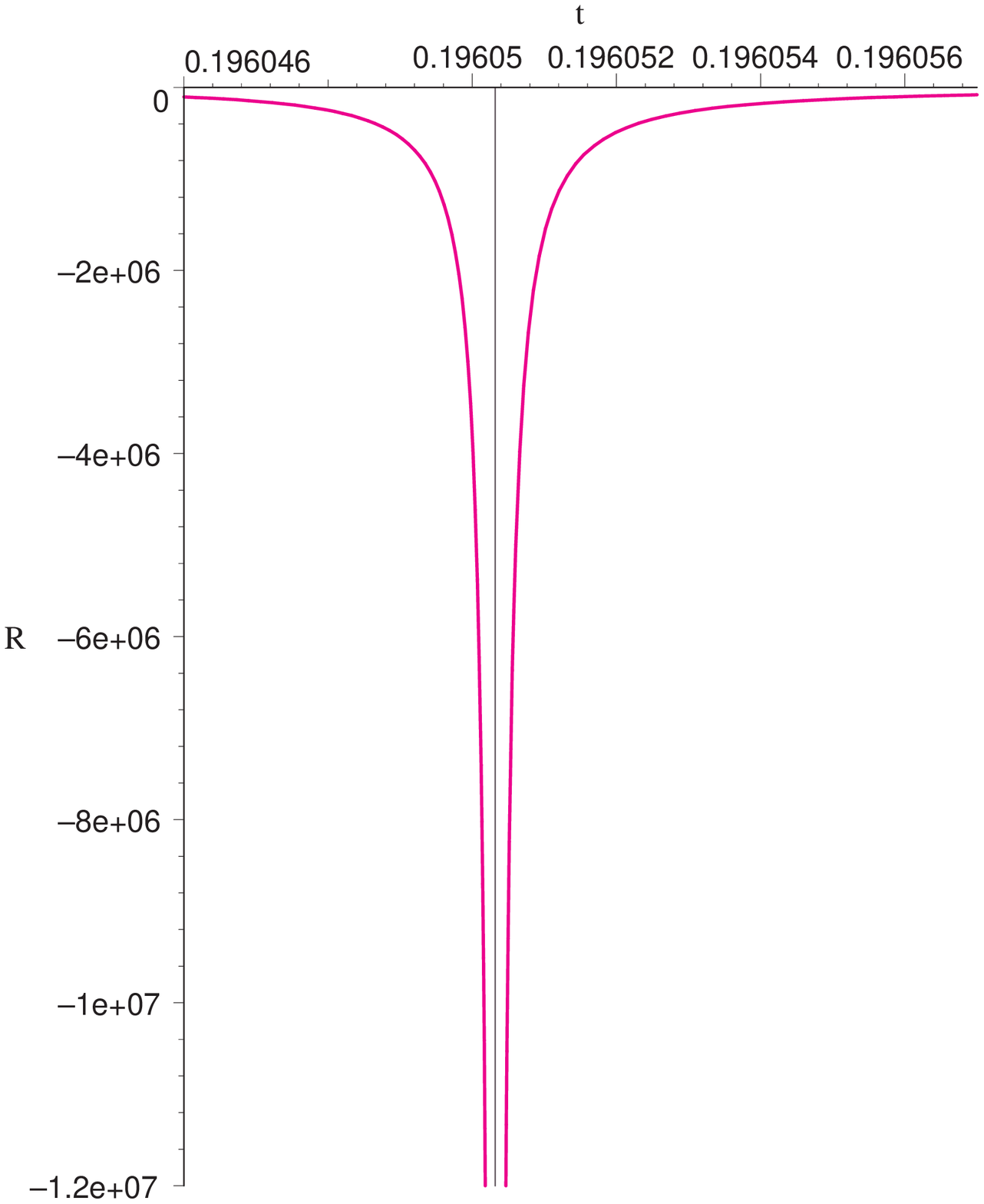}
\caption{Plot of $R$ vs $t$ in $j=0.011$ ensemble for critical value of $\phi=0.6872707$. }
\label{kn21}
\end{minipage}
\hspace{0.6cm}
\begin{minipage}[b]{0.5\linewidth}
\centering
\includegraphics[width=2.5in,height=2.5in]{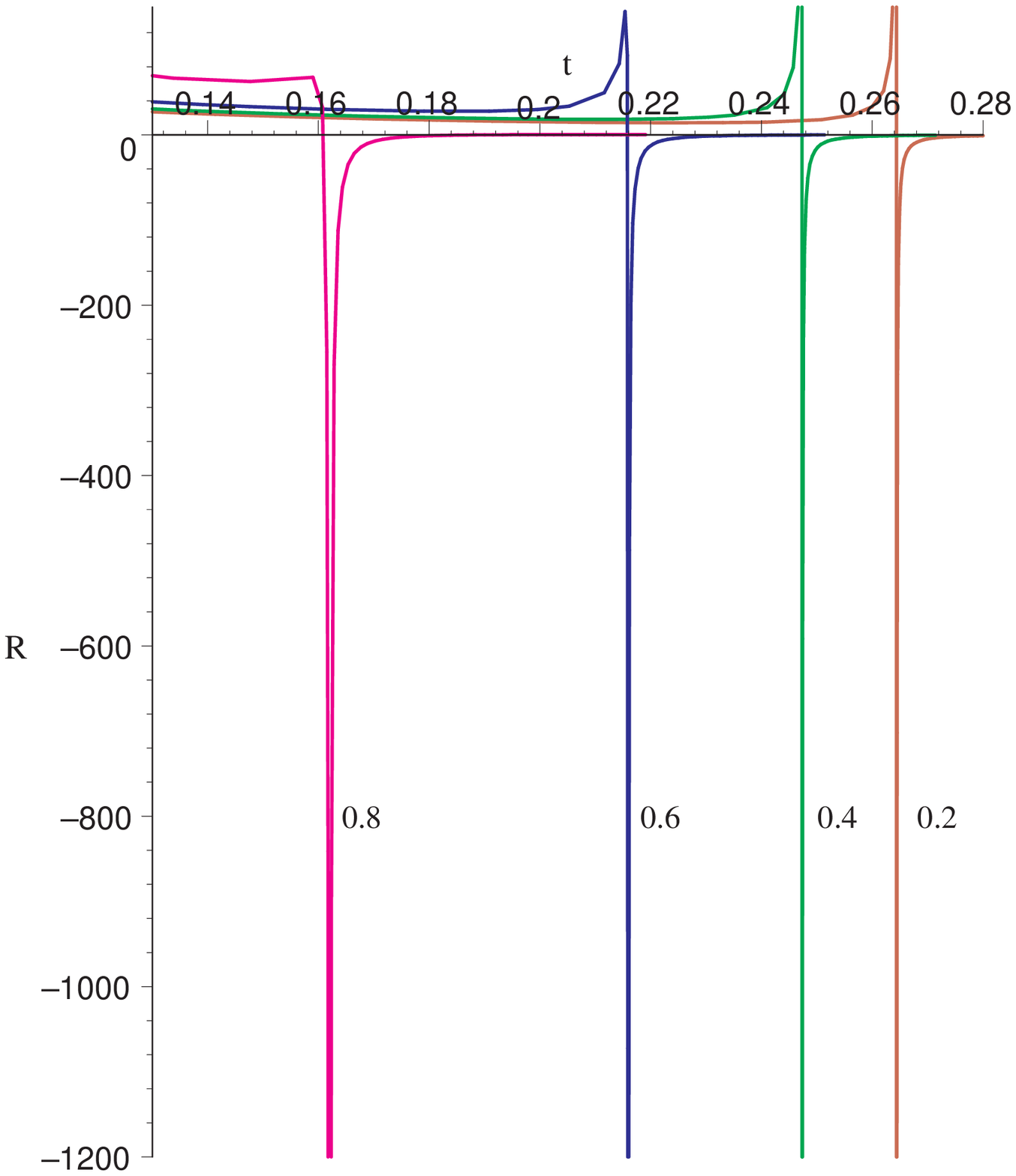}
\caption{Divergence of Ruppeiner curvature at critical values of $j$, for different values of $\phi$}
\label{kn22}
\end{minipage}
\end{figure}

This feature becomes more pronounced in the negative window. Fig.(\ref{kn19}) is a plot of $R$ vs $t$ for $\phi=0.675$ (in the negative $\phi$ window). 
We have not colour coded or labelled the plots as by now the multivaluedness and general flow of the curves is apparent. Here there is no need to consider the 
magnitude of $R$ since both its divergences are on the negative side, even though for this particular value of $\phi$, Ruppeiner curvature still jumps from positive
to negative at phase transition.  The reader can easily notice the similarity with the Van der Waals case of fig(\ref{fig0c}). The subsequent behaviour of these plots, 
shown in figs(\ref{kn20}) and (\ref{kn21}) exactly mirrors the Van der Waals case  depicted in fig(\ref{fig0f}), culminating in a second order phase transition along the 
$\phi_c=0.68727$ curve at  $t_c=0.196$, the temperture at which the free energy curve shows an inflection. Similar to the Van der Waals case, the 
unphysical branch gets pushed to infinity for the ${\phi}_c$ plot and $R$ becomes a single valued function of temperature. In fig. (\ref{kn21}),
we have increased the accuracy of the numerical value of $\phi$ to show the exact coincidence of the divergence with the value of the temperature at 
which the free energy shows an inflection.

\begin{figure}[!ht]
\centering
\includegraphics[width=3.5in,height=2.5in]{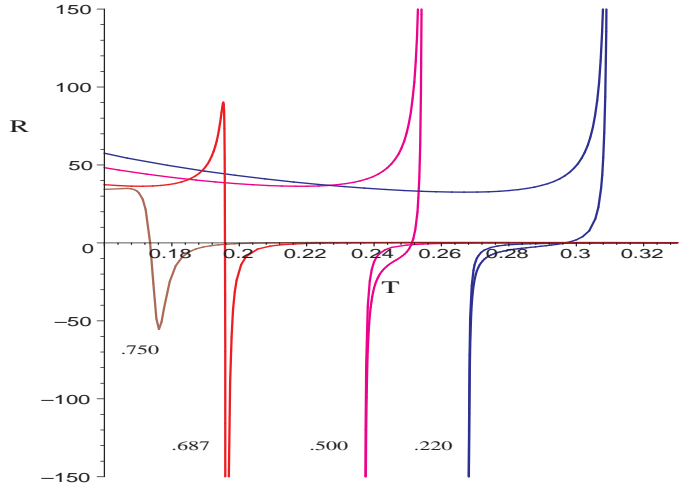}
\caption{Ruppeiner curvature in the fixed $j = 0.011$ ensemble for different values of $\phi$}
\label{kn23}
\end{figure}

Finally, in fig. (\ref{kn23}), we have plotted a snapshot of the Ruppeiner curvature for different values of the potential $\phi$ for a fixed value of 
$j = 0.011$. As expected, the two branches of $R$ separate out at the critical value of $\phi$.

In fig. (\ref{kn22}), we have shown the behaviour of the
the Ruppeiner  curvature as a function of the temperature for the critical values of the angular momentum, corresponding to
$\phi = 0.2$, $0.6$ and $0.8$ (the red, blue and green lines in the figure respectively). The critical values
of $j$ for the given value of $\phi$  can be obtained from fig. (\ref{kn9}). Qualitatively, we can
understand these plots as follows. As we increase the value of the entropy from the extremal value, the system moves along a stable branch, on which
the Ruppeiner curvature diverges at a certain critical value of the entropy. Beyond this, the curvature comes down along the second stable branch,
when we further increase the entropy. The unstable branch, as discussed before, is pushed to infinity, and the divergence occurs at the end of the 
``vapour pressure'' curve. Note also that the critical temperature at which $R$ diverges decreases as we increase the value of the electric potential.

To summarize, in this subsection, we have studied the Ruppeiner geometry of the KN-AdS black hole in the fixed $j$ ensemble. We have calculated
and analysed the Ruppeiner curvature $R$ and seen that
it closely resembles our discussion on the Van der Waals system, in that $R$ captures the behaviour of the first order phase transitions
along (the analogue of) the vapour pressure curve in the latter, that culminates in a second order phase transition where it shows a negative
divergence. At the first order phase transition, the
Ruppeiner curvature jumps from one stable branch to another, at a temperature dictated by the free energy. We have interpreted this in terms of
the correlation volume. As emphasized at the end of
the previous section, the Van der Waals system is an approximate model, but in the case of black holes, with a well defined fundamental relation in all
the phases, our analysis has a more direct physical significance. We can say with some certainty that even for a  non extensive system such as the
KN-AdS black hole being considered here, there is a change in correlation volume at a first order phase transition. These are the main results of
this subsection.

\subsection{KN-AdS black hole in the fixed $q$ ensemble}

Finally, we come to the fixed $q$ ensemble mentioned in the beginning of this section. This case is markedly different from the fixed $j$ ensemble
that we have discussed. Note that in this case, we can restrict our analysis to the case $\omega < 1$. 
To begin with, we express $j$ in eq. (\ref{potkn}) in terms of $\omega$, $s$ and $q$. The physical solution (obtained by
matching with the known $q=0$ case) is
\begin{equation}
j = 2\omega\pi\left(s^2 + s\pi + \pi^2q^2\right)\frac{\left[s\pi\left(s^2\left(1 - \omega^2\right) + s\left(2\pi - \pi \omega^2\right)+\pi^2\right)\right]^{\frac{1}{2}}}
{\left[4\pi^3s^2\left(1 - \omega^2\right)+ 4\pi^4s\left(2-\omega^2\right)+ 4\pi^5\right]}
\end{equation}
In analogy with the previous case, we calculate the specific heats $c_{\omega}$ and $c_j$ where
\begin{equation}
c_{\omega} = t\left(\frac{\partial s}{\partial t}\right)_{\omega},~~~c_j = t\left(\frac{\partial s}{\partial t}\right)_j
\end{equation}

\begin{figure}[!ht]
\centering
\includegraphics[width=2.5in,height=2.5in]{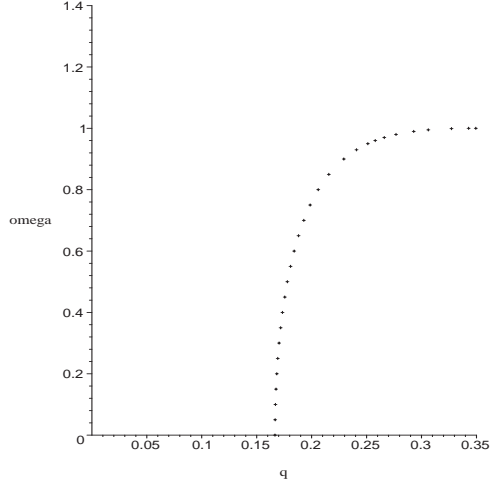}
\caption{Critical value of $q$ vs $\omega$ for the KN-AdS black hole in the fixed $q$ ensemble.}
\label{kn24}
\end{figure}

The expressions are too lengthy, but we will find it convenient to use these specific heats to describe the phase structure, as we had
done for the fixed $j$ ensemble. First, as in the previous case, we show the curve of critical points in the $q-\omega$ plane,  fig. (\ref{kn24}). Starting at a critical charge $q_c=1/6$, as appropriate for the RN-AdS black hole, it asymptotes to the $\omega=1$ line, with the regions of coexistence lying outside the curve. Namely, for all values of $q$ beyond $q_c$, there is a critical value of $\omega$ below which there is no phase coexistence. On the other hand, we see that below $q_c$, there is
phase coexistence but no second order phase transition.

\begin{figure}[!ht]
\begin{minipage}[b]{0.5\linewidth}
\centering
\includegraphics[width=2.7in,height=2.5in]{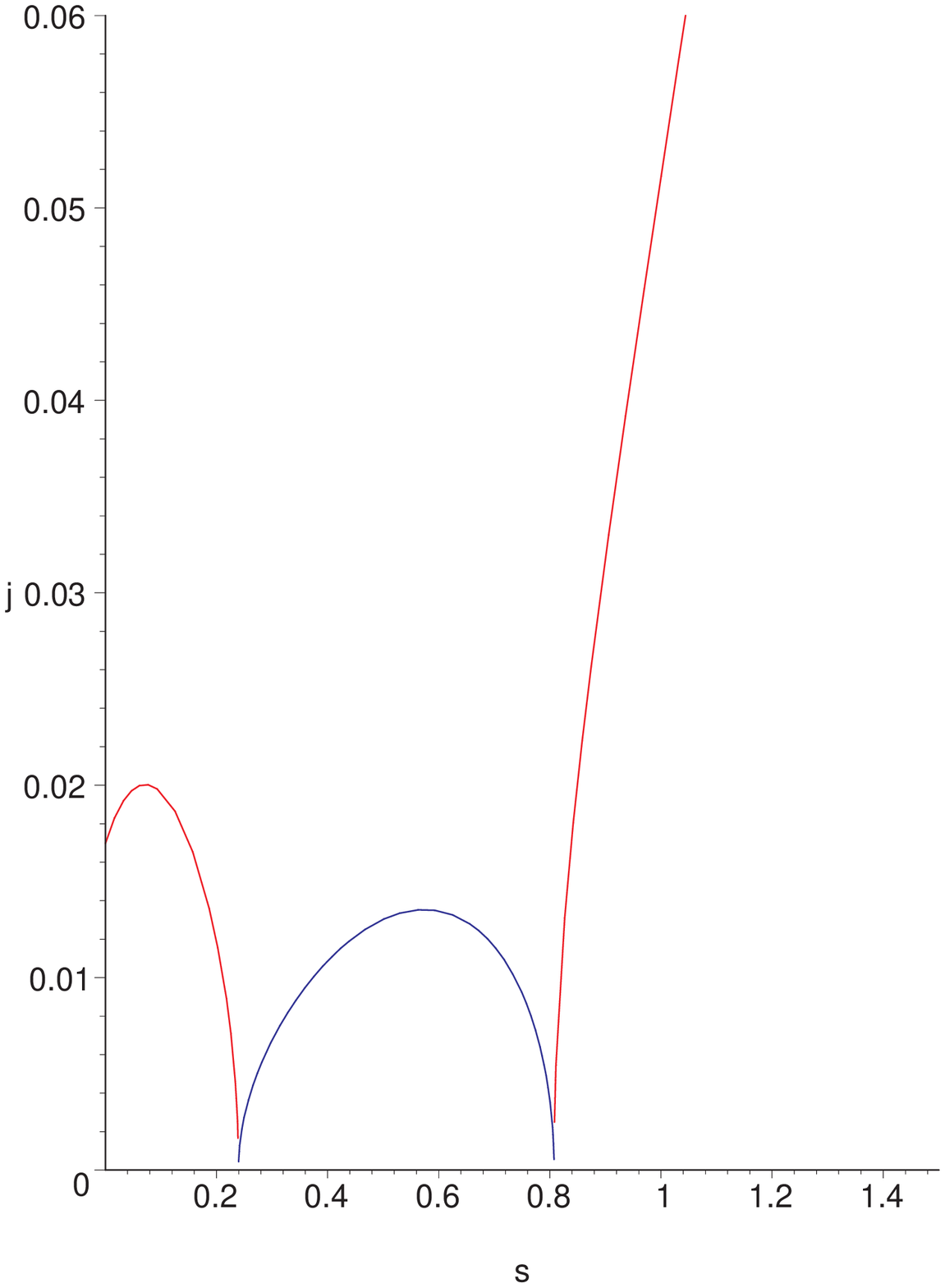}
\caption{Divergence curves of $c_{\omega}$ (red lines) and $c_j$ (blue line) in the $j - s$ plane for $q = 0.14$. }
\label{kn25}
\end{minipage}
\hspace{0.6cm}
\begin{minipage}[b]{0.5\linewidth}
\centering
\includegraphics[width=2.5in,height=2.5in]{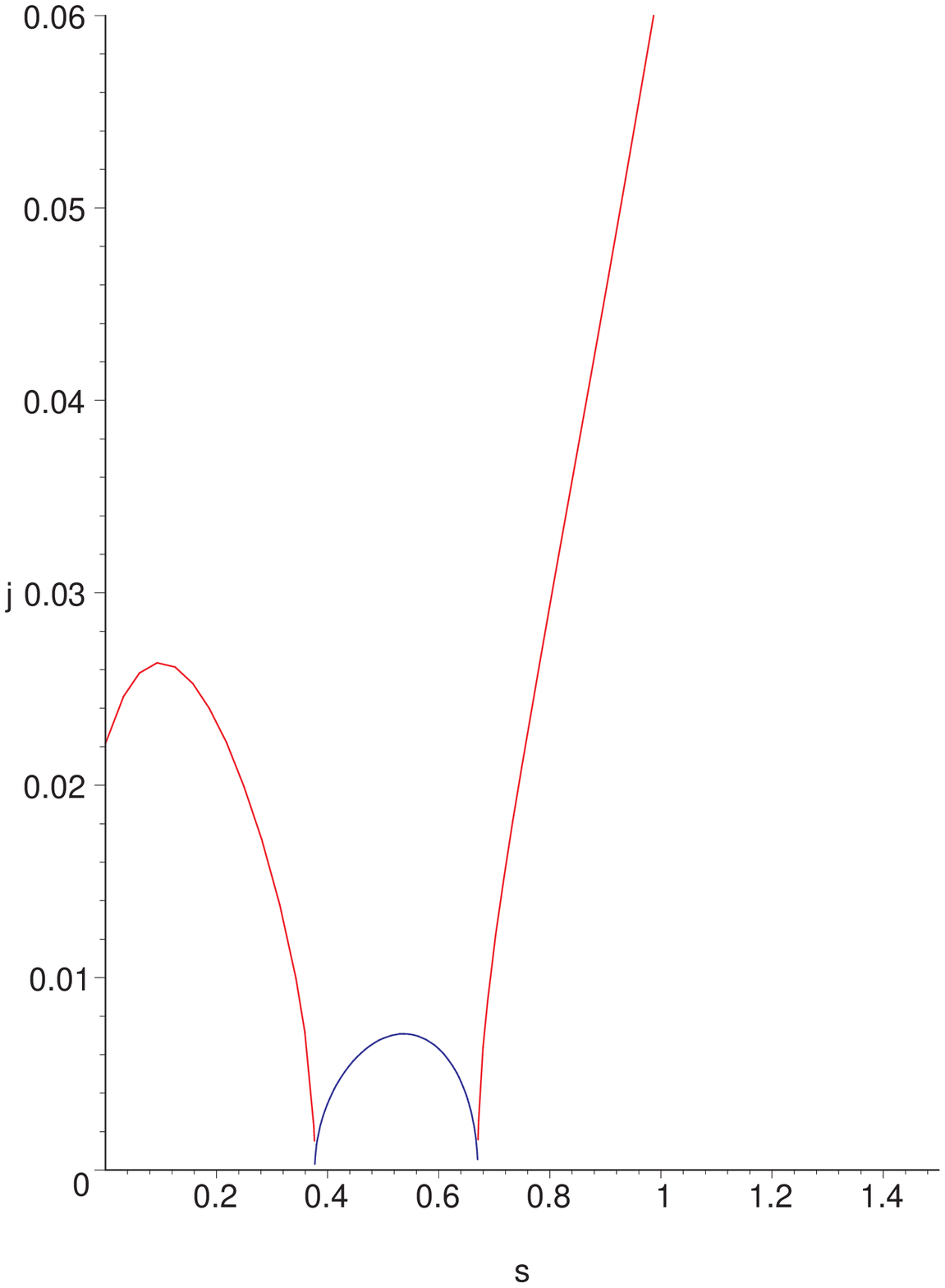}
\caption{Divergence curves of $c_{\omega}$ (red lines) and $c_j$ (blue line) in the $j - s$ plane for $q = 0.16$.}
\label{kn26}
\end{minipage}
\end{figure}

In order to understand the phase structure better, we have, in figs. (\ref{kn25}) and (\ref{kn26}), plotted the curves along which the specific heats
$c_{\omega}$ and $c_j$ diverge, in the $j - s$ plane, for $q = 0.14$ and $q = 0.16$, both values less than $q_c=1/6$. In these figures, the red lines and the blue line correspond to the
divergence of $c_{\omega}$ and $c_j$ respectively. We will call the (collection of) red lines as the $c_{\omega}$ curve and the blue line as the
$c_j$ curve. From the figures, it is clear that as we increase the charge, the area bounded by the $c_j$ curve shrinks, and we find that at $q = q_c$,
it disappears. This is the critical value of $q$ alluded to in fig. (\ref{kn18}). Beyond this point the two disjoint $c_{\omega}$ curves meet up, which brings into existence the second order critical points. Our analysis for the phases of the system
in this fixed $q$ ensemble will closely follow that of the fixed $j$ ensemble. In this case, we find by direct calculation that (apart from some unimportant factors)
\begin{equation}
\left(\frac{\partial j}{\partial \omega}\right) \simeq \frac{c_{\omega}}{c_j}
\end{equation}
As for the behaviour of the specific heats, we have the following results (refer to fig. (\ref{kn26})). $c_j$ is negative inside the blue semicircular region,
and is positive outside this region. $c_{\phi}$ is negative in the region inbetween the two red curves, and is positive everywhere else (i.e to the left and
right of the left red curve and the right red curve respectively). Hence, we can now present a clear picture of the phase structure of the system as follows.

\begin{figure}[!ht]
\begin{minipage}[b]{0.5\linewidth}
\centering
\includegraphics[width=2.7in,height=3.0in]{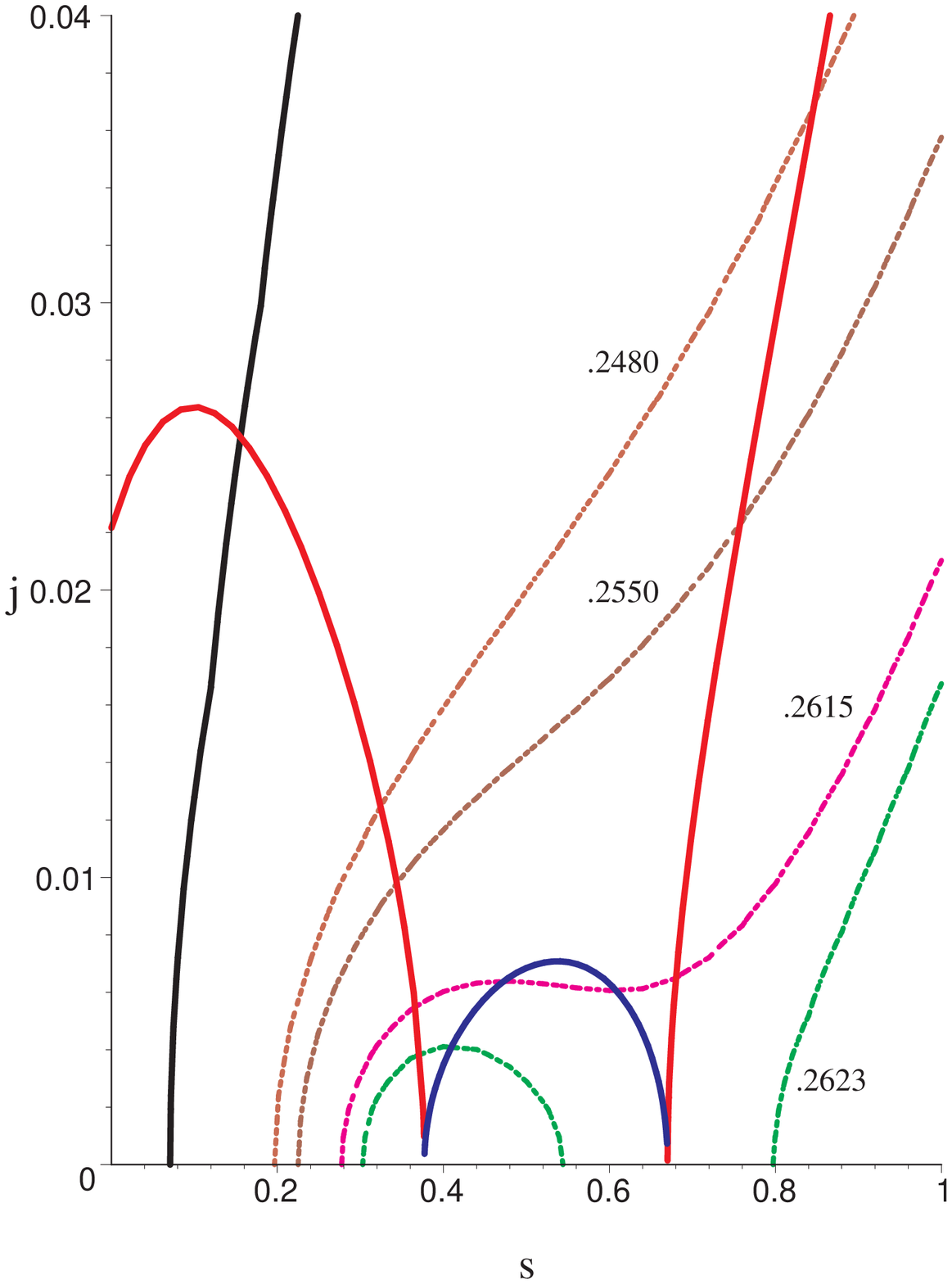}
\caption{Phase plot of isotherms in the $j-s$ plane for fixed $q$ ensemble,with $q=0.16$. The numbers denote the various values
of temperature. }
\label{kn27}
\end{minipage}
\hspace{0.6cm}
\begin{minipage}[b]{0.5\linewidth}
\centering
\includegraphics[width=2.5in,height=3.0in]{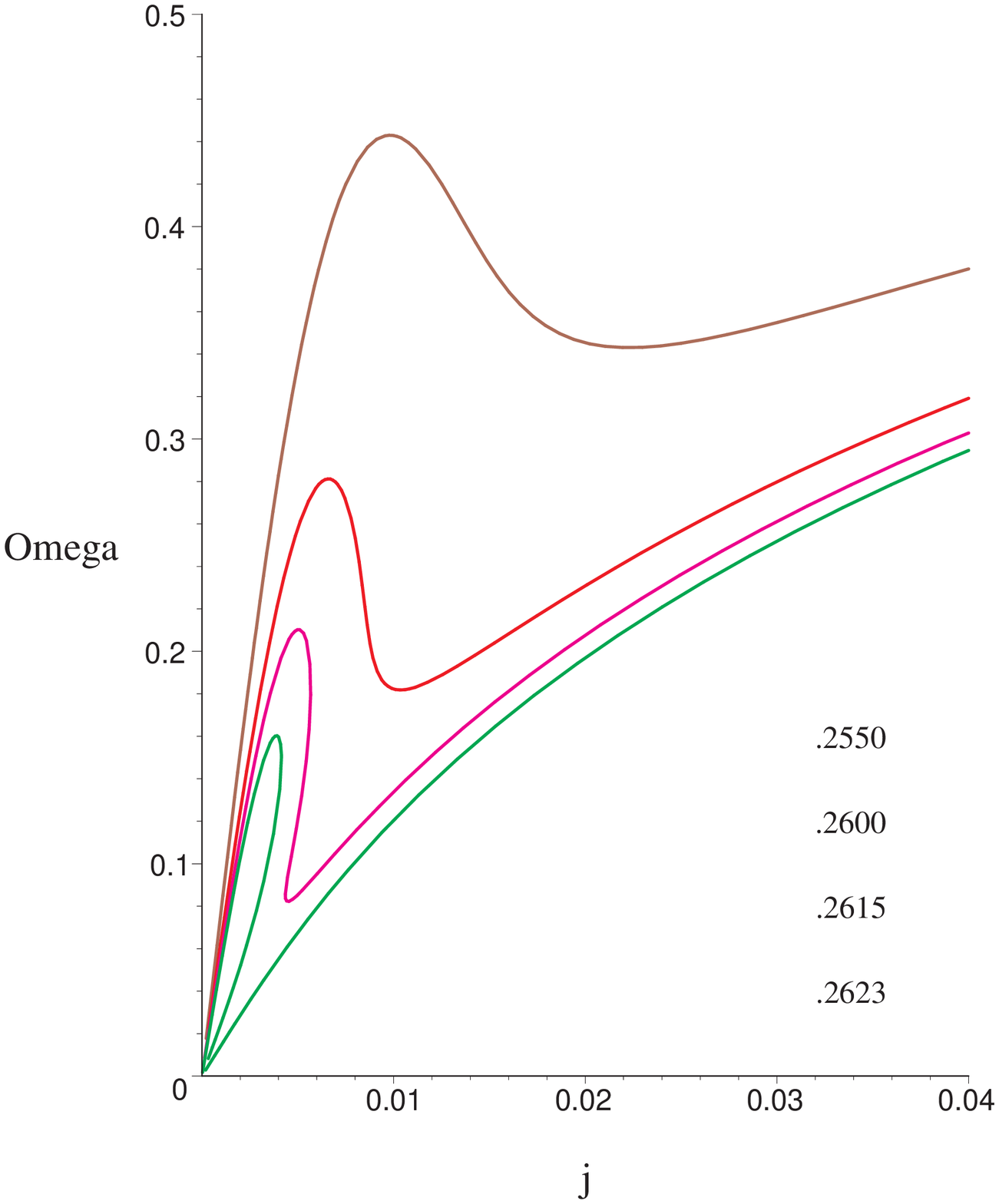}
\caption{Isotherms in the $\omega - j$ plane in the fixed $q$ ensemble, with $q=0.16$. The numbers denote some values of $\phi$ in fig. (\ref{kn21}).}
\label{kn28}
\end{minipage}
\end{figure}

\begin{figure}[!ht]
\begin{minipage}[b]{0.5\linewidth}
\centering
\includegraphics[width=2.7in,height=3.0in]{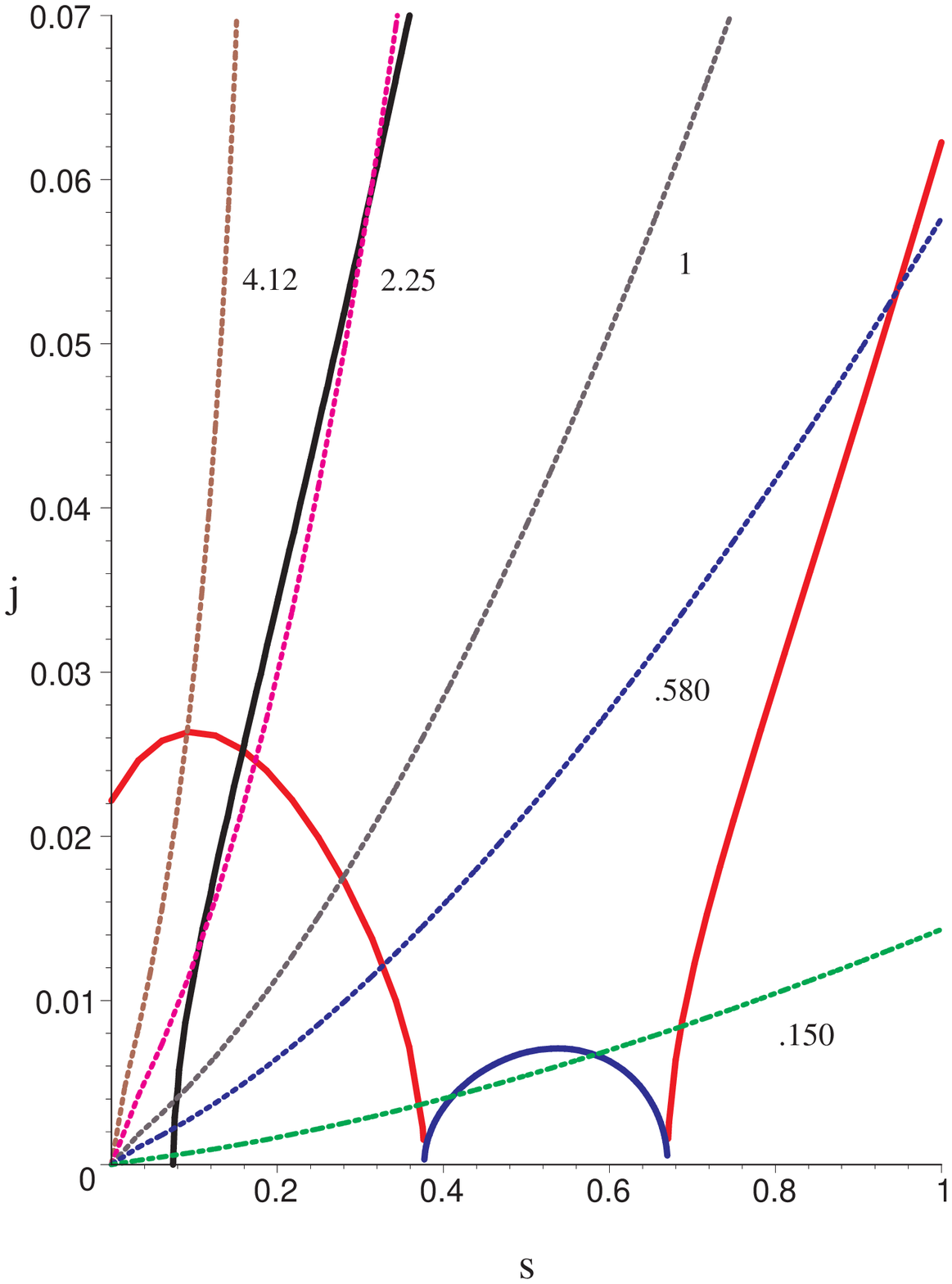}
\caption{Isopotential phase plot in the $j-s$ plane in the fixed $q$ ensemble, with $q=0.16$, for various values of $\omega$.}
\label{kn29}
\end{minipage}
\hspace{0.6cm}
\begin{minipage}[b]{0.5\linewidth}
\centering
\includegraphics[width=2.7in,height=3.0in]{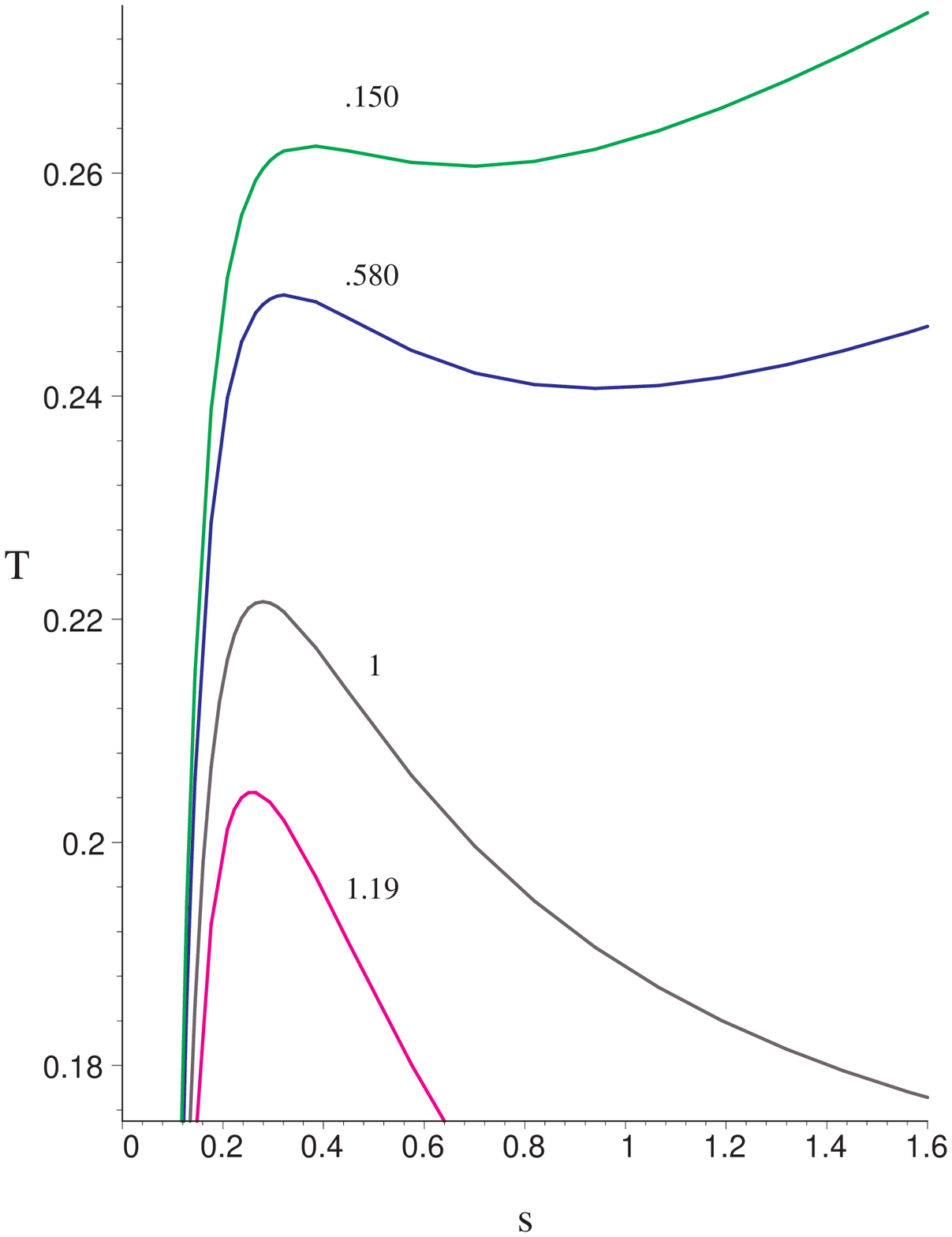}
\caption{Isopotential plots for $t$ vs $s$ for some fixed values of $\omega$ of fig. (\ref{kn23}), in the fixed $q$ ensemble with $q= 0.16$}
\label{kn30}
\end{minipage}
\end{figure}

In fig. (\ref{kn27}), we have plotted the isotherms in the $q - s$  plane for $q = 0.16$, which is at a value less than $q_c$, (see \ref{kn18}) . Let us focus on the pink line. For this isotherm, the system
starts from the extremal value of the entropy and is initially in a region where $c_{\omega}$ and $c_j$ are both positive, and hence so
is $\left(\frac{\partial j}{\partial \omega}\right)$. It crosses the $c_{\omega}$ curve as the entropy is increased, after which $c_{\omega}$
becomes negative, although $c_j$ remains positive, and hence $\left(\frac{\partial j}{\partial \omega}\right)$ changes sign. By the same
logic, it changes sign three more times before becoming positive again in the region to the right of the $c_{\omega}$ curve. The behaviour of
the other isotherms are similar. In fig. (\ref{kn28}), we have plotted the isotherms in the $\omega - j$ plane, that shows exactly the behaviour discussed
above. Thus, it is easy to see that, along such isotherms, the black hole will undergo a first order phase transition at an appropriate pressure decided by the Gibbs free energy for this ensemble. Expectedly, the `critical' isotherm is missing. In fig. (\ref{kn29}), we have shown plots of constant $\omega$ in the $j- s$ plane, for various values of $\omega$. Again, we point out that there is an absence of a critical isopotential curve. Besides, we also make the observation that for curves with $\omega>1$ there is no large black hole branch.

\begin{figure}[!ht]
\begin{minipage}[b]{0.5\linewidth}
\centering
\includegraphics[width=2.7in,height=3.0in]{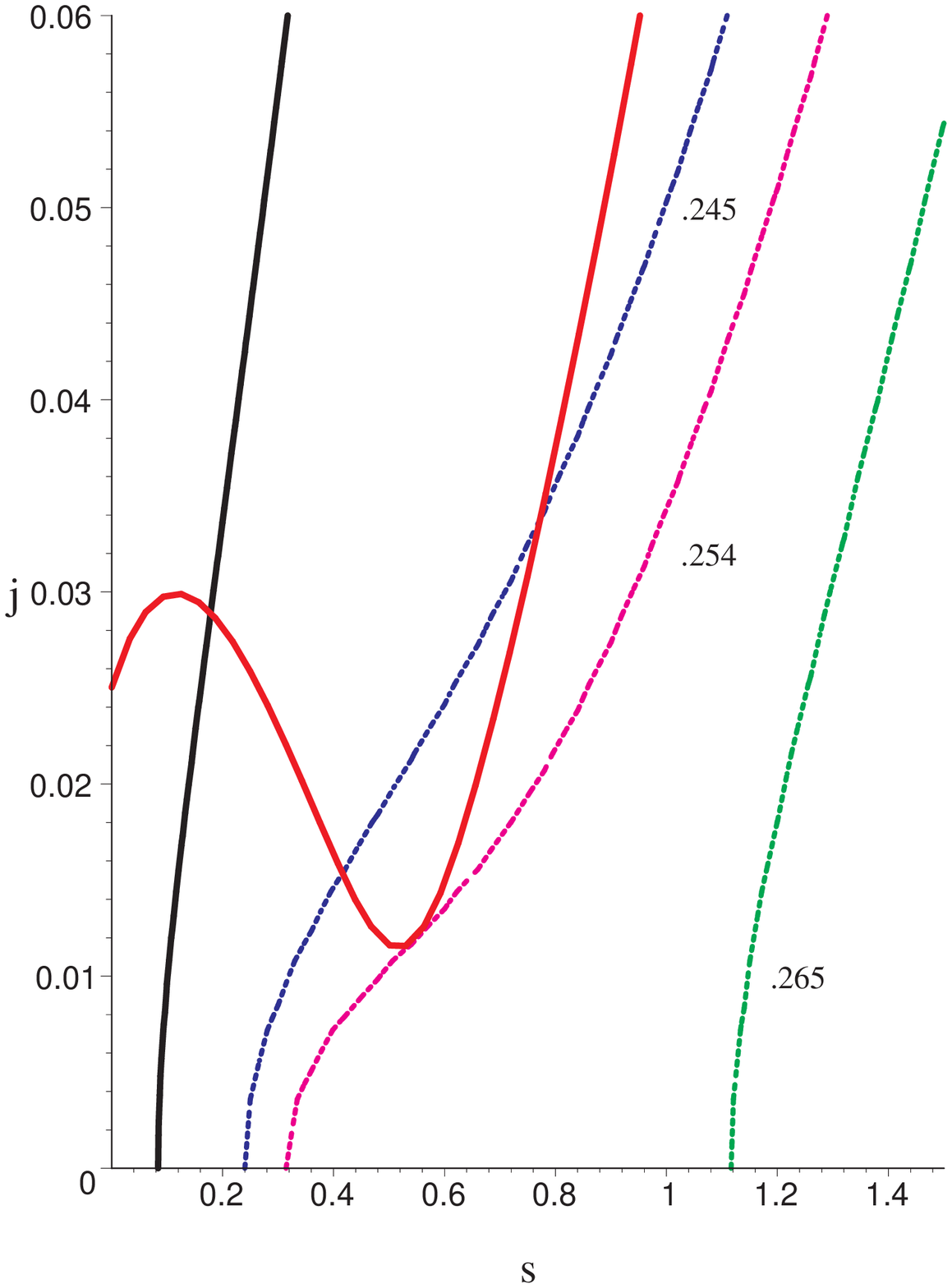}
\caption{Phase plot of isotherms in the $q-s$ plane for fixed $q$ ensemble,with $q=0.17$. The numbers denote the various values
of temperature. }
\label{kn31}
\end{minipage}
\hspace{0.6cm}
\begin{minipage}[b]{0.5\linewidth}
\centering
\includegraphics[width=2.5in,height=3.0in]{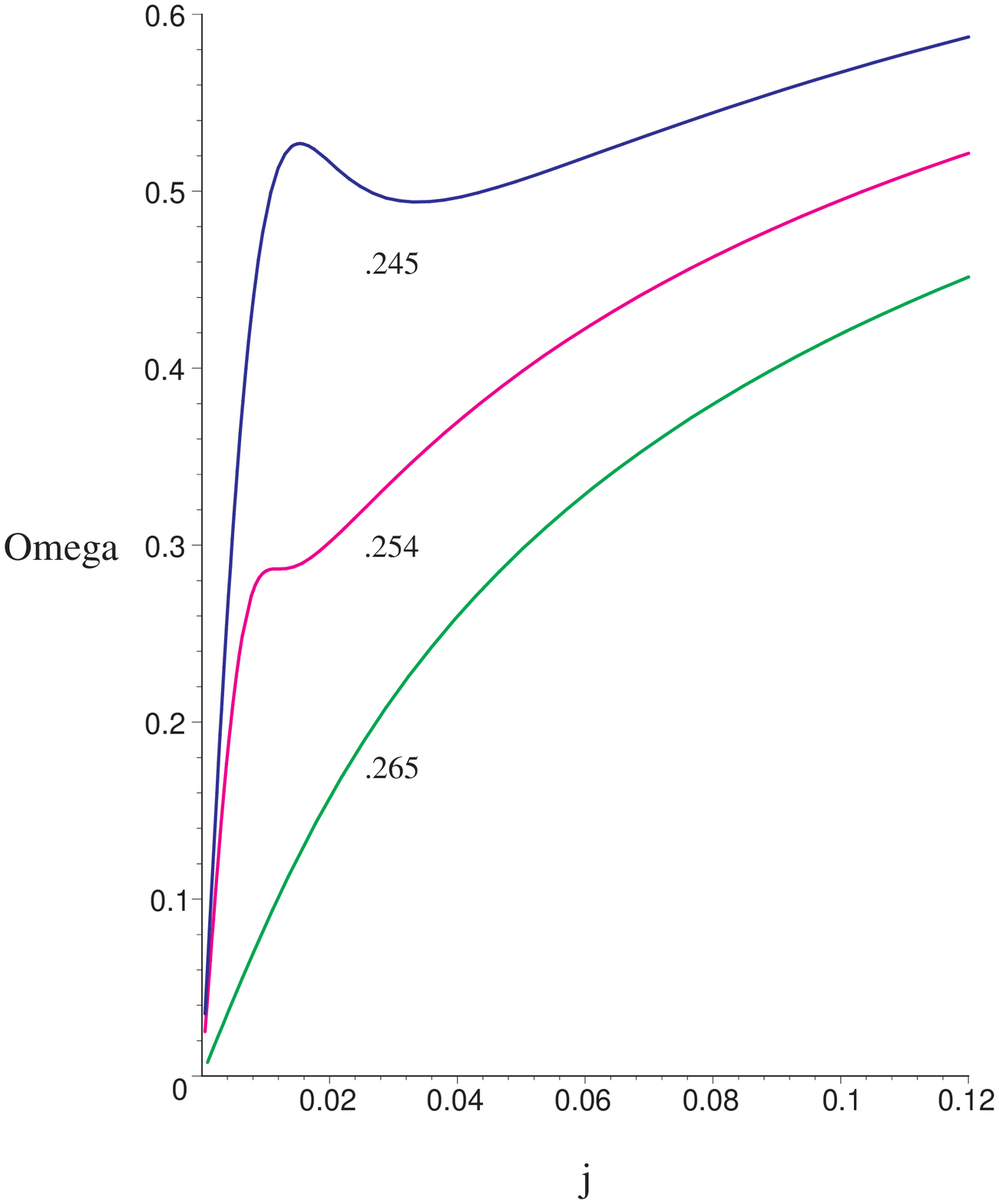}
\caption{Isotherms in the $\omega - j$ plane in the fixed $q$ ensemble, with $q=0.17$. The numbers denote some values of $t$ in fig. (\ref{kn25}).}
\label{kn32}
\end{minipage}
\end{figure}

Finally, we need to look at the case $q > \frac{1}{6}$, where as discussed previously, we can expect to find a second order phase transition. Figs. (\ref{kn31}) and (\ref{kn32}) show the isotherms for $q = 0.17$, and the corresponding plots of the isotherms in the $\omega - j$ plane. It is easy to make out the critical isotherm above which the black hole has a phase co-existence regime.

\begin{figure}[!ht]
\begin{minipage}[b]{0.5\linewidth}
\centering
\includegraphics[width=2.7in,height=3.0in]{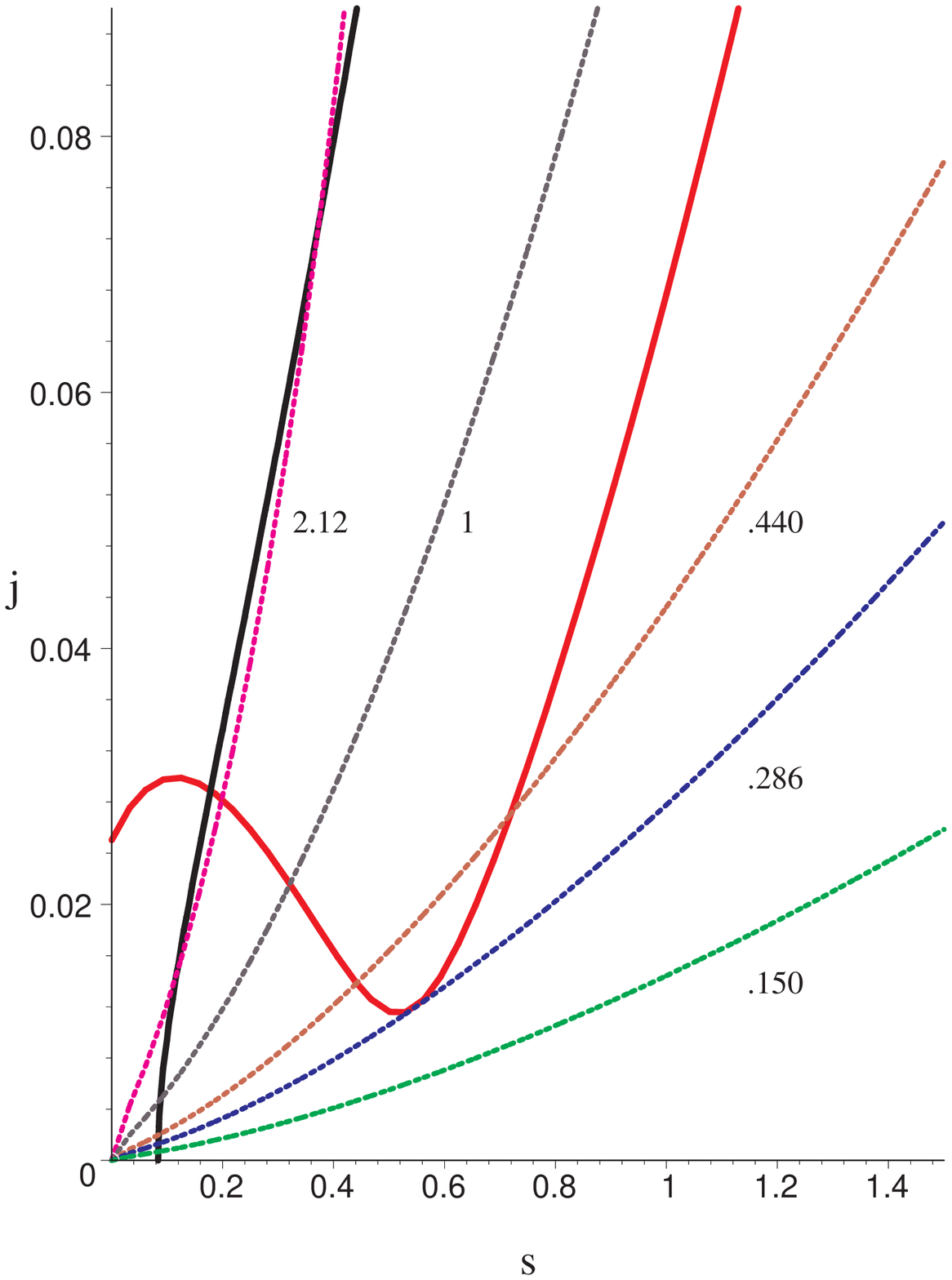}
\caption{Isopotential phase plot in the $q-s$ plane in the fixed $q$ ensemble, with $j=0.17$, for various values of $\omega$.}
\label{kn33}
\end{minipage}
\hspace{0.6cm}
\begin{minipage}[b]{0.5\linewidth}
\centering
\includegraphics[width=2.7in,height=3.0in]{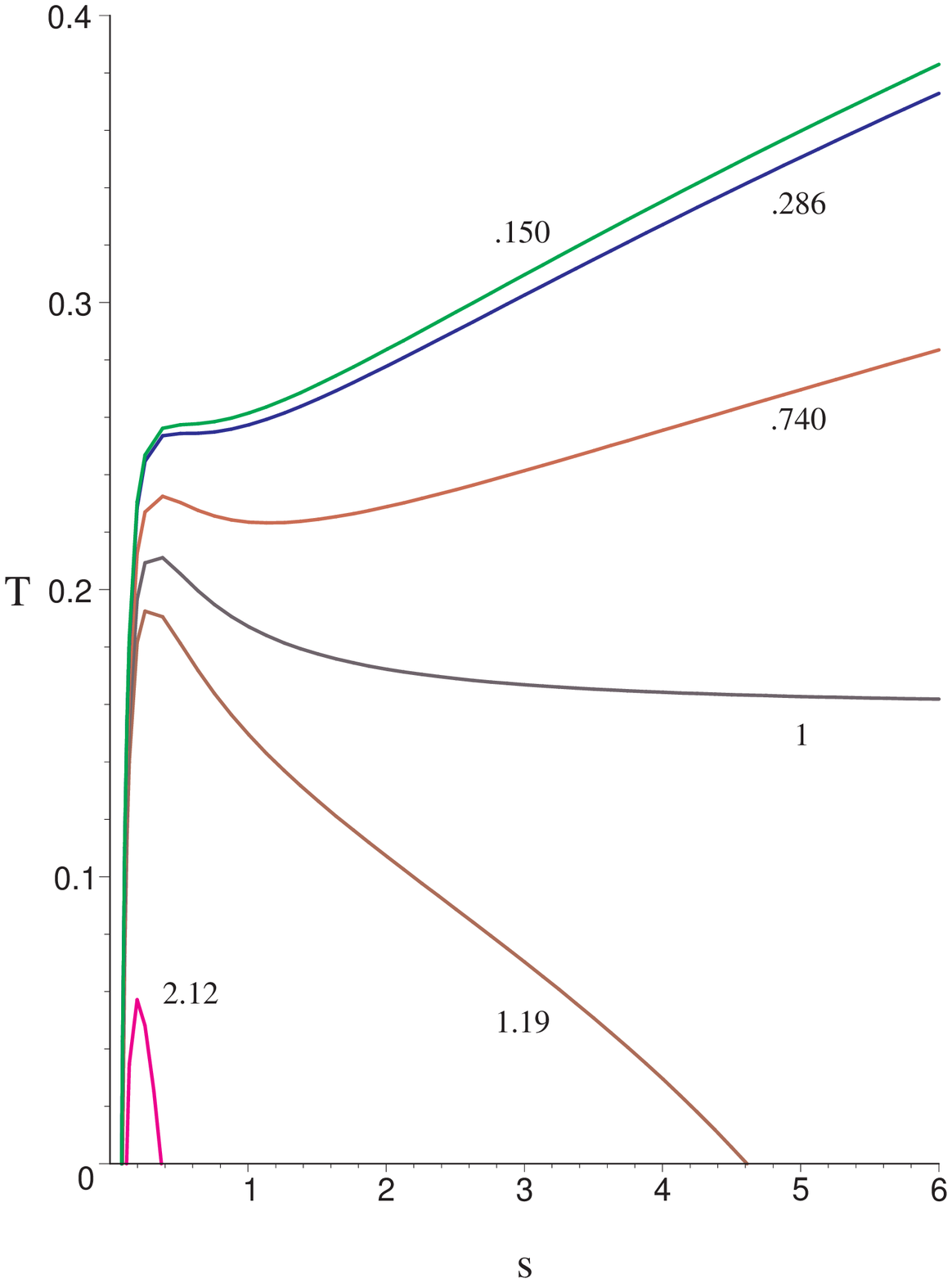}
\caption{Isopotential plots for $t$ vs $s$ for some fixed values of $\omega$ of fig. (\ref{kn27}), in the fixed $q$ ensemble with $q = 0.17$}
\label{kn34}
\end{minipage}
\end{figure}

Constant $\omega$ plots for the same value of $q = 0.17$ are shown in figs. (\ref{kn33}) and (\ref{kn34}). Once again, we notice the critical isopotential and an absence of large black hole branch for $\omega$ greater than one.

\begin{figure}[!ht]
\begin{minipage}[b]{0.5\linewidth}
\centering
\includegraphics[width=2.7in,height=2.7in]{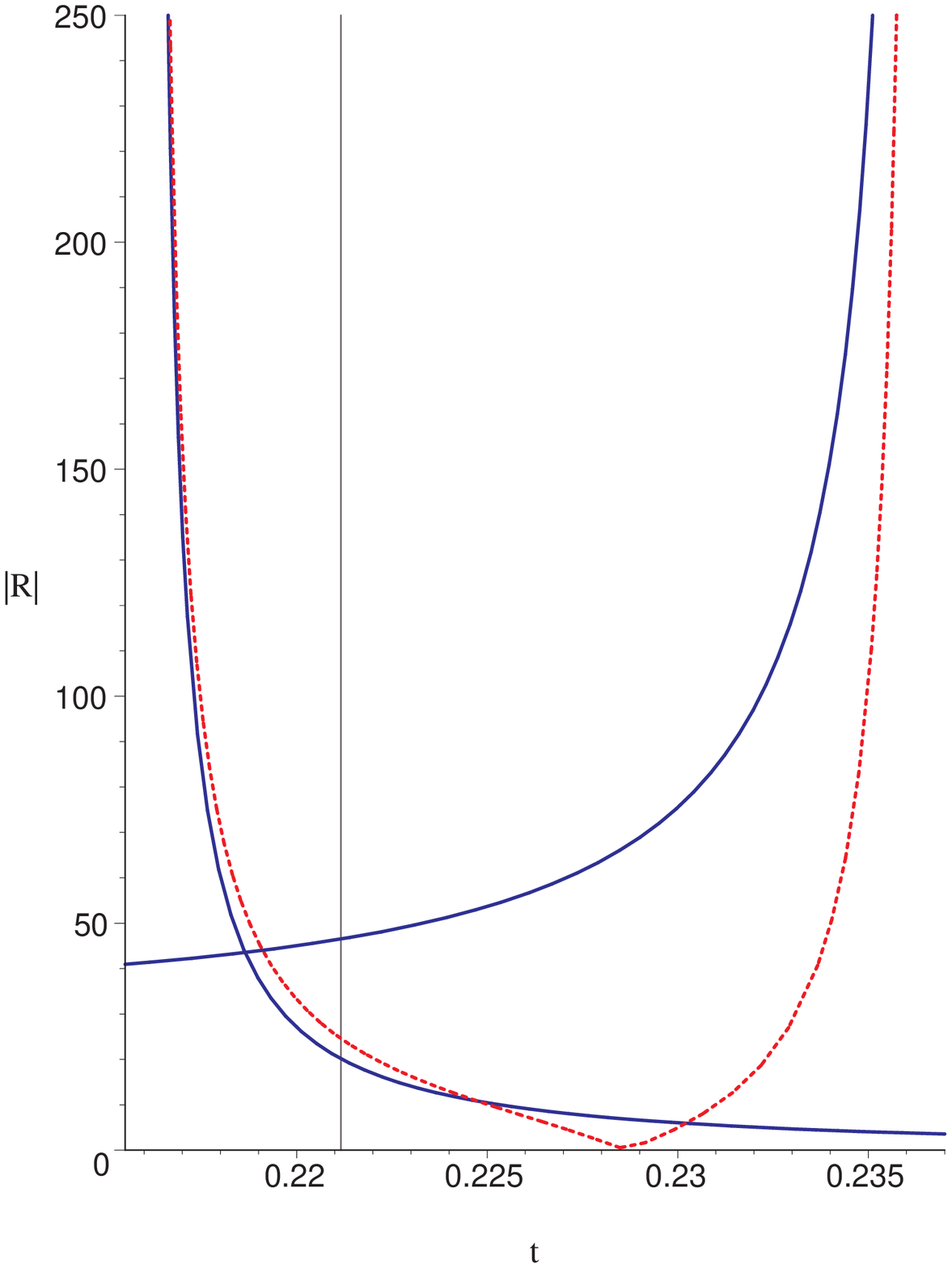}
\caption{Behaviour of $|R|$ for $q=0.16$ and $\omega = 0.8$ in the fixed $q$ ensemble.}
\label{kn35}
\end{minipage}
\hspace{0.6cm}
\begin{minipage}[b]{0.5\linewidth}
\centering
\includegraphics[width=2.7in,height=2.7in]{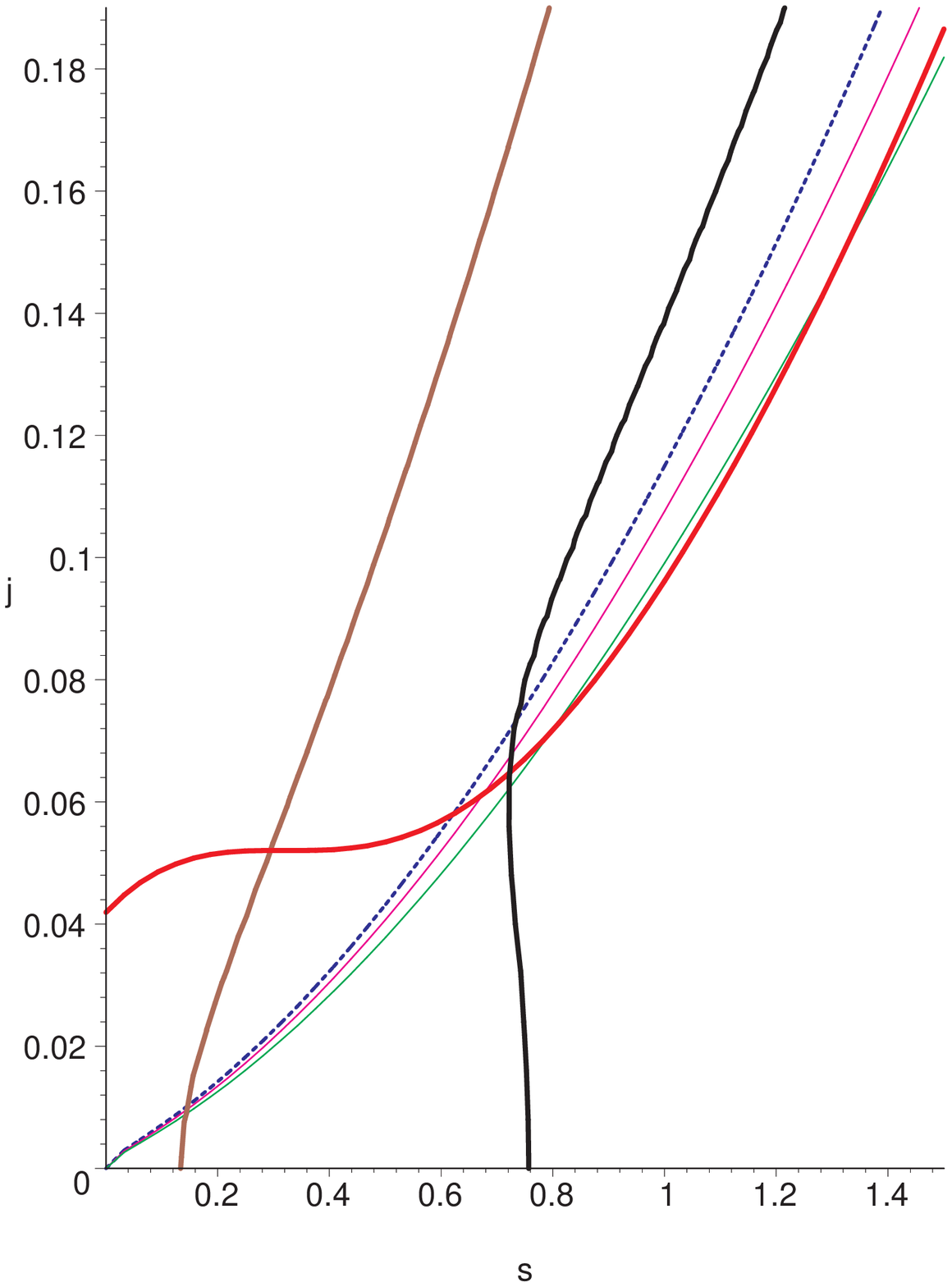}
\caption{Plot of zeroes and infinities of $R$ for $q=0.22$ ensemble.}
\label{kn36}
\end{minipage}
\end{figure}

\begin{figure}[!ht]
\begin{minipage}[b]{0.5\linewidth}
\centering
\includegraphics[width=2.7in,height=2.7in]{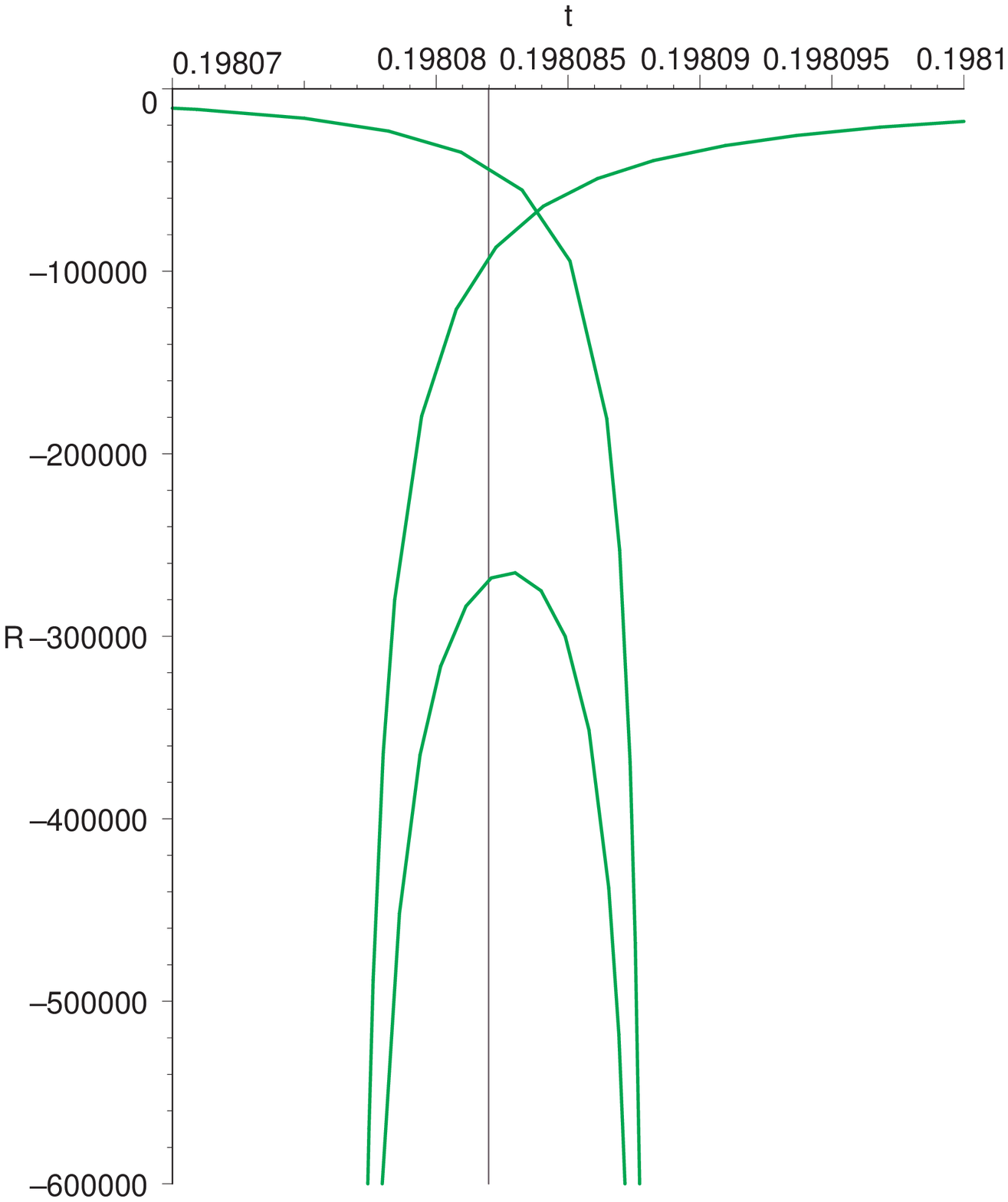}
\caption{Behaviour of $|R|$ for $q=0.16$ and $\omega = 0.8$ in the fixed $q$ ensemble.}
\label{kn37}
\end{minipage}
\hspace{0.6cm}
\begin{minipage}[b]{0.5\linewidth}
\centering
\includegraphics[width=3.0in,height=2.7in]{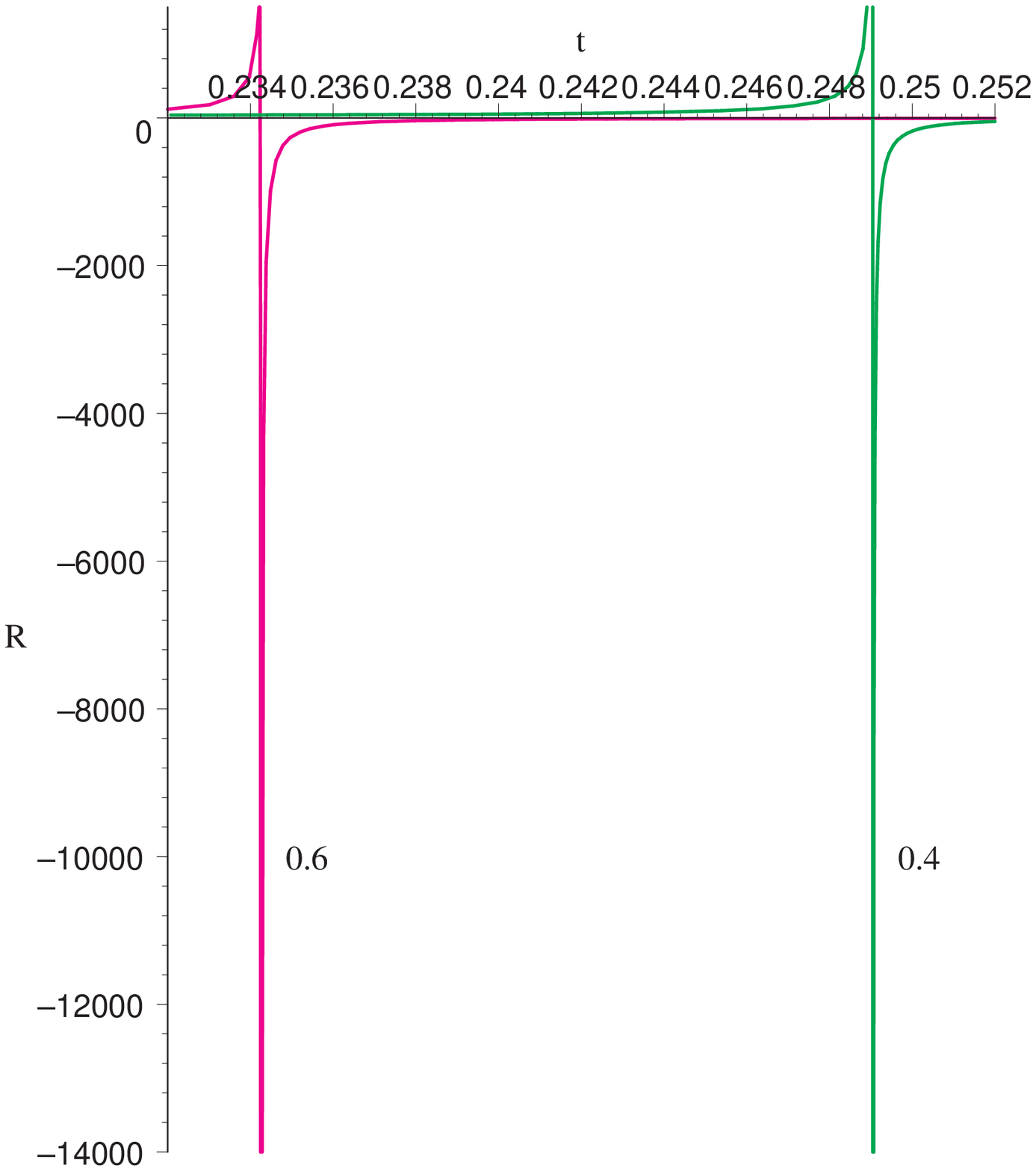}
\caption{Divergence of $|R|$ at critical values of $q$ for $\omega = 0.6$ and $0.4$.}
\label{kn38}
\end{minipage}
\end{figure}

It now remains to calculate the Ruppeiner curvature $R$ for this ensemble.
It can once again be verified that the positivity of the line element is ensured in our region of interest. 
Once again the curvature is a lengthy expression, but it has features similar to the constant $j$ case. Thus, it diverges at extremality by changing its sign and along 
the $c_{\omega}$ curve without change in signature. Besides, it goes to zero in the limit of $s$, and therefore temperature, going to infinity. Fig (\ref{kn35}) is a plot 
of $|R|$ vs $t$ in for $q=0.16$. It can serve as a representative plot for all points in the $q<q_c$ region. We observe that the free energy line corresponding to the 
transition temperature too at a small relative distance on the right side of the intersection of the two stable branches. As a result, this seems to tell us that at the first 
order phase transition point the large black hole branch has a lower `correlation volume' than the small black hole branch. In order to investigate the $q>q_c$ region 
we first make a plot of zeroes and infinities of $R$ similar to fig(\ref{kn11}) for the constant $q$ ensemble. Thus, in fig(\ref{kn36}) for $q=0.22$ ensemble, the 
curvature is positive between the extremal (brown) line and the line $R=0$ (black line), while negative on the right side. Three isopotentials have been shown. with 
the blue dotted one having $\omega=1$. We can notice that the lowest green colored isopotential, which is slightly above criticality, crosses the $c_{\omega}$ 
curve both the times to the right of $R=0$ line. Thus, just like in the constant $j$ ensemble, we once again recover a range of potentials ($\omega$ here) near 
the critical isoptential, such that the behavior of the scalar curvature $R$ closely parallels that  of the Van der Waals gas near the transition region. We draw one 
such plot in fig(\ref{kn37}).

In the figure we clearly see that the free energy line is now to the left of the intersection, so that the correlation volume increases from the small black holes to the large black holes. As we approach the critical point, the unstable and metastable branches get pushed away to infinity and $R$ becomes a single valued function with a divergence at the critical temperature.

Finally, in fig. (\ref{kn38}), we show the divergence of the Ruppeiner curvature for $\phi = 0.6$ and $0.4$ at the critical value of $q$. We see that as in the
fixed $j$ ensemble, the critical temperature reduces with increasing $\omega$. 

\section{Conclusions and Summary}

In this paper, we have examined in details various aspects of thermodynamic geometry at phase transitions and critical points in the thermodynamics of 
4-D KN-AdS black holes. Admittedly, in the absence of complete analytical control, given the algebraic complexity of the expressions involved in the analysis, 
much of our results have been illustrated graphically. However, this has, we believe, brought out a host of new features not studied before in the literature. 
Let us first recall our main results. 

To begin with, we had a relook at the thermodynamic geometry of the Van der Waals gas. Our motivation was that, as has been known, many features of 
asymptotically AdS black hole thermodynamics closely resemble this system, and we find a confirmation of the same from the point of view of Ruppeiner geometry 
as well. In this study, we came across some novel results in thermodynamic geometry, in particular in relation to first order phase transitions. We
studied these in details for the Van der Waals example, and established that the Ruppeiner curvature encodes in a very specific way, the behaviour of the
system at a first order phase transition along the vapour pressure curve, that culminates in a critical point. 

Thereafter, we studied the KN-AdS black hole in the grand canonical, as well as the mixed ensembles where one of the thermodynamic charges were 
held fixed. These latter cases also, to the best of our knowledge, have not been reported in the literature. We have uncovered novel phase behaviour in
these mixed ensembles, and we have seen that in these, 
the system undergoes a first order phase transition along the analogue of the vapour pressure curve in the Van der Waals system and
that it culminates at a critical point. 

\begin{figure}[!ht]
\centering
\includegraphics[width=3in,height=2.5in]{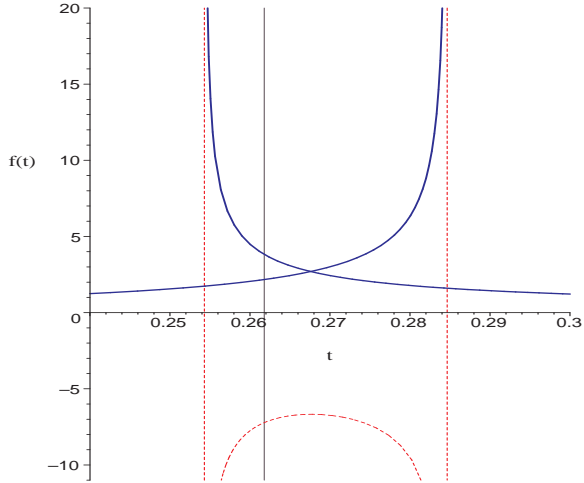}
\caption{Relative mean fluctuation of energy as a function of temperature in the fixed $j$ ensemble. 
for $j = 0.011$ and $\phi = 0.375$}
\label{final}
\end{figure}

One of the most striking features of our analysis is the behaviour of the Ruppiner curvature at first order phase transitions. We have seen that, given the
interpretation of the curvature as a correlation volume, one can make interesting predictions about the stability of the system.  
For example, the Ruppiener curvature $R$ seemingly captures more than just the Gaussian fluctuations in the system. Note that for a conventional
thermodynamic system, $R$ is always negative. A larger (less negative) value of $R$ is then interpreted in terms of greater stability of the system \cite{rupp}. 
However, for black holes, we have seen that the curvature usually changes sign. Given the interpretation of $|R|$ as a correlation volume, it would therefore seem 
more reasonable that a more stable system would correspond to a smaller value of $|R|$. This would seem to imply that in fig. (\ref{kn17}) for example,
a first order phase transition drives the system from a small black hole to a more stable large black hole branch. Taking into account only second order
moments of the fluctuation in the energy seem to give a somewhat different picture. In fig. (\ref{final}), we have plotted the relative mean fluctuation
of the energy $\frac{\delta\left(E^2\right)}{E^2}$ as a function of the temperature $t$ in the fixed $j$ ensemble. In the figure, the two blue lines represent the two 
stable black hole branches. The red curves indicates $\frac{\delta\left(E^2\right)}{E^2}$ in the unstable region, which is negative indicating unphysicality. 
From this figure, we see that the relative mean fluctuation increases as the black hole makes a jump in first order phase transition, i.e the black hole jumps
to a less stable branch (along the vertical free energy line). This discrepancy is probably due to the fact that the Ruppeiner curvature captures more 
information than just the Gaussian fluctuations. We defer this issue for future investigations. 

Also, note that our analysis conforms to the behaviour of the scalar curvature for critical phenomena in standard thermodynamics. In the latter, 
$R$ is always negative in the neighborhood of the critical point. We have been able to establish that in the mixed ensembles that we have considered
in this paper, there exists a finite neighborhood of the critical point where $R$ is negative and it diverges to $-\infty$ at the critical point. 
Finally, it should be emphasized that characterisation of first order phase transitions is an important issue in the thermodynamics of condensed matter
systems. We believe that our qualification of Ruppeiner's original assertion vis a vis first order phase transitions should find applications in this exciting
branch of research. 

\begin{center}
{\bf Acknowledgements}
\end{center}
\noindent
The authors would like to thank V. Subrahmanyam for discussions. A.S would like to acknowledge useful discussions with
Andrew Cabral and Pankaj Mishra.


\begin{thebibliography}{99}
\bibitem{td1}
  R.~M.~Wald,
  ``The thermodynamics of black holes,''
  Living Rev.\ Rel.\  {\bf 4}, 6 (2001)
  [arXiv:gr-qc/9912119].
\bibitem{td2}
D.~N.~Page,
  ``Hawking radiation and black hole thermodynamics,''
  New J.\ Phys.\  {\bf 7}, 203 (2005)
  [arXiv:hep-th/0409024].
  \bibitem{td3}
  R.~Brout, S.~Massar, R.~Parentani and Ph.~Spindel,
  ``A Primer for Black Hole Quantum Physics,''
  Phys.\ Rept.\  {\bf 260}, 329 (1995)
  [arXiv:0710.4345 [gr-qc]].
\bibitem{td4}
See, e.g. R. M. Wald, ``Black Hole Entropy is Noether Charge,'' Phys. Rev. {\bf D 48} (1993), 3427,
[arXiv:gr-qc/9307038]
 \bibitem{td5}
 V. Iyer, R. M. Wald, ``Some Properties of Noether Charge and  Proposal for Dynamical Black Hole Entropy,''
 Phys. Rev. {\bf D50} (1994) 846, [arXiv: gr-qc/9403028]
 \bibitem{td6}
  P. C. W Davies, Rep. Prog. Phys {\bf 41} (1977) 1313, Class. Quant. Gravity {\bf 6} (1989) 1909.
  \bibitem{calda}
M.~M.~Caldarelli, G.~Cognola and D.~Klemm,
``Thermodynamics of Kerr-Newman-AdS black holes and conformal field
theories,''
Class.\ Quant.\ Grav.\  {\bf 17}, 399 (2000)
[arXiv:hep-th/9908022].
  \bibitem{td7}
C.~O.~Lousto,
  ``The Fourth law of black hole thermodynamics,''
  Nucl.\ Phys.\  B {\bf 410}, 155 (1993)
  [Erratum-ibid.\  B {\bf 449}, 433 (1995)]
  [arXiv:gr-qc/9306014].
\bibitem{td8}
  D.~Pavon and J.~M.~Rubi,
  ``Nonequilibrium Thermodynamic Fluctuations of Black Holes,''
  Phys.\ Rev.\  D {\bf 37}, 2052 (1988).
\bibitem{td9}
A. Curir, ``Rotating black holes as dissipative spin-thermodynamical systems,''
Gen. Rel. Grav. {\bf 13} (1981), 417.
\bibitem{td10}
C.~O.~Lousto,
  ``The Emergence of an effective two-dimensional quantum description from the
  study of critical phenomena in black holes,''
  Phys.\ Rev.\  D {\bf 51}, 1733 (1995)
  [arXiv:gr-qc/9405048].
\bibitem{td11}
R.~G.~Cai and Y.~S.~Myung,
  ``Critical behavior for the dilaton black holes,''
  Nucl.\ Phys.\  B {\bf 495}, 339 (1997)
  [arXiv:hep-th/9702159].
\bibitem{td12}
R.~G.~Cai, Z.~J.~Lu and Y.~Z.~Zhang,
  ``Critical behavior in 2+1 dimensional black holes,''
  Phys.\ Rev.\  D {\bf 55}, 853 (1997)
  [arXiv:gr-qc/9702032].
\bibitem{td12a}
G.~Arcioni and E.~Lozano-Tellechea,
 ``Stability and critical phenomena of black holes and black rings,''
  Phys.\ Rev.\  D {\bf 72}, 104021 (2005)
  [arXiv:hep-th/0412118].
\bibitem{td13}
A.~Chamblin, R.~Emparan, C.~V.~Johnson and R.~C.~Myers,
``Charged AdS black holes and catastrophic holography,''
Phys.\ Rev.\  D {\bf 60}, 064018 (1999)
[arXiv:hep-th/9902170].
\bibitem{td14}
A.~Chamblin, R.~Emparan, C.~V.~Johnson and R.~C.~Myers,
  ``Holography, thermodynamics and fluctuations of charged AdS black holes,''
  Phys.\ Rev.\  D {\bf 60}, 104026 (1999)
  [arXiv:hep-th/9904197].
\bibitem{malda}
O.~Aharony, S.~S.~Gubser, J.~M.~Maldacena, H.~Ooguri and Y.~Oz,
  ``Large N field theories, string theory and gravity,''
  Phys.\ Rept.\  {\bf 323}, 183 (2000)
  [arXiv:hep-th/9905111].
\bibitem{tisza}
L. Tisza, ``Generalized Thermodynamics,'' Pub. MIT Press, Cambridge, MA (1966)
\bibitem{callen}
H. B. Callen, ``Thermodynamics and an Introcution to Thermostatitics,'' Pub. Wiley, New York (1985)
\bibitem{weinhold}
F. Weinhold, J. Chem Phys. {\bf 63} (1075) 2479, {\it ibid} J. Chem Phys. {\bf 63} (1975) 2484.
\bibitem{rupp}
G. Ruppeiner, Rev. Mod. Phys. {\bf 67} (1995) 605, erratum {\it ibid} {\bf 68} (1996) 313.
\bibitem{ferrara}
S.~Ferrara, G.~W.~Gibbons and R.~Kallosh,
  ``Black holes and critical points in moduli space,''
  Nucl.\ Phys.\  B {\bf 500}, 75 (1997)
  [arXiv:hep-th/9702103].
\bibitem{aman}
J.~E.~Aman, I.~Bengtsson and N.~Pidokrajt,
  ``Geometry of black hole thermodynamics,''
  Gen.\ Rel.\ Grav.\  {\bf 35}, 1733 (2003)
  [arXiv:gr-qc/0304015].
\bibitem{caibtz}
R.~G.~Cai and J.~H.~Cho,
  ``Thermodynamic curvature of the BTZ black hole,''
  Phys.\ Rev.\  D {\bf 60}, 067502 (1999)
  [arXiv:hep-th/9803261].
\bibitem{cai1}
J.~y.~Shen, R.~G.~Cai, B.~Wang and R.~K.~Su,
 ``Thermodynamic geometry and critical behavior of black holes,''
  Int.\ J.\ Mod.\ Phys.\  A {\bf 22}, 11 (2007)
  [arXiv:gr-qc/0512035].
\bibitem{ts1}
T.~Sarkar, G.~Sengupta and B.~Nath Tiwari,
  ``On the thermodynamic geometry of BTZ black holes,''
  JHEP {\bf 0611}, 015 (2006)
  [arXiv:hep-th/0606084].
\bibitem{ts2}
T.~Sarkar, G.~Sengupta and B.~N.~Tiwari,
  ``Thermodynamic Geometry and Extremal Black Holes in String Theory,''
  JHEP {\bf 0810}, 076 (2008)
  [arXiv:0806.3513 [hep-th]].
\bibitem{diosi}
L. Diosi, B. Lukacs, A. Racz, ``Mapping the Van der Waals State Space,'' J. Chem. Phys. {\bf 91} (1989) 3061, 
L. Diosi, B. Lukacs, ``Spatial Correlation in Diluted Gases from the Viewpoint of the Metric of the Thermodynamic
State Space,'' J. Chem. Phys. {\bf 84} (1986) 5081.
\bibitem{ruppvw}
G. Ruppeiner, ``Thermodynamic curvature : origin and meaning,'' in {\it Advances in Thermodynamics, Vol. 3},
Edited by Sieniutycz and Salamon (1990).\\
H. Janyszek, ``Riemannian geometry and stability of thermodynamical equilibrium systems,''
J. Phys. A, {\bf 23}, 477 (1990).
\bibitem{rupp2}
G. Ruppeiner, ``Thermodynamics : a Riemannian Geometric Model,'' Phys. Rev. {\bf A20} (1979) 1608,\\
``Thermodynamic Critical Fluctuation Theory ?'' Phys. Rev. Lett. {\bf 50} (1983) 287,\\
``Thermodynamic Curvature of the Multicomponent Ideal Gas,'' Phys. Rev. {\bf A41} (1990) 2200.
\bibitem{santoro}
M. Santoro, A. S. Benight, ``On the Geometrical Thermodynamics of Chemical Reactions,'' [arXiv: math-ph/0507026]
\bibitem{janyszek1}
H. Janyszek, R. Mrugala, ``Geometrical Structure of the State Space in Classical Statistical and Phenomenological Thermodynamics,''
Rep. Math. Phys. {\bf 27} (1989) 145.
\bibitem{landau}
L. Landau and E. M. Lifshitz, ``Statistical Physics,'' Pergamon Press (1988)
\bibitem{bala}
 V.~Balasubramanian and P.~Kraus,
  ``A stress tensor for anti-de Sitter gravity,''
  Commun.\ Math.\ Phys.\  {\bf 208}, 413 (1999)
  [arXiv:hep-th/9902121].
 \bibitem{gibbons}
 G.~W.~Gibbons and S.~W.~Hawking,
 ``Action Integrals And Partition Functions In Quantum Gravity,''
  Phys.\ Rev.\  D {\bf 15}, 2752 (1977).
  \bibitem{carlip}
S.~Carlip and S.~Vaidya,
  ``Phase transitions and critical behavior for charged black holes,''
  Class.\ Quant.\ Grav.\  {\bf 20}, 3827 (2003)
  [arXiv:gr-qc/0306054].
\bibitem{rupp3}
G.~Ruppeiner,
  ``Stability And Fluctuations In Black Hole Thermodynamics,''
  Phys.\ Rev.\  D {\bf 75}, 024037 (2007).
  \bibitem{rupp4}
  G.~Ruppeiner,
  ``Thermodynamic curvature and phase transitions in Kerr-Newman black holes,''
  Phys.\ Rev.\  D {\bf 78}, 024016 (2008)
  [arXiv:0802.1326 [gr-qc]].
\bibitem{mirza}
 B.~Mirza and M.~Zamani-Nasab,
``Ruppeiner Geometry of RN Black Holes: Flat or Curved?,''
JHEP {\bf 0706}, 059 (2007)
 [arXiv:0706.3450 [hep-th]].


\end{thebibliography}
\end{document}